\documentclass[longauth]{aa}  
\usepackage{graphicx}
\usepackage{subcaption}

\usepackage{txfonts}
\usepackage{multirow}
\usepackage{makecell}

\usepackage[colorlinks=true,
            linkcolor=blue,
            citecolor=blue,
            urlcolor=blue]{hyperref}
\begin{document}

   \titlerunning{Beyond traditional emission-line diagnostics: using autoencoders to uncover AGN in DESI spectra}
   \title{Beyond traditional emission-line diagnostics: using autoencoders to uncover active galactic nuclei in DESI spectra}

   %\subtitle{}

\authorrunning{Alcolea, J.A. et al.}

\author{%
Alcolea, J. A.\inst{1} \corrauth{joseantalcolea@gmail.com}
\and
Siudek, M.\inst{2} \corrauth{msiudek@iac.es}
\and
Eriksen, M.\inst{3,4} \email{eriksen@pic.es}
\and
Mezcua, M.\inst{1,5} \corrauth{mezcua@ice.csic.es}
\and
Pucha, R.\inst{6,7} \email{dr.raga.pucha@gmail.com}
\and
Juneau, S.\inst{8} \email{stephanie.juneau@noirlab.edu}
\and
Gontcho A Gontcho, S.\inst{9} \email{satya@virginia.edu}
\and
Panda, S.\inst{10} \email{swayamtrupta.panda@noirlab.edu}
\and
Aguilar, J.\inst{11} \email{jaguilar@lbl.gov}
\and
Ahlen, S.\inst{12} \email{ahlen@bu.edu}
\and
Bianchi, D.\inst{13,14} \email{davide.bianchi1@unimi.it}
\and
Brodzeller, A.\inst{11} \email{allysonbrodzeller@lbl.gov}
\and
Brooks, D.\inst{15} \email{david.brooks@ucl.ac.uk}
\and
Castander, F. J.\inst{1,5} \email{fjc@ice.csic.es}
\and
Claybaugh, T.\inst{11} \email{tmclaybaugh@lbl.gov}
\and
Cuceu, A.\inst{11} \email{acuceu@lbl.gov}
\and
de la Macorra, A.\inst{16} \email{macorra@fisica.unam.mx}
\and
Dey, B.\inst{17,18} \email{b.dey@utoronto.ca}
\and
Doel, P.\inst{15} \email{apd@star.ucl.ac.uk}
\and
Ferraro, S.\inst{11,19} \email{sferraro@lbl.gov}
\and
Font-Ribera, A.\inst{3,20} \email{afont@ifae.es}
\and
Forero-Romero, J. E.\inst{21,22} \email{je.forero@uniandes.edu.co}
\and
Gaztañaga, E.\inst{1,5,23} \email{gaztanaga@gmail.com}
\and
Gutierrez, G.\inst{24} \email{gaston@fnal.gov}
\and
Hahn, C.\inst{25} \email{changhoon.hahn@utexas.edu}
\and
Herrera-Alcantar, H. K.\inst{26,27} \email{herreraa@iap.fr}
\and
Joyce, D.\inst{8} \email{richard.joyce@noirlab.edu}
\and
Kehoe, R. \inst{28} \email{kehoe@physics.smu.edu}
\and
Kirkby, D.\inst{29} \email{dkirkby@uci.edu}
\and
Kisner, T.\inst{11} \email{tskisner@lbl.gov}
\and
Kremin, A.\inst{11} \email{akremin@lbl.gov}
\and
Lahav, O.\inst{15} \email{o.lahav@ucl.ac.uk}
\and
Lamman, C.\inst{30} \email{lamman.1@osu.edu}
\and
Landriau, M.\inst{11} \email{mlandriau@lbl.gov}
\and
Le Guillou, L.\inst{31} \email{llg@lpnhe.in2p3.fr}
\and
Meisner, A.\inst{8} \email{aaron.meisner@noirlab.edu}
\and
Miquel, R.\inst{3,20} \email{rmiquel@ifae.es}
\and
Moustakas, J.\inst{32} \email{jmoustakas@siena.edu}
\and
Nadathur, S.\inst{23} \email{seshadri.nadathur@port.ac.uk}
\and
Percival, W.\inst{33,34,35} \email{will.percival@uwaterloo.ca}
\and
Prada, F.\inst{36} \email{fprada@iaa.es}
\and
P\'erez-Ráfols, I.\inst{37} \email{ignasi.perez.rafols@upc.edu}
\and
Rossi, G.\inst{38} \email{graziano@sejong.ac.kr}
\and
Sanchez, E.\inst{39} \email{eusebio.sanchez@ciemat.es}
\and
Schlafly, E.\inst{40} \email{eschlafly@stsci.edu}
\and
Schlegel, D.\inst{11} \email{djschlegel@lbl.gov}
\and
Schubnell, M.\inst{41} \email{schubnel@umich.edu}
\and
Silber, J.\inst{11} \email{jhsilber@lbl.gov}
\and
Sprayberry, D.\inst{8} \email{david.sprayberry@noirlab.edu}
\and
Tarlé, G.\inst{41} \email{gtarle@umich.edu}
\and
Weaver, B. A.\inst{8} \email{benjamin.weaver@noirlab.edu}
\and
Zou, H.\inst{42} \email{zouhu@nao.cas.cn}
}
 
\institute{ 
\inst{1} Institute of Space Sciences (ICE, CSIC), Campus UAB, Carrer de Magrans, 08193 Barcelona, Spain\\ 
\inst{2} Instituto de Astrofísica de Canarias (IAC), Departamento de Astrofísica, Universidad de La Laguna (ULL), 38200, La Laguna, Tenerife, Spain \\ 
\inst{3} Institut de Física d’Altes Energies (IFAE), The Barcelona Institute of Science and Technology, 08193 Bellaterra (Barcelona), Spain \\ 
\inst{4} Port d'Informaci\'{o} Cient\'{i}fica (PIC), Campus UAB, C. Albareda s/n, 08193 Bellaterra (Barcelona), Spain \\ 
\inst{5} Institut d’Estudis Espacials de Catalunya (IEEC), Edifici RDIT, Campus UPC, 08860 Castelldefels (Barcelona), Spain \\ 
\inst{6} Department of Physics and Astronomy, University of Utah, 115 South 1400 East, Salt Lake City, UT 84112, USA \\ 
\inst{7} Steward Observatory, University of Arizona, 933 North Cherry Avenue, Tucson, AZ 85721, USA \\
\inst{8} NSF NOIRLab, 950 N. Cherry Ave., Tucson, AZ 85719, USA\\
\inst{9} University of Virginia, Department of Astronomy, Charlottesville, VA 22904, USA\\
\inst{10}  International Gemini Observatory/NSF NOIRLab, Casilla 603, La Serena, Chile
\inst{11} Lawrence Berkeley National Laboratory, 1 Cyclotron Road, Berkeley, CA 94720, USA \\
\inst{12} Department of Physics, Boston University, 590 Commonwealth Avenue, Boston, MA 02215 USA\\
\inst{13} Dipartimento di Fisica ``Aldo Pontremoli'', Universit\`a degli Studi di Milano, Via Celoria 16, I-20133 Milano, Italy\\
\inst{14} INAF-Osservatorio Astronomico di Brera, Via Brera 28, 20122 Milano, Italy\\
\inst{15} Department of Physics \& Astronomy, University College London, Gower Street, London, WC1E 6BT, UK\\
\inst{16} Instituto de F\'{\i}sica, Universidad Nacional Aut\'{o}noma de M\'{e}xico,  Circuito de la Investigaci\'{o}n Cient\'{\i}fica, Ciudad Universitaria, Cd. de M\'{e}xico  C.~P.~04510,  M\'{e}xico\\
\inst{17} Department of Astronomy \& Astrophysics, University of Toronto, Toronto, ON M5S 3H4, Canada\\
\inst{18} Department of Physics \& Astronomy and Pittsburgh Particle Physics, Astrophysics, and Cosmology Center (PITT PACC), University of Pittsburgh, 3941 O'Hara Street, Pittsburgh, PA 15260, USA\\
\inst{19} University of California, Berkeley, 110 Sproul Hall \#5800 Berkeley, CA 94720, USA\\
\inst{20} Instituci\'{o} Catalana de Recerca i Estudis Avan\c{c}ats, Passeig de Llu\'{\i}s Companys, 23, 08010 Barcelona, Spain \\
\inst{21} Departamento de F\'isica, Universidad de los Andes, Cra. 1 No. 18A-10, Edificio Ip, CP 111711, Bogot\'a, Colombia\\
\inst{22} Observatorio Astron\'omico, Universidad de los Andes, Cra. 1 No. 18A-10, Edificio H, CP 111711 Bogot\'a, Colombia\\
\inst{23} Institute of Cosmology and Gravitation, University of Portsmouth, Dennis Sciama Building, Portsmouth, PO1 3FX, UK\\
\inst{24} Fermi National Accelerator Laboratory, PO Box 500, Batavia, IL 60510, USA \\
\inst{25} Department of Astronomy, University of Texas at Austin, 2515 Speedway, TX 78712, USA \\
\inst{26} Institut d'Astrophysique de Paris. 98 bis boulevard Arago. 75014 Paris, France \\
\inst{27} IRFU, CEA, Universit\'{e} Paris-Saclay, F-91191 Gif-sur-Yvette, France\\
\inst{28} Department of Physics, Southern Methodist University, 3215 Daniel Avenue, Dallas, TX 75275, USA\\
\inst{29} Department of Physics and Astronomy, University of California, Irvine, 92697, USA\\
\inst{30} The Ohio State University, Columbus, 43210 OH, USA\\
\inst{31} Sorbonne Universit\'{e}, CNRS/IN2P3, Laboratoire de Physique Nucl\'{e}aire et de Hautes Energies (LPNHE), FR-75005 Paris, France\\
\inst{32} Department of Physics and Astronomy, Siena University, 515 Loudon Road, Loudonville, NY 12211, USA\\
\inst{33} Department of Physics and Astronomy, University of Waterloo, 200 University Ave W, Waterloo, ON N2L 3G1, Canada\\
\inst{34} Perimeter Institute for Theoretical Physics, 31 Caroline St. North, Waterloo, ON N2L 2Y5, Canada\\
\inst{35} Waterloo Centre for Astrophysics, University of Waterloo, 200 University Ave W, Waterloo, ON N2L 3G1, Canada\\
\inst{36} Instituto de Astrof\'{i}sica de Andaluc\'{i}a (CSIC), Glorieta de la Astronom\'{i}a, s/n, E-18008 Granada, Spain\\
\inst{37} Departament de F\'isica, EEBE, Universitat Polit\`ecnica de Catalunya, c/Eduard Maristany 10, 08930 Barcelona, Spain\\
\inst{38} Department of Physics and Astronomy, Sejong University, 209 Neungdong-ro, Gwangjin-gu, Seoul 05006, Republic of Korea\\
\inst{39} CIEMAT, Avenida Complutense 40, E-28040 Madrid, Spain\\
\inst{40} Space Telescope Science Institute, 3700 San Martin Drive, Baltimore, MD 21218, USA\\
\inst{41} University of Michigan, 500 S. State Street, Ann Arbor, MI 48109, USA\\
\inst{42} National Astronomical Observatories, Chinese Academy of Sciences, A20 Datun Road, Chaoyang District, Beijing, 100101, P.~R.~China
}

   \date{Received X; }

\abstract
   {The growing volume of spectroscopic data in modern surveys motivates the development of data-driven approaches that complement traditional emission-line diagnostics for active galactic nuclei (AGN) identification.}
   {We aim to develop and validate a machine learning (ML) framework for galaxy and AGN classification that exploits the full optical spectrum, using unsupervised learning to capture spectral features within a semi-supervised classification framework that complements classical approaches.}
   {We adapt the \software{SPENDER} autoencoder architecture to compress Dark Energy Spectroscopic Instrument (DESI) galaxy spectra into a low-dimensional latent space that encodes physically meaningful, label-agnostic information. We introduce a quantitative classification method via k-d tree nearest-neighbor search in latent space. The model is trained on 50,222 spectra from the DESI Main Survey, corresponding to the internal \textit{Guadalupe} dataset and released as part of Data Release 1 (DR1), and restricted to $z \leq 0.5$. We validate the performance using labels derived from \software{FastSpecFit}'s emission line measurements defining seven galaxy classes: AGN, broad-line (BL), composite, star-forming, passive, retired, and Other.}
    {The model achieves high performance in identifying AGN and BL sources, with accuracies of 0.952 and 0.965, respectively. The model reliably recovers AGN and BL sources even in low signal-to-noise ratio (S/N) spectra and identifies AGN missed by single-diagnostic standard methods. Our classification metrics are benchmarked against traditional diagnostics and we show they represent lower limits of the model’s true performance. We find that the latent space learned by the autoencoder not only separates AGN and non-AGN galaxies, but also correlates with key physical properties such as stellar mass and star-formation rate (SFR), demonstrating that the method captures astrophysical meaningful features. }
   {This work demonstrates that unsupervised spectral representation learning, implemented within a semi-supervised classification framework, offers a powerful and scalable pathway toward more complete and reliable AGN catalogs for current and upcoming large spectroscopic surveys.}


   \maketitle
%
%-------------------------------------------------------------------

\section{Introduction}
\label{sec:Intro}

%-------------------------------------------------------------
% 1. Active Galactic Nuclei (AGN): What they are and why they matter
%-------------------------------------------------------------

Active galactic nuclei (AGN) are among the most powerful and energetic sources in the Universe \citep{AGN}. They are powered by accretion of material onto supermassive black holes (SMBHs; $M_{\rm BH} \geq 10^6 M_{\odot}$) located at the center of galaxies \citep{salpeter, zel}. AGN are found in $\sim$1-10\% of all galaxies \citep{beckmann} and emit across the entire electromagnetic spectrum, from radio to gamma-rays. 
Studying AGN is crucial for understanding both galaxy evolution and black hole growth \citep{coev, harrison}. For instance, AGN feedback plays a crucial role in regulating, and possibly halting altogether, star formation processes within massive galaxies \citep{feedback}. This phenomenon is commonly incorporated into cosmological simulations to ensure that the simulated galaxy properties align with observations \citep{feedsim, Vogelsberger_2014}. Moreover, studying AGN hosted in dwarf galaxies can provide insights into understanding the formation of seed black holes in the early Universe \citep{Mezcua_2017}. The study of AGN can even constrain cosmological parameters and test whether dark energy density varies with time \citep{agnDM}.

Despite the large variety of AGN classes defined over the years — the so-called AGN zoo — the unified model of AGN proposes a common physical structure \citep{Unified1, Unified2, beckmann}, with observational differences arising mainly from the orientation, accretion rate, presence of jets, and properties of the host galaxy \citep{AGN}. A particularly interesting type of AGN are the broad-line (BL) AGN, whose spectra exhibit broad emission lines (e.g., $H\alpha$, $H\beta$). These lines originate in the broad-line region (BLR), a compact and fast-moving gas region located close to the SMBH \citep{AGN}. The BLR not only enables SMBH mass estimates through virial techniques \citep{greeneBH, MoranBH, suhBH}, but also offers valuable information on accretion processes, the structure of the inner AGN, and the co-evolution of SMBHs with their host galaxies \citep{netzer2015, shen2019}. It is worth noting that transient broad emission lines can also be produced by stellar processes such as Type II supernovae, luminous blue variables, or Wolf-Rayet stars, particularly in low-mass galaxies, and that follow-up spectroscopy is often required to distinguish between AGN activity and these transient stellar phenomena \citep{Baldassare_2016}.

%-------------------------------------------------------------
% 2. Classical AGN identification: spectroscopy and its limitations
%-------------------------------------------------------------
AGN identification methods exploit their multi-wavelength emission. While X-ray, radio, infrared (IR), or gamma-ray observations can effectively reveal AGN activity \citep{AGN}, optical spectroscopy remains the primary tool for AGN selection in large extragalactic surveys and has been shown to provide a more complete AGN census \citep{Juneau2013}. The standard approach to spectroscopic AGN classification relies on emission-line diagnostics such as the Baldwin–Phillips–Terlevich (BPT) diagram \citep{bpt}, which uses ratios of prominent emission lines (e.g., $[\ion{O}{III}]/H\beta$ vs. $[\ion{N}{II}]/H\alpha$) to distinguish between ionization mechanisms driven by AGN, star formation, or a combination of both. However, these methods typically focus on a limited number of emission lines, requiring prior spectral fitting and often missing the broader spectral diversity of galaxies. Moreover, establishing a uniform selection method remains challenging, as different techniques for identifying AGN often yield distinct and only partially overlapping samples \citep{Juneau2013, AGN, AGN_obscure}.

Large spectroscopic surveys have transformed our view of galaxy populations over the last two decades. Projects such as the Sloan Digital Sky Survey (SDSS; \citealp{dlsdss}) and, more recently, the Dark Energy Spectroscopic Instrument\footnote{\href{https://www.desi.lbl.gov/}{https://www.desi.lbl.gov/}} (DESI; \citealp{DR1}) are delivering optical spectra for millions of galaxies, enabling statistical studies of unprecedented scope.

Even with only the Early Data Release (EDR; \citealp{EDR}), DESI has already led to major scientific advances using classical analysis techniques, based on a sample of approximately 1.8 million unique targets. For instance, \citet{Siudek_2024} produced a value-added catalog with physical properties for more than 1.3 million galaxies using \software{CIGALE} spectral energy distribution fitting \citep{cigale_boquien}, and explored its use as a proxy for AGN identification \citep{Siudek_2025}. Similarly, \citet{raga2025} tripled the census of dwarf AGN candidates through emission-line fitting, while \citet{guo_CLAGN} increased the known population of changing-look AGN by approximately 30\%. Additional studies revealed new connections between dust reddening and radio emission in quasars \citep{fawcett} and provided improved single-epoch black hole mass estimates for over 55,000 DESI quasars \citep{Pan2025}.  These results highlight the scientific power of DESI even when exploited with traditional, model-based approaches.

DESI Data Release 1 (DR1; \citealp{DR1}) has already dramatically increased both the volume and diversity of available spectra, including more than 18.7 million unique targets. Forthcoming releases, such as Data Release 2 (DR2; \citealp{DR2}),  are expected to expand the dataset to more than 40 million sources in the coming years. This growth will be further amplified by future massive spectroscopic surveys such as 4MOST \citep{4most}, the Prime Focus Spectrograph (PFS; \citealp{pfs}) and WEAVE \citep{WEAVE}, which will collectively provide spectroscopic observations for millions of additional sources.

Extracting physical information from galaxy spectra at this scale is inherently challenging. Classical model-fitting and line-based analysis techniques are computationally intensive and can become major bottlenecks when applied to datasets of this size and complexity \citep{SpecPT}. These limitations naturally motivate the adoption of machine learning (ML) approaches for spectroscopic analysis in astronomy \citep{MMU, Siudek_2025}, which offer the scalability and flexibility required to exploit the full information content of galaxy spectra beyond a limited set of emission-line measurements \citep{huertas_23}. 

%-------------------------------------------------------------
% 3. Machine learning for spectral analysis: unsupervised methods and open challenges
%-------------------------------------------------------------
Briefly, ML algorithms fall into two main categories: supervised methods, which require labeled data, and unsupervised methods, which do not. There are categories in between, such as semi-supervised learning, which combines elements of both supervised and unsupervised learning, typically by learning representations from unlabeled data and subsequently using labeled information to guide classification. Several recent studies have demonstrated the potential of these methods (e.g. \citealp{VIMOS_a,VIMOS_b}, \citealp{Hviding2024}, \citealp{perez-diaz}), highlighting in particular the power of unsupervised approaches, which allow for the analysis of the full spectrum without prior assumptions or labels. Applying unsupervised ML to galaxy spectra offers several advantages: (i) it enables the exploitation of multiple AGN signatures simultaneously across the full spectral range, (ii) it uncovers natural groupings in a high-dimensional data-driven space, and (iii) it avoids biases introduced by relying on predefined labels or selection criteria \citep{VIMOS_a}. By using the entire spectral information without imposing strict thresholds, unsupervised methods not only enhance AGN identification but also open the door to discovering diverse AGN types based on their intrinsic spectral features.

% Citas otros trabajos en ML a spectros

Several recent studies have demonstrated the potential of ML methods applied to galaxy spectra. For instance, \cite{VIMOS_a, VIMOS_b} applied a Fisher Expectation-Maximization algorithm to the VIMOS Public Extragalactic Redshift Survey, categorizing over 50,000 VIPERS galaxies into 12 classes based on shared physical and spectral properties. This approach led to AGN identification that exceeded the mid-IR classification \citep{Siudek_springer} and allowed the discovery of rare sources missed by standard methods \citep{lisiecki, siudek_22, Mezcua_2023}.

Focusing on dimensionality reduction, \cite{Portillo} showed that an autoencoder—a neural network that compresses data into a low-dimensional representation (encoding) and then reconstructs it (decoding)— can condense SDSS spectra into just six latent parameters while retaining enough information to accurately reconstruct the original spectra. This nonlinear approach outperforms traditional methods like Principal Component Analysis. \cite{Teimoorinia} expanded on this by adding convolutional layers to extract correlated features from the MaNGA survey galaxy spectra, while \cite{Pat} employed a Probabilistic Autoencoder \citep{PAE} on SDSS DR16 to improve spectral reconstruction and detect outliers in the latent space. More recently, \citet{Nicolaou} applied a Variational Autoencoder to approximately 200,000 DESI spectra, demonstrating that it can compress spectral dimensionality by a factor of 100 while retaining enough information to accurately reconstruct features, and successfully identifying anomalous spectra both with artifacts and with unusual physical characteristics.
 
\software{SPENDER} \citep{SPENDER,SPENDER2}—the basis for our work—is an autoencoder designed to analyze and reconstruct galaxy spectra. By encoding observed-frame spectra and applying redshift transformations, it produces a latent representation that enables consistent comparison of spectral features across galaxies at different redshifts. This approach reduces redshift-driven biases and is robust to common artifacts, allowing spectra from large surveys to be ingested directly without extensive preprocessing.

\software{SPENDER} has been successfully applied to a variety of spectroscopic analyses, including accurate reconstructions and noise suppression in SDSS-II data \citep{SPENDER}, outlier detection \citep{SPENDER2}, and the identification of anomalies in the DESI Bright Galaxy Survey, such as a redshift bias in the \software{Redrock}\footnote{\href{https://redrock.readthedocs.io}{https://redrock.readthedocs.io}} algorithm (Bailey et al. (in preparation)) caused by M-dwarf contamination \citep{SPENDERDESI}. It has also been adapted to reconstruct intrinsic quasar (QSO) spectra and to measure the Ly$\alpha$ forest \citep{spenderQ}. By encoding spectra into a few latent variables, this method facilitates data exploration and interpretation, as demonstrated in studies linking optical and IR emission \citep{Spender_wise}. These applications illustrate the growing impact of ML in spectroscopic research, enabling analyses that are difficult to achieve with traditional techniques.

%%This work

In this work, we take a step toward automated AGN identification and galaxy classification by adapting the \software{SPENDER} architecture within a semi-supervised workflow. The autoencoder learns a low-dimensional spectral representation without using labels, while class assignments are derived from reference classifications based on emission-line diagnostics and propagated according to the proximity of similar spectra in the learned latent space. We apply this method to a representative sample of 50,241 galaxies from DESI DR1 \citep{DR1}, demonstrating its ability to recover AGN candidates and highlighting its potential for scalable AGN searches in upcoming large spectroscopic datasets.

This paper is organized as follows. Section~\ref{sec:data} describes the DESI data, sample selection, and labeling strategy. Section~\ref{sec:methods} presents the \software{SPENDER}-based methodology and the k-d tree classification framework. The results are presented in Section~\ref{sec:results}, including AGN identification and analysis of discrepant classifications. We interpret these results in Section~\ref{sec:discussion} and summarize our main conclusions in Section~\ref{sec:conclusions}.

\section{Data}\label{sec:data}

\subsection{The DESI Survey}

The DESI spectrograph, which is robotically fiber-positioned and reconfigurable in 3 minutes, is situated at the Mayall 4-meter telescope at the Kitt Peak National Observatory \citep{DESI_Collaboration_2022, poppett}. DESI's advanced optics and 5,000 robotic positioners enable swift, high-resolution spectral measurements of galaxies across 14,000 square degrees, with coverage extending from 360 nm to 980 nm. This allows the exploration of redshifts up to $z \sim 1.7$ for emission-line galaxies, while QSOs are observed to significantly higher redshifts, with Lyman-$\alpha$ forest science enabled up to $z \sim 3.5$ \citep{DESI2016b.Instr, Silber_2023, Miller_2024}.

\subsection{Guadalupe Dataset}
Data from the first two months of the DESI Main Survey---referred to internally as \textit{Guadalupe}---are employed in this work. This dataset comprises 43 of the 261 observation nights of the Main Survey \citep{schlafly} and is part of DR1 \citep{DR1}. \textit{Guadalupe} was designed to follow the standard DESI survey strategy and target selection, and is therefore considered broadly representative of the main galaxy sample \citep{DR1}. The dataset contains more than 3 million galaxies, of which 1,178,007 are galaxies and QSOs in our redshift range of interest: $z \in [0.001, 0.5]$. The lower threshold avoids contamination from sources within the Milky Way \citep{raga2025}, while the upper limit ensures that key emission lines such as H$\alpha$ remain within the DESI spectral range. Future work will extend the analysis to higher redshifts. Each spectrum consists of 7,781 spectral measurements spanning the observed wavelength range $\lambda_{\rm obs} = [3600, 9824]$\,\r{A}.

\subsection{Preprocessing and quality cuts}
DESI spectra are automatically processed through the DESI spectroscopic pipeline \citep{Guy_2023}, followed by spectral classification and redshift determination using the \software{Redrock} fitting pipeline \citep{Anand_2024_redrock}. \software{Redrock} assigns a spectral type (GALAXY, QSO, or STAR), redshift ($\tt z$), and a warning bitmask ($\tt ZWARN$), among other parameters. We retain only entries with a galaxy or QSO spectral type, with $\tt ZWARN = 0$ and no fiber issues (${\tt COADD\_FIBERSTATUS} = 0$), following DESI recommendations \citep{EDR, DR1}.

\subsection{Spectral Classification and Labeling}
\label{sub:class}
Since \software{SPENDER} does not produce direct class predictions, external labels are required to assign spectral classes and evaluate the classification performance. In this work, we use reference labels derived from emission-line measurements obtained with \software{FastSpecFit} v3.1 \citep{FastSpecFit23}, a spectrophotometric fitting algorithm that provides emission-line measurements and broadband photometry for DESI spectra. \software{FastSpecFit} decomposes permitted Balmer lines into narrow and broad components only when the inclusion of a broad component leads to a statistically significant improvement of the fit. Using these measurements, galaxy classes are assigned according to the following criteria, summarized in Table~\ref{tab:class_counts}:
\begin{itemize}
    \item BL sources\footnote{BL emission is characteristic of AGN, so these sources are expected to be AGN-dominated, although their BPT classification was not examined} are labeled when the broad component of the H$\alpha$ emission is significantly detected, with an amplitude-over-noise ratio (AON) $\ge 3$ and a S/N $\ge 5$, while the narrow component of the H$\alpha$ line has ${\rm S/N} \ge 3$.  
    \item The remaining galaxies, those with ${\rm S/N} > 3$ in H$\alpha$, H$\beta$, [OIII]$\lambda5007$, and [NII]$\lambda6584$, are classified using the BPT diagram with the criteria from Kewley \citep{kewely} and Kauffman \citep{kaufman}—hereafter the [NII]-BPT diagram—into narrow-line (NL) AGN, star-forming, or composite.  
    \item Galaxies not meeting these thresholds are further categorized by the H$\alpha$ equivalent width (EW) into passive ($0 < \rm EW_{H\alpha} < 0.5$ \AA), retired ($0.5 \le \rm EW_{H\alpha} \le 3$ \AA). 
    \item Galaxies that do not satisfy any of the above criteria are classified as Other. This is a heterogeneous category that includes sources whose nature cannot be robustly determined based on the available emission-line measurements, such as peculiar galaxies, spectral outliers, or objects with ambiguous or low-S/N features. In many cases, this Other class arises as a consequence of the S/N thresholds applied to emission-line measurements when defining the remaining galaxy classes.

\end{itemize}

Apart from these seven galaxy sub-classes, we categorize the dataset into three main galaxy classes: AGN (including NL AGN and BL), star-forming (including star-forming and composite), and Other/passive, the latter encompassing galaxies classified as retired, passive and Other.

\subsection{Sample Selection}
The sample of 50,241 galaxies used in this work was constructed by retaining all NL and BL AGN, which are much less numerous than the other classes. For the remaining classes, galaxies were randomly selected in bins of absolute magnitude and redshift, following the procedure of \cite{Siudek_2024}. This selection ensures balanced class representation and uniform coverage across the magnitude–redshift plane.

\subsection{Flux Normalization and Masking}
For flux normalization, the median flux is computed over a flat region in the rest-frame range $\lambda_{\rm rest} = [5300, 5850]$\,\r{A}, and the spectrum is divided by this value, centering it at unity. This suppresses overall amplitude as a dominant feature and mitigates redshift-dependent scaling \citep{SPENDER}. Sources containing infinity or NaN values in the normalized flux, as well as those with a negative normalization factor, are excluded. After these cuts, the final dataset contains 50,222 sources, as summarized in Table~\ref{tab:class_counts}.

\subsection{Value-Added Catalogs}
Additionally, the Value Added Catalog\footnote{\href{https://data.desi.lbl.gov/doc/releases/edr/vac/cigale/}{https://data.desi.lbl.gov/doc/releases/edr/vac/cigale/}} from \cite{Siudek_2024} is utilized to obtain additional variables, including star formation rate (SFR) and stellar mass. These parameters are derived by applying the \software{CIGALE} spectral energy distribution fitting code \citep{cigale_boquien} to the multiwavelength photometry of DESI galaxies \citep{Dey_2019}.

To further assess the reliability of our model's AGN predictions, the AGN/Galaxy Classification Value-Added Catalog (hereafter AGN/Galaxy VAC) from DESI DR1\footnote{\href{https://data.desi.lbl.gov/doc/releases/dr1/vac/agnqso/}{https://data.desi.lbl.gov/doc/releases/dr1/vac/agnqso/}} is also used. This comprehensive catalog incorporates AGN diagnostics based on a variety of optical, ultraviolet, and IR criteria, utilizing emission-line measurements from \software{FastSpecFit} and mid-IR photometry from WISE. A detailed description of the catalog will be provided in Juneau et al. (in preparation).

\begin{table*}[ht]
  \centering
  \begin{tabular}{ccccc}
    \toprule
    Galaxy class & N (\%) & Galaxy sub-population & N(\%) & Criterion \\
    \midrule\midrule
    \multirow{2}{*}{AGN}  & \multirow{2}{*}{12,217 (24.3\%)} & BL & 5,592 (11.1\%) &$\mathrm{S/N}_{H\alpha,\text{n}} \ge 3$, $\mathrm{S/N}_{H\alpha,\text{b}} \ge 5$, $\mathrm{AON}_{H\alpha,\text{b}} \ge 3$

\\

        &        & NL  & 6,625 (13.2\%) & [N II]-BPT diagram \& $\mathrm{S/N}_{\mathrm{lines}} \geq 3$ \\
    \cmidrule{1-5}
    \multirow{2}{*}{Star-forming} & \multirow{2}{*}{14,791 (29.5\%)} & Composite & 7,271 (14.5\%) & [N II]-BPT diagram \& $\mathrm{S/N}_{\mathrm{lines}} \geq 3$ \\
                 &        & Star-forming & 7,520 (15.0\%) & [N II]-BPT diagram \& $\mathrm{S/N}_{\mathrm{lines}} \geq 3$ \\
    \cmidrule{1-5}
    \multirow{3}{*}{Other/passive} & \multirow{3}{*}{23,214 (46.2\%)} & Passive & 7,473 (14.9\%) & $0 < \rm EW_{H\alpha} < 0.5$ \AA \\
                  &        & Retired & 7,908 (15.7\%) & $0.5 \le \rm EW_{H\alpha} \le 3$ \AA \\
                  &        & Other & 7,833 (15.6\%) & All above criteria fails \\
    \hline \\
    \textbf{Total} & 50,222 & & & \\
    \bottomrule
    
  \end{tabular}
  \caption{Galaxy class, sub-population counts, and selection criteria for the representative Guadalupe dataset after masking. Here, 'n' refers to the narrow component and 'b' to the broad component of H$\alpha$.}
  \label{tab:class_counts}

\end{table*}

\section{Methods}
\label{sec:methods}
\subsection{Model}
The \software{SPENDER} architecture \citep{SPENDER, SPENDERDESI, SPENDER2} is adapted to compress observed galaxy spectra into a low-dimensional latent representation.

The encoder operates on the observed-frame spectra and produces an initial representation. Then, a multi-layer perceptron \citep{DL_goodfellow} explicitly incorporates the redshift to transform this representation into a latent space in the rest-frame. Finally, the decoder reconstructs the spectra in the rest-frame. However, the flexibility of the autoencoder introduces degeneracy in the latent space, allowing similar spectra to be represented in multiple ways. Consequently, while the latent space itself does not need to explicitly encode redshift information because reconstruction occurs in the rest-frame, spectra at different redshifts may still be mapped to distinct latent representations \citep{SPENDER2}. To address this, a total loss function is employed, combining three components: fidelity loss, which quantifies reconstruction accuracy using inverse-variance weighted mean squared error; similarity loss, which ensures proximity of similar spectra in latent space; and consistency loss, which enforces closeness between original and augmented spectra. These losses are defined following \cite{SPENDER2}.

Following \citet{SPENDER}, we adopt a latent space dimensionality of ten. That work showed that increasing the latent space from six to ten dimensions leads to small but consistent improvements in reconstruction performance, as quantified by the total loss function, without introducing signs of overfitting. This choice therefore provides a good compromise between compact representation and accurate spectral reconstruction.

For every spectrum a mask is implemented on top of the 100 strongest telluric lines, where zero weights are assigned in the 5 \r{A} region around line centers \citep{SPENDER}.  When a mask is implemented, the spectra remain unchanged; only the weights are set to zero. This ensures that these regions are not contributing to the final loss during training.

\subsubsection{Training}
Several preliminary training runs were performed to optimize the training pipeline, focusing on learning rate adjustments. The final training scheme ensures loss function convergence over 60 epochs by employing a cyclic learning rate scheduler, specifically PyTorch's OneCycleLR \citep{OneCycleLR}. In this setup, the learning rate follows a cyclical pattern within each epoch, with a peak value that is manually decreased by an order of magnitude every 20 epochs, from $10^{-3}$ to $10^{-5}$. To mitigate stochastic effects, the training procedure was initially repeated across five independent runs; however, we observed negligible differences in both the loss function and evaluation metrics between runs. Consequently, only a single representative training run is reported in this work.

\subsection{UMAP}
For visualization purposes, the Uniform Manifold Approximation and Projection (UMAP) method \citep{umap} was implemented. This dimensionality reduction technique takes the 10-dimensional latent space as input and projects it into a 2-dimensional space, where the two parameters are subsequently plotted. The different galaxy classes are then color-coded a posteriori to examine their distribution and clustering in the latent space. 

The \software{umap} package \citep{umap-software} was used for this implementation with \software{Python} 3.10, adopting the default hyperparameters (n\_neighbors=15, min\_dist=0.1, and metric=euclidean). Since UMAP projections are non-unique and depend on the adopted hyperparameters and stochastic initialization, we assessed the robustness of the resulting visual structures by repeating the projection with different random seeds and alternative parameter configurations. Although the exact geometry of the 2-dimensional embedding varies between realizations, as expected from the stochastic nature of the algorithm, the main structures and class separations discussed in this work remain stable.

Plotting the 2-dimensional UMAP projection of the latent space enables straightforward visual inspection while approximately preserving the topological structure. However, information is inevitably lost during the dimensionality reduction process, and distances or apparent separations in the projected space should not be interpreted quantitatively. Therefore, all quantitative analyses and classification metrics presented in this work are computed directly within the original 10-dimensional latent space rather than from the 2-dimensional UMAP projection, which is used solely for visualization purposes.

\subsection{Classification: k-d tree}

The \software{SPENDER} model does not produce direct class predictions, so class assignments must be derived from the learned spectral representation. To this end, we implement a k-d tree model \citep{kdtree} within the trained 10-dimensional latent space. Galaxy classes are predicted via a majority vote among the 10 nearest neighbors, exploiting the proximity of spectra with similar spectral features in the learned representation. The neighbor labels are derived from \software{FastSpecFit} emission-line measurements (Section~\ref{sec:data}) and are used as reference classifications for both class assignment and metric evaluation. This nearest-neighbor approach accommodates complex data distributions without imposing predefined decision boundaries. This step completes the semi-supervised workflow: the spectral representation is learned without labels, while class assignments are derived from reference classifications propagated through proximity in latent space.

To determine the hyperparameters of the nearest-neighbor classifier using the k-d tree algorithm, we performed iterative runs exploring several distance metrics, including Minkowski/Euclidean, Manhattan, and Chebyshev distances, and leaf sizes between 10 and 60 in steps of 10. The model performance was highly stable across the explored hyperparameter space. Variations in leaf size produced no measurable impact on the classification metrics, while changing the distance metric led to variations below $\sim0.4\%$ in global accuracy and macro F1 score (see Section~\ref{sub:metrics} and Appendix~\ref{sup:metrics} for metric definitions). The Minkowski metric yielded the best overall AGN performance, although the improvement relative to other metrics was marginal ($<0.15\%$ in AGN accuracy). We therefore adopted the default \software{scikit-learn} parameters  \citep{scikit-learn}, using a leaf size of 40 and the Minkowski metric with $p=2$, implemented in \software{Python} 3.10.

\subsection{Performance metrics}\label{sub:metrics}

The class predictions from the k-d tree, along with the reference labels derived from \software{FastSpecFit} line measurements, enable the evaluation of the model's performance through various metrics. In multi-class classification, several strategies exist for computing the evaluation metrics. The one-vs-one considers all pairwise combinations of classes, while one-vs-rest binarizes the problem by comparing each class against all others \citep{scikit-learn}. In our analysis, we adopt the latter strategy.

The definitions of all performance metrics used in this work are provided in Appendix ~\ref{sup:metrics}. All metrics were computed using the \software{scikit-learn} package \citep{scikit-learn}, with \software{Python} 3.10. The \texttt{multi\_class} parameter was set to one-vs-rest.

\section{Results}
\label{sec:results}
\subsection{Identification of AGN}
\label{sec:agn_identification}

To assess the performance of the model in identifying AGN, we first use the three main galaxy classes: AGN (including NL AGN and BL), star-forming (including star-forming and composite), and Other/passive, the latter encompassing galaxies classified as Other, retired, and passive. Table~\ref{tab:results_1} summarizes the classification performance of the \software{SPENDER} model based on these categories.

The model achieves the highest accuracy and Area Under the Curve (AUC) for AGN classification (0.952 and 0.976, respectively), as also reflected in the Receiver Operating Characteristic (ROC) curves in Fig.~\ref{fig:roc_panels} panel (a), where the AGN curve lies slightly above those of the other classes, indicating a marginally better trade-off between sensitivity and specificity across thresholds. The performance for star-forming and Other/passive galaxies is slightly lower but remains robust, with accuracy values above 90\% and AUC scores exceeding 0.97. The star-forming class shows the highest recall (0.904), suggesting the model is particularly effective at recovering most of these sources. However, the recall values are very similar across the three classes, indicating balanced sensitivity. The Other/passive class attains the highest precision (0.963) and a strong F1-score (0.930), reflecting a high degree of purity and consistency in the predicted labels. Overall, the results highlight the model’s ability to classify galaxies reliably across all categories.

\begin{table*}[ht]
  \centering
\begin{tabular}{lcccccc}
  \toprule
  Galaxy class & Accuracy & Precision & Recall & FPR & F1-score & AUC \\
  \midrule\midrule
  AGN (BL + NL AGN) & 0.952 & 0.907 & 0.893 & 0.030 & 0.900 & 0.976 \\
  star-forming (star-forming + composite) & 0.910 & 0.813 & 0.904 & 0.087 & 0.856 & 0.956 \\
  Other/passive (passive + retired + Other) & 0.938 & 0.963 & 0.901 & 0.030 & 0.930 & 0.971 \\
  \bottomrule
\end{tabular}
\caption{Classification performance for the three galaxy classes. Metric definitions are provided in Appendix~\ref{sup:metrics}.}
  \label{tab:results_1}
\end{table*}

\begin{figure}[ht]
\centering
\includegraphics[width=\hsize]{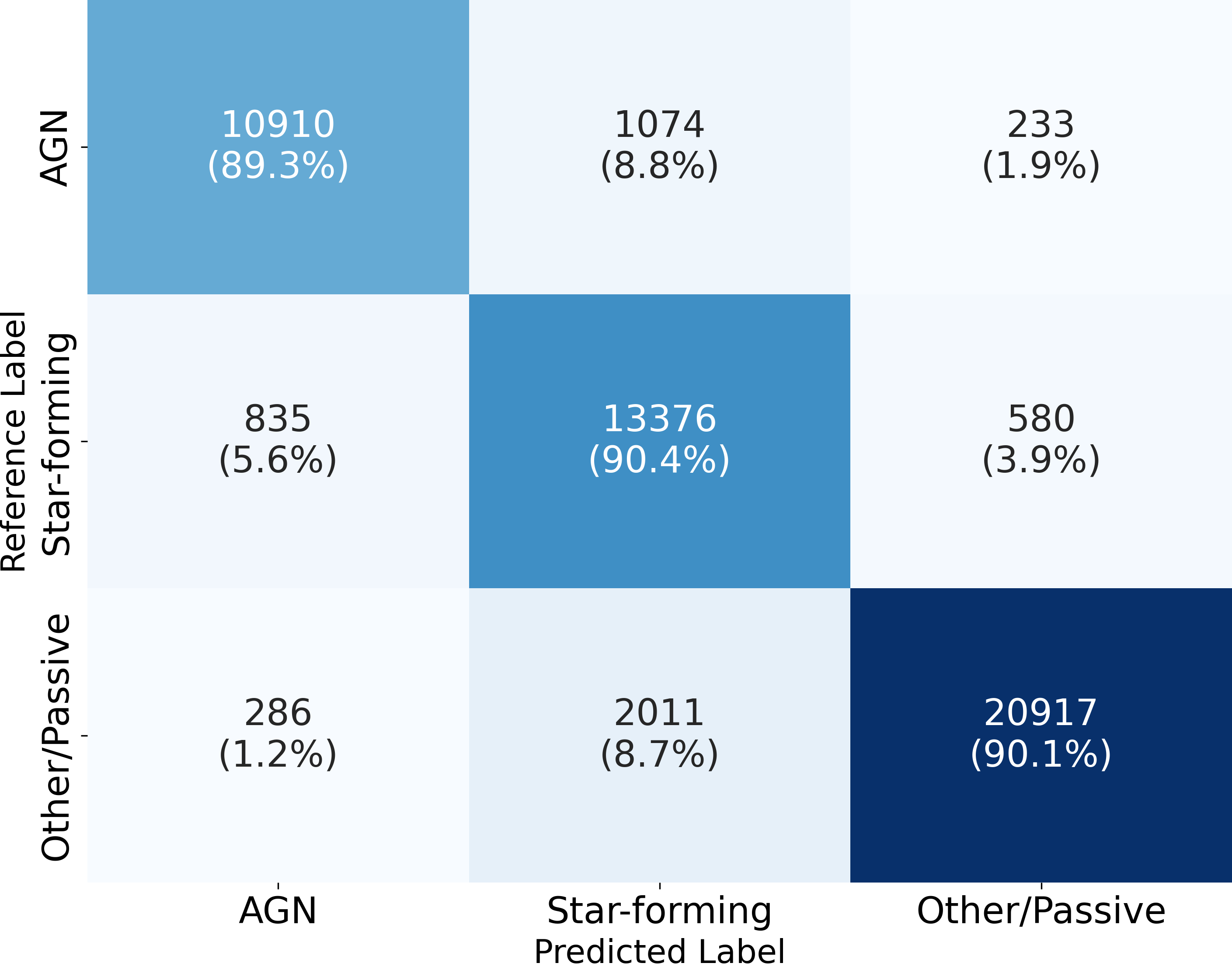}
\caption{Confusion matrix for the three-class galaxy classification. Percentages are computed with respect to the reference labels.}
\label{pic:conf_3}
\end{figure}

Figure ~\ref{pic:conf_3} presents the confusion matrix for the three-class classification. The model performs  well in AGN identification; only 835 and 286 sources predicted as AGN are labeled as star-forming and Other/passive, respectively. These cases account for 9.3\% of all AGN predictions, while 10,910 (90.7\%) sources are consistently identified as AGN.

It is important to note that the metrics and confusion matrix values presented rely on classifications derived from the emission-line measurements provided by \software{FastSpecFit}. These classifications are used here as reference labels for the evaluation of the model. However, as discussed in Section~\ref{sec:out} and Appendix~\ref{sec:S/N}, some sources exhibit weak emission lines (e.g., H$\alpha$, H$\beta$, [OIII]$\lambda$5007, or [NII]$\lambda$6584) with S/N below 3, leading to their classification as Other despite possibly being genuine AGN or star-forming galaxies. Moreover,  as discussed in Section~\ref{sec:out}, some of these sources might still be AGN according to alternative diagnostics beyond the [NII]-BPT diagram. Therefore, the reported accuracy values should be treated as lower bounds on the true model performance.

\begin{figure}[ht]
\centering
\includegraphics[width=\hsize]{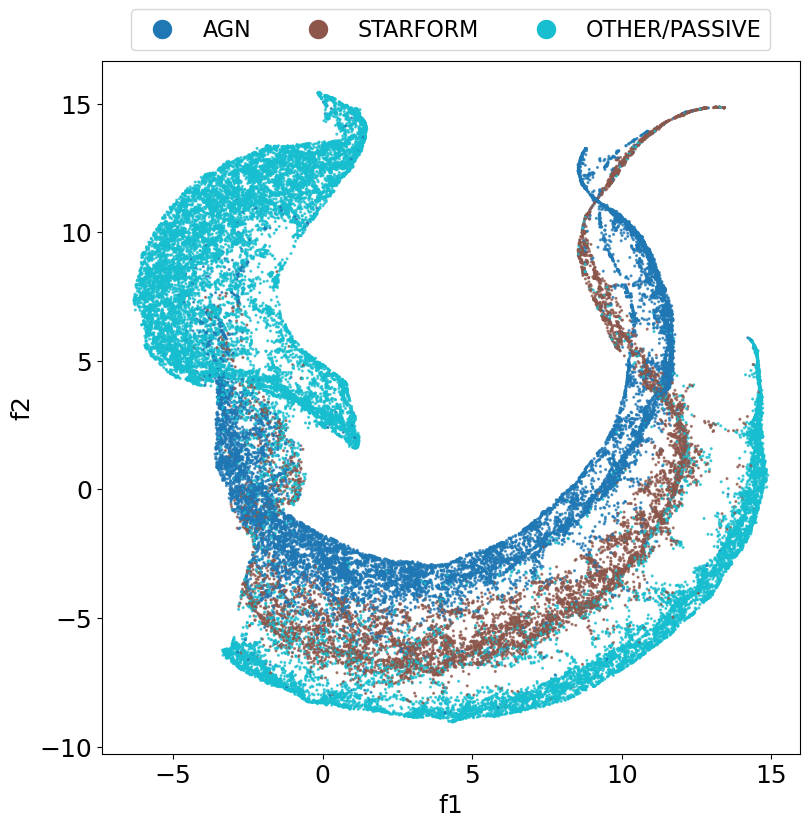}
\caption{UMAP projection of the 10-dimensional latent space into 2-dimensions. Galaxies are color-coded a posteriori according to their \software{FastSpecFit} classification: AGN (blue), star-forming (brown), and Other/passive (cyan).}
\label{fig:umap3}
\end{figure}

To further explore how the \software{SPENDER} model encodes spectral information, we project the 10-dimensional latent space onto a 2D plane using UMAP \citep{umap} (Fig. ~\ref{fig:umap3}). The three galaxy classes, which are not used in the training, are color-coded a posteriori. The clustering of these categories suggests that the latent space retains meaningful physical information, grouping galaxies with similar spectral properties together.

The UMAP embedding exhibits a curved, bifurcated morphology, with three distinct branches that become narrower toward the left end. The AGN population (blue) follows a well-defined arc that remains relatively separated from the Other/passive class (cyan). The Other/passive galaxies are distributed across two regions: a cluster in the upper left and a more compact branch in the lower right. The star-forming galaxies (brown) are interspersed between the AGN and Other/passive populations, forming a diffuse region on the left side with additional mixing with the AGN class and a sharper, more compact branch on the right.

\subsection{Identification of galaxy sub-populations}
\label{sec:7class_results}

Beyond distinguishing between AGN, star-forming, and Other/passive galaxies, this section evaluates the model’s performance in classifying an extended set of seven galaxy classes: NL AGN, BL, star-forming, composite, passive, retired, and Other.

Table~\ref{tab:class_tots} summarizes the classification performance for this extended galaxy taxonomy. The model achieves its best performance for BL, with the highest accuracy (0.965) and AUC (0.973), also evident in Fig.~\ref{fig:roc_panels} panel (b), where the BL ROC curve remains well above those of the other classes across most thresholds. NL AGN are also well identified, with an accuracy of 0.937 and an AUC of 0.946. star-forming and composite galaxies show solid performance, with accuracy values of 0.904 and 0.875, and AUCs of 0.936 and 0.897, respectively. passive and retired galaxies are more challenging to classify, achieving lower accuracy (0.848 and 0.833) and AUC scores (0.890 and 0.875). The Other class shows the lowest recall (0.327) and F1-score (0.400), indicating that this heterogeneous group is the most difficult to distinguish reliably. This difficulty is also reflected in Fig.~\ref{fig:roc_panels} panel (b), where its ROC curve lies closest to the diagonal. 

\begin{table*}[ht]
  \centering
\begin{tabular}{lcccccc}
  \toprule
  Galaxy class & Accuracy & Precision & Recall & FPR & F1-score & AUC \\
  \midrule\midrule
  BL & 0.965 & 0.839 & 0.845 & 0.020 & 0.842 & 0.973 \\
  NL AGN & 0.937 & 0.768 & 0.746 & 0.034 & 0.757 & 0.946 \\
  composite & 0.875 & 0.561 & 0.630 & 0.083 & 0.594 & 0.897 \\
  star-forming & 0.904 & 0.669 & 0.716 & 0.062 & 0.692 & 0.937 \\
  passive & 0.848 & 0.491 & 0.597 & 0.108 & 0.539 & 0.890 \\
  retired & 0.833 & 0.468 & 0.468 & 0.099 & 0.468 & 0.875 \\
  Other & 0.847 & 0.516 & 0.327 & 0.057 & 0.400 & 0.787 \\
  \bottomrule
\end{tabular}
\caption{Performance metrics for classification of the extended galaxy population into seven classes. Metric definitions are provided in Appendix~\ref{sup:metrics}.}
\label{tab:class_tots}
\end{table*}

\begin{figure}[ht]
\centering
\includegraphics[width=\hsize]{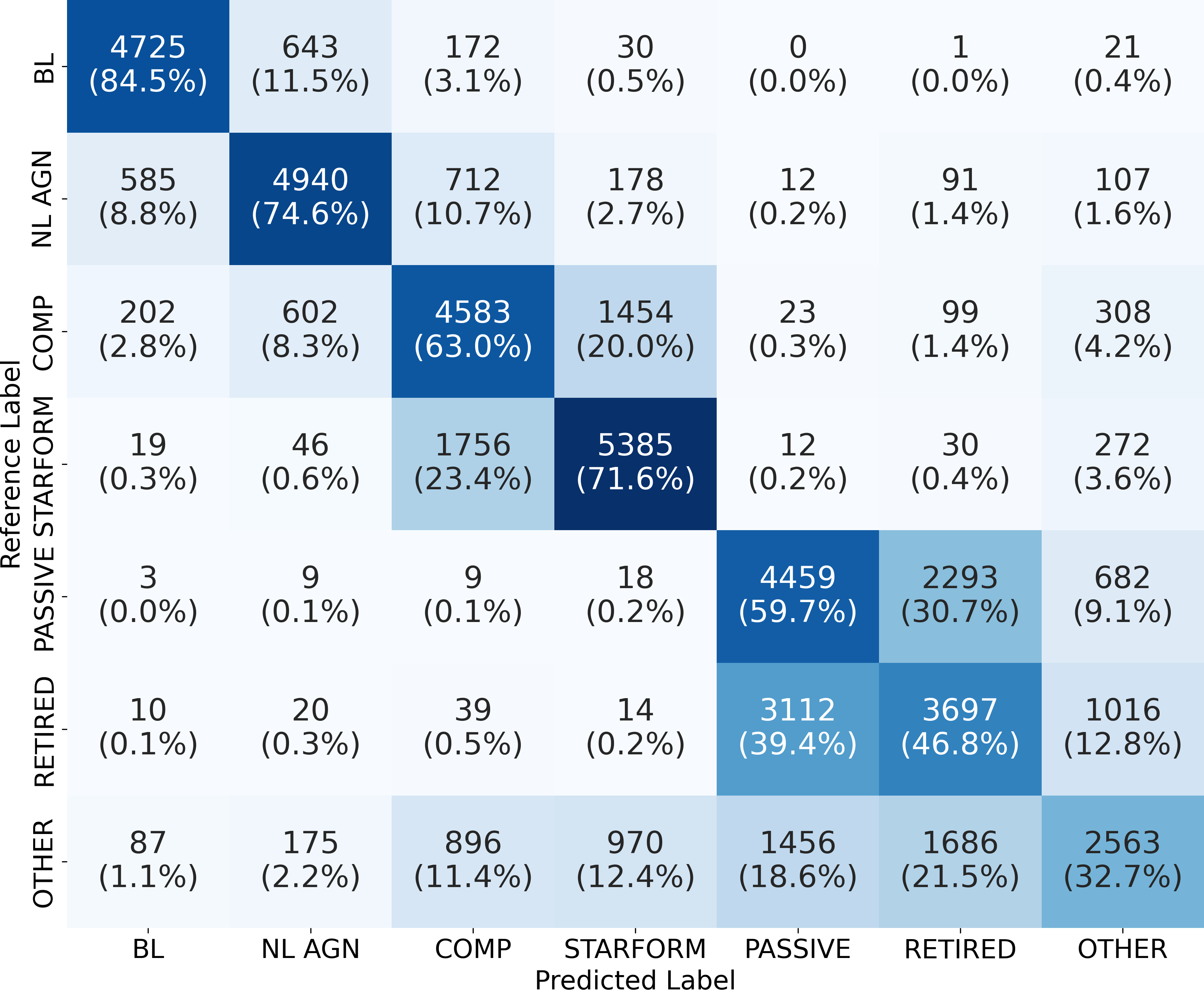}
\caption{Confusion matrix for extended galaxy classification.  Percentages are computed with respect to the reference labels.}
\label{pic:conf_7}
\end{figure} 

\begin{figure}
\centering
 \includegraphics[width=\hsize]{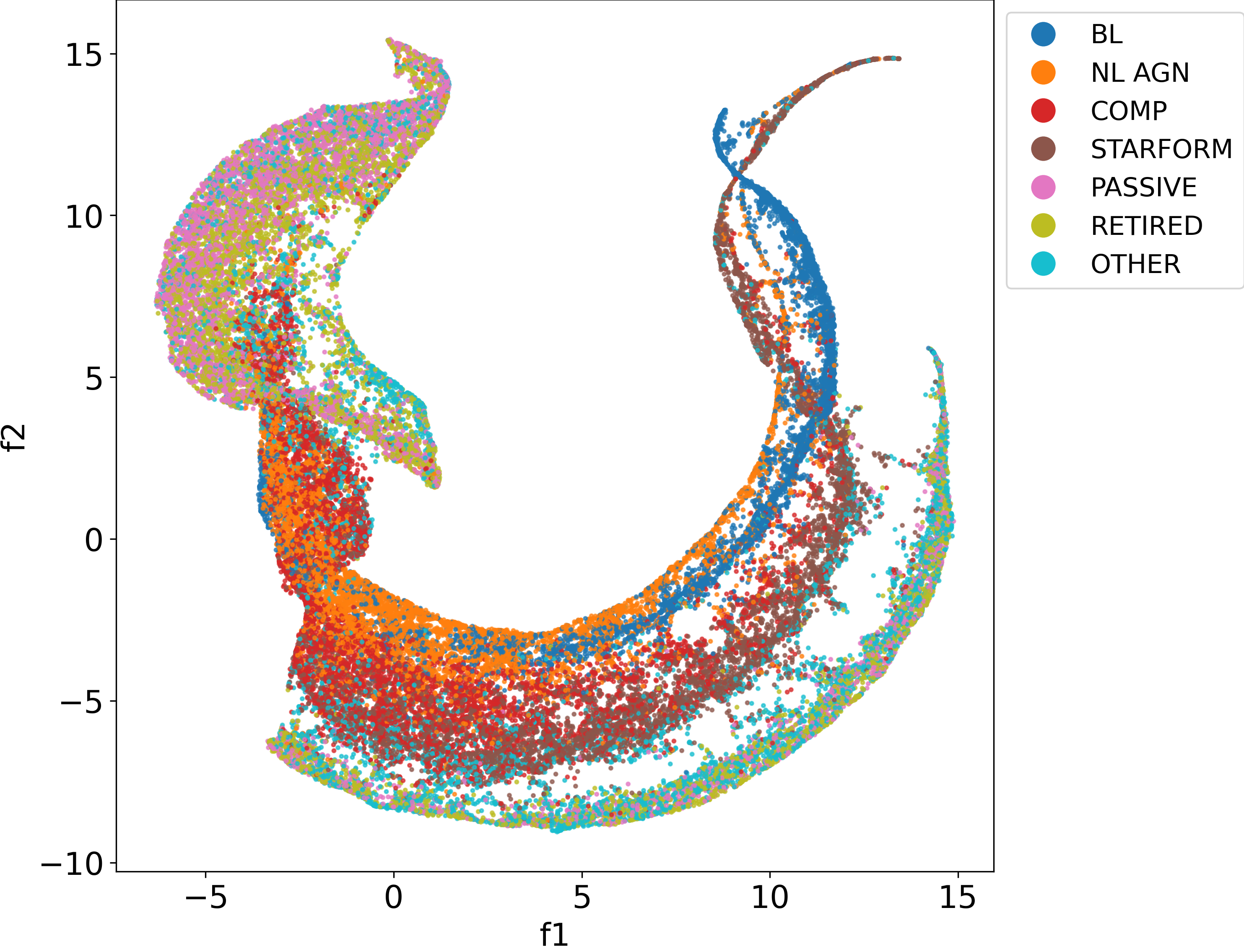}
\caption{UMAP embedding of the \software{SPENDER} model’s 10 latent features. The points are colored a posteriori according to seven galaxy classes: NL AGN, BL, composite, star-forming, passive, retired, and Other.}
 \label{fig:umap_tots}
\end{figure}

As shown in the confusion matrix (Figure ~\ref{pic:conf_7}), disagreements in the NL AGN class predominantly involve BL and composite galaxies. In the UMAP embedding of the 10 latent features (Figure ~\ref{fig:umap_tots}), colored by these extended galaxy classes, we observe that NL AGN, BL sources, and a mixture of star-forming and composite galaxies occupy distinct regions of the latent space. Notably, the BL sources and NL AGN regions are well separated, reinforcing the model's ability to distinguish these two classes. However, the relationship between star-forming and composite galaxies is more complex, with significant overlap between them, especially on the left side of the latent space, where the branches are closer together. This reflects the inherent ambiguity in their spectral features, as recent studies using dimensionality reduction or latent space clustering show that the distinction between star formation-dominated and composite sources is gradual rather than sharp \citep{Pat, Tous2025}.

The classification challenges are most evident for passive and retired galaxies, which frequently misclassify into one another, as observed in the confusion matrix (Figure \ref{pic:conf_7}). As shown in Figure \ref{fig:umap_tots}, these two classes tend to cluster together in the latent space. Retired, passive, and other galaxies form a distinct structure in the UMAP embedding, appearing split into two separate regions: a bulge in the upper-left corner and a second cluster located at lower f2 UMAP values. As discussed in Section~\ref{sec:phys}, this separation reflects underlying physical differences between these galaxy populations, rather than being solely a projection artifact.

\subsection{Latent distribution of physical properties}
\label{sec:phys}
\begin{figure*}
\centering
\includegraphics[width=\hsize]{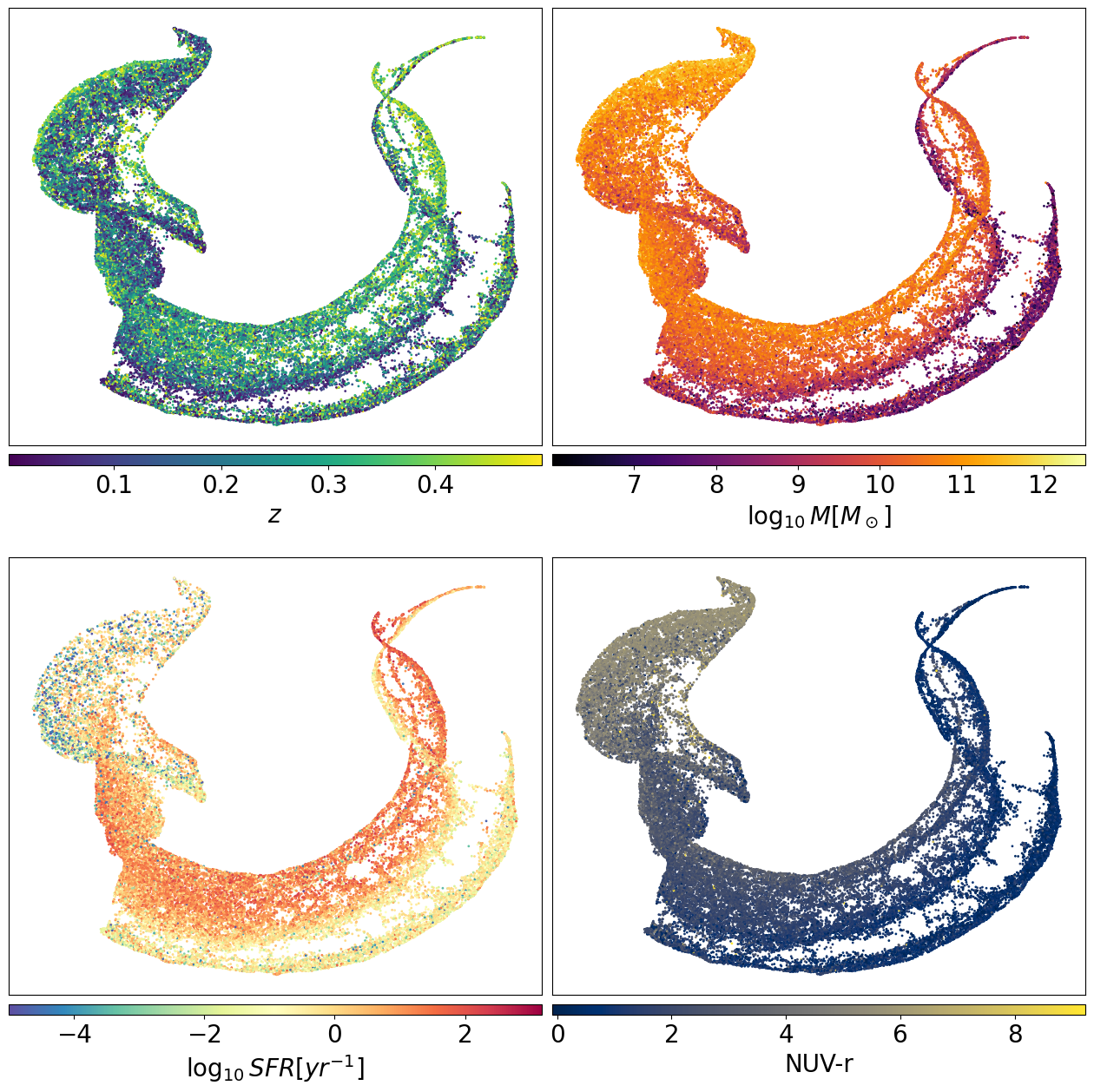}
\caption{UMAP embedding of the 10-dimensional latent space, color-coded by redshifts (top left), stellar masses (top right), specific SFRs (lower left), and NUV-$r$ color (lower right).}
\label{pic:physs}
\end{figure*}

The UMAP embedding of the latent space shows clear visual trends when color-coded by galaxy properties from the Value Added Catalog presented in \cite{Siudek_2024} (Figure~\ref{pic:physs}). Although the \software{SPENDER} architecture incorporates additional loss terms to discourage the encoding of redshift information, redshift trends can still appear in the latent representation. As discussed in \cite{SPENDER2}, this effect arises primarily from selection effects in magnitude-limited samples rather than from an explicit encoding of redshift. In particular, high-redshift galaxies are preferentially massive, while low-mass galaxies are observed only at low redshift, naturally inducing correlated trends between redshift and stellar mass in the latent space. The NL AGN and BL branches show distinct redshift gradients, particularly in BL where the extreme right segment is dominated by $z\in[0.4,0.5]$. By contrast, stellar masses show comparatively weak variation along both branches. Star-forming galaxies display a vertical gradient in both redshift and mass, with the most massive objects preferentially located toward the left end of the sequence.

Objects with ongoing or recent star formation within the last Gyr typically have NUV–$r$ colors $<5.5$, as reported in \cite{NUV2, NUV}. The observed NUV–$r$ gradient therefore suggests that galaxies in the top-right cluster are consistent with little or no star formation within this period, whereas those in the bottom-right branch show signatures of more recent activity. This interpretation is supported by the strong correlation between NUV–$r$ color and SFR shown in the lower-left panel of Figure~\ref{pic:physs}.

Regarding the structures dominated by passive, retired, and other galaxies (see Figure~\ref{fig:umap_tots}), the upper-left cluster tends to contain more massive systems with lower SFRs than the second cluster located at lower UMAP f2 values. This suggests that the observed separation reflects underlying physical differences between these populations, and that stellar population properties may capture distinctions not fully described by the current $EW_{\mathrm{H}\alpha}$-based separation between passive and retired galaxies. In particular, the passive population appears substantially more complex and heterogeneous than implied by the standard spectroscopic classification. This interpretation is consistent with previous studies highlighting the diversity of passive systems \citep{VIMOS_a, VIMOS_b, Portillo, Siudek_springer}.

\subsection{Sources with discrepant classifications}
\label{sec:out}

In this section, we analyze sources for which the model prediction and the reference classification do not agree, with a particular focus on the identification of AGN. This analysis combines stacked spectra of off-diagonal populations in the confusion matrix with independent validation using the DESI DR1 AGN/Galaxy VAC.

\subsubsection{Assessing AGN predictions with stacked spectra}
\label{subsec:bpt}

To further assess the reliability of our model, we constructed stacked spectra for each off-diagonal cell of the confusion matrix (Fig.~\ref{pic:conf_3}), corresponding to sources for which the model prediction and the reference classification do not agree. The stack spectra were computed using the \software{desigal} package \citep{desigal}, applying an inverse-variance weighted mean with bootstrap resampling to ensure robustness.

\begin{figure}
\centering
\includegraphics[width=\hsize]{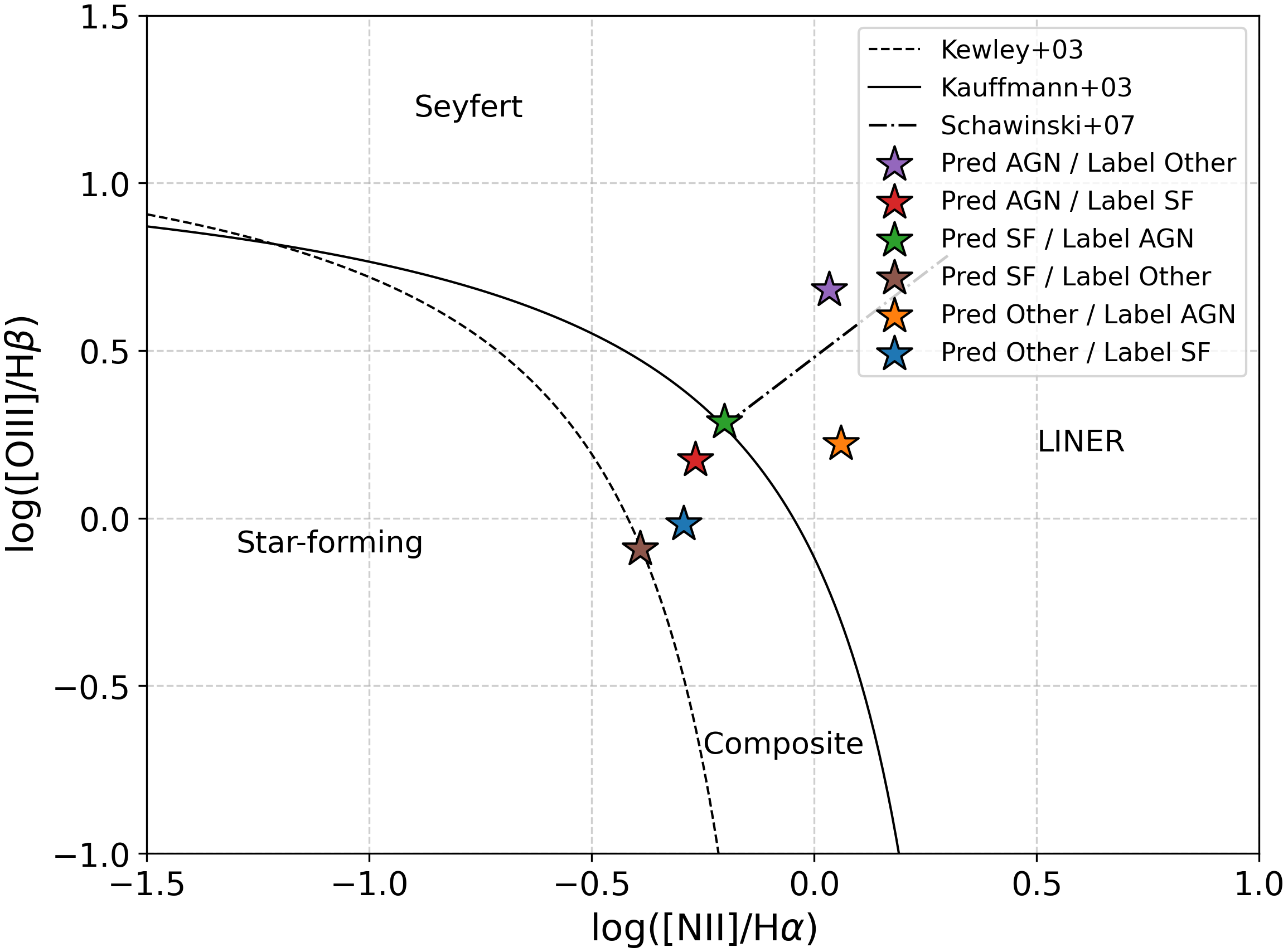}
\caption{Location of the stacked spectra corresponding to sources with discrepant classifications on the [NII]-BPT diagram.
Each symbol represents the emission-line ratios measured from the \software{FastSpecFit} model of the corresponding stacked spectrum. The classical demarcation lines separating star-forming, composite, and AGN (Seyfert and LINER) regions are shown for reference.}
\label{pic:bpt}
\end{figure}

We focus first on the 286 sources labeled as Other/passive by classical diagnostics but predicted as AGN by our model. The emission-line measurements derived from the stacked spectrum using \software{FastSpecFit} are shown in the [NII]-BPT diagram (Fig.~\ref{pic:bpt}), including the Seyfert/LINER division from \citet{Schawinski2007}, where Seyfert galaxies exhibit high-ionization emission lines and LINERs lower-ionization spectra \citep{AGN}. The stack lies in the Seyfert region, indicating a high-ionization nucleus consistent with AGN activity.

Although many individual spectra lack sufficient S/N to be classified using standard diagnostic diagrams, the stacked spectrum demonstrates that, on average, this population exhibits line ratios consistent with AGN excitation. This is consistent with the fact that 144 out of 286 sources (50.3\%) fail the S/N $>3$ requirement in at least one of the BPT lines, with 121 of these (84.0\%) failing in H$\beta$ and/or [OIII], highlighting the dominant role of the vertical BPT axis. A detailed breakdown of the limiting emission lines is provided in Appendix~\ref{sec:S/N}.

Moreover, among these 286 sources, 98 (34.3\%) are classified as BL sources within the extended classification scheme, despite not being initially labeled as such due to a low S/N in the H$\alpha$ broad component. The stacked spectrum reveals a clear broad H$\alpha$ component with a FWHM exceeding $\sim$1200 km\,s$^{-1}$, providing additional evidence for their BL nature (see Appendix~\ref{sec:broad_stack}).

A comprehensive analysis of all discrepant prediction--label combinations, including additional diagnostic diagrams, is presented in Appendix~\ref{sec:stack}.

\subsubsection{Using the AGN/Galaxy VAC for DESI DR1}
\label{subsec:vac}

To further validate our model’s AGN predictions, we cross-matched the 1121 sources predicted as AGN but labeled as star-forming (835, 74.5\%) or Other/passive (286, 25.5\%) (see Fig.~\ref{pic:conf_3}) with the AGN/Galaxy VAC of DESI DR1, which compiles AGN identifications based on multiple diagnostics beyond the [NII]-BPT. Of these 1121 sources, 1048 (93.5\%) are identified as AGN by at least one diagnostic method in the AGN/Galaxy VAC. Among these, 885 (78.9\%) are supported by at least two independent diagnostics, and 523 (46.6\%) by three or more, highlighting the robustness of the classification. A breakdown of the contributing diagnostic methods is provided in Table~\ref{tab:agn_methods}.

The majority of identified AGN, 974 (86.9\%), correspond to non-BPT optical diagnostics. Among these, 822 (73.3\%) are classified through the mass–excitation diagram \citep{MEx} and 792 (70.6\%) as strong AGN according to the WHAN method \citep{WHAN}. Additional optical diagnostics include the kinematics–excitation diagram \citep{KE}, the Blue diagram \citep{blue1, blue2}, the [Ne V] line diagnostic \citep{NeV}, and the He II BPT diagram \citep{heII}. Beyond the classical [N II]-BPT diagram, the VAC also incorporates [SII]- and [OI]-based BPT diagnostics following \citealp{sIIbpt1, LINER}. In the infrared, 317 (28.3\%) sources are identified through WISE color criteria, satisfying at least one of the AGN selection schemes proposed in \citealp{WISE1, WISE2, WISE3, WISE4, WISE5, WISE6}. QuasarNet \citep{quasarnet, Chaussidon2023}, a ML classifier originally developed for SDSS spectra and subsequently adapted for DESI, is used to identify QSOs and to reclassify ambiguous targets. Finally, 216 (19.3\%) sources exhibit broad emission lines.

\begin{table}[ht]
  \centering
  \begin{tabular}{l r}
    \toprule
    AGN Diagnostic Method & Number of Sources (\%) \\
    \midrule
    Total unique AGN detected & 1048 (93.5\%) \\
    Detected by $\geq$ 2 methods & 885 (78.9\%) \\
    Detected by $\geq$ 3 methods & 523 (46.6\%) \\
    Optical AGN (non-BPT) & 974 (86.9\%) \\
    BPT: Seyfert (robust) & 176 (15.7\%) \\
    \midrule
    Mass-excitation AGN & 822 (73.3\%) \\
    WHAN: Strong AGN & 792 (70.6\%) \\
    WISE color diagnostic & 317 (28.3\%) \\
    Kinematics-excitation AGN & 294 (26.2\%) \\
    BPT (Seyfert/LINER) & 239 (21.3\%) \\
    Broad emission lines & 216 (19.3\%) \\
    QuasarNet reclassifies as QSO & 167 (14.9\%) \\
    BLUE diagram: AGN & 97 (8.6\%) \\
    Ne V AGN & 85 (7.6\%) \\
    \software{Redrock} QSO template & 81 (7.2\%) \\
    MgII broad line (afterburner) & 46 (4.1\%) \\
    WHAN: Weak AGN & 34 (3.0\%) \\
    He II BPT AGN & 22 (2.0\%) \\
    \bottomrule
  \end{tabular}
  \caption{Number and percentage (relative to 1121 total sources classified as AGN by our model but labeled as star-forming or Other/passive) of AGN detected by each diagnostic method from the DESI DR1 AGN/Galaxy VAC. Overlaps between methods are expected; summary counts for multi-diagnostic identifications are included, along with details of AGN detected by non-BPT methods and robust BPT.}
  \label{tab:agn_methods}
\end{table}

\section{Discussion}
\label{sec:discussion}
\subsection{AGN identification}
Multiwavelength AGN samples selected in the X-ray, infrared, optical/UV, or radio occupy partially disjoint regions of classical optical diagnostic diagrams, so neither a single criterion nor combinations of criteria can capture the full diversity of AGN populations \citep{Juneau2013, trump, AGN, sanchez}. In particular, BPT-based classifications—the most widely used—can miss a substantial fraction of bona fide AGN due to weak or obscured emission lines or intrinsically low accretion states \citep{cid, AGN, sanchez}; for instance, up to $\sim$40\% of X-ray–selected AGN cannot be placed on the BPT diagram because of weak lines \citep{Agostino_2019}. As a result, traditional selections tend to underestimate the AGN fraction and motivate approaches that exploit the full spectral information rather than a limited set of emission-line ratios.

Our model shows strong performance in AGN identification, achieving an accuracy of 0.952 and an AUC of 0.976 in the three-class definition (AGN, star-forming, and Other/passive; Section~\ref{sec:agn_identification}). Direct comparisons with previous work are challenging due to differences in datasets, class definitions, and evaluation strategies, as most autoencoder-based approaches rely primarily on qualitative inspection of latent spaces. Among quantitative studies, the supervised GaSNet-II model \citep{GasNetII} reports an AGN accuracy of 0.93 on SDSS spectra. While our semi-supervised approach attains slightly higher accuracy, comparisons should be interpreted with caution given the differing training setups and datasets.

Moreover, our approach recovers AGN signatures from low-S/N spectra that are missed by standard BPT-based classifications. This is illustrated by the stacked spectrum of sources predicted as AGN but labeled as Other/passive, which lies firmly in the AGN region of the [NII]-BPT diagram (Fig.~\ref{pic:bpt}; see also Section~\ref{subsec:bpt}). Although individual spectra often lack sufficient S/N for classification (Appendix~\ref{sec:S/N}), the stacked spectrum reveals clear AGN-like line ratios. Moreover, 34.3\% of these sources are classified as BL AGN in the extended scheme, and their stacked spectra (Figure~\ref{fig:broad_fit}) show clear broad Balmer components, demonstrating that the model can recover both NL and BL AGN that are missed by traditional line-based selection due to S/N limitations.

This interpretation is reinforced by the cross-match with the VAC (Section~\ref{subsec:vac}), where 93.5\% of the 1121 sources predicted as AGN but labeled as star-forming or Other/passive are identified as AGN by at least one independent diagnostic, and nearly half by three or more. Most of these detections rely on optical diagnostics beyond the [NII]-BPT diagram, which is known to be sensitive to metallicity effects \citep{StorchiBergmann1998, Carvalho}. This confirms that many of these sources are genuine AGN missed by single-diagram approaches, and implies that our reported performance metrics represent conservative lower limits on the model’s true AGN identification capability. The high precision of our model’s AGN predictions indicates that when a source is classified as AGN, there is strong confidence in its validity, suggesting that the resulting AGN sample is both reliable and more complete than those produced by traditional approaches.

Conversely, for sources labeled as AGN but predicted as Other/passive (Section~\ref{sup:passive-AGN}), the stacked spectrum lies in the LINER region of the [NII]-BPT diagram and in the weak-AGN regime of the WHAN diagram, close to the retired boundary (${\rm EW(H\alpha)} \simeq 3$ \AA). This suggests that a fraction of these systems may be ionized by evolved stellar populations rather than by AGN, a regime where BPT ratios alone can misclassify retired galaxies as AGN \citep{stasinska, sanchez}. In this sense, the model prediction reduces contamination of the AGN class introduced by line-ratio-based labeling.

In the seven-class experiment (Section~\ref{sec:7class_results}), the model also shows strong performance in identifying BL sources. Most disagreements occur between BL and NL AGN (Figure~\ref{pic:conf_7}), highlighting the intrinsically ambiguous nature of defining broad emission lines through fixed width thresholds. Consistently, 62\% of sources labeled as NL AGN but predicted as BL by our model are identified as BL in the AGN/Galaxy VAC of DESI DR1, which adopts broader line-width criteria (Table~\ref{tab:bl_misclass}). This result highlights the flexibility of our approach in recovering BL AGN beyond restrictive single-line definitions; further details are provided in Appendix~\ref{app:bl_vac}.

From a methodological perspective, our approach can be regarded as semi-supervised, as it combines unsupervised spectral representation learning with label-based classification within a single workflow. The autoencoder operates in a fully data-driven manner, learning a low-dimensional representation of the full optical spectrum without using labels. Once trained, a k-d tree is constructed in the latent space to exploit local proximity and identify similar spectra. The workflow becomes semi-supervised at the classification stage, where external labels—derived here from \software{FastSpecFit} emission-line measurements—are used as reference classifications to assign class predictions and compute quantitative performance metrics. Importantly, the learned latent representations are not explicitly optimized to reproduce any specific diagnostic classification, and therefore do not inherit the biases associated with a given labeling scheme, unlike supervised models trained directly on labeled data. As demonstrated by the stacking analysis and the cross-matching with independent diagnostics, this separation allows the resulting predictions to mitigate known limitations of classical methods, such as incompleteness at low S/N or contamination by retired galaxies. This design also makes it possible to project alternative classification schemes onto the same latent space, including those based on different optical diagnostics or multiwavelength AGN selections, without retraining the representation itself. Moreover, once trained on a large spectroscopic sample, class assignments can be calibrated using a limited, high-quality labeled subset and propagated to the remaining sources through proximity in latent space, reducing the reliance on traditional line-based diagnostics and enabling scalable AGN classification for future DESI data releases.
\subsection{Latent Space Structure and Physical Interpretation}

In Figure~\ref{fig:umap_tots}, the learned latent space clearly separates NL AGN and BL sources, validating the model’s discriminative power. Some overlap exists, particularly in specific regions such as the left side. A notable overlap is also observed between star-forming and composite galaxies. While we expect this overlap to be significantly reduced when scaling the model to larger datasets from future DESI releases, part of it will inevitably persist due to physical causes, such as the smooth transition between AGN-dominated and host-galaxy-dominated spectra \citep{cid}.

The UMAP projection also shows that the Other class is widely distributed and overlaps with several other classes (see Figure~\ref{fig:umap_tots}). This behavior, together with the relatively low performance of the model for this category (e.g., a recall of 0.327 in the 7 galaxy sub-population scheme; see Table~\ref{tab:class_tots}), is expected, since these sources do not share a well-defined or homogeneous spectroscopic signature in the latent space. Objects are labeled as Other when their nature cannot be robustly determined from the available emission-line measurements, often because of low-S/N diagnostic lines, ambiguous spectral features, or incomplete measurements. Since the model learns directly from the full optical spectrum rather than from a limited set of emission-line ratios, many of these galaxies can be associated with physically similar populations in the latent space. This is particularly evident for AGN sources recovered within the dispersed Other population (Section~\ref{sec:out}), supporting the interpretation that latent representation can recover physically meaningful structures beyond the discrete categories defined by emission-line diagnostics.

In the case of passive and retired galaxies—whose classification metrics are comparatively poor—these classes are typically defined using thresholds in emission-line strengths or equivalent widths, particularly H$\alpha$ as adopted in this work, making them highly sensitive to observational quality \citep{cid2}. In the latent space, these populations separate into two coherent clusters that do not align with their reference labels, suggesting underlying physical differences not captured by the current classification scheme. Galaxies in the upper-left cluster tend to be more massive and exhibit lower SFRs (see Figure~\ref{pic:physs}), indicating that stellar population properties may provide a more physically meaningful basis for classification than the $EW_{H_\alpha}$-based definition of passive and retired galaxies. A similar separation has been reported in other dimensionality-reduction studies \citep{VIMOS_a, VIMOS_b, Portillo, Siudek_springer}.

 Visual inspection of the latent space is a powerful feature of these models, helping to break the "black box" nature commonly associated with other ML applications \citep{andrianomena2023}. The latent space structure provides astrophysical insight into the spectral and population diversity of galaxies, enabling the exploration of transitions and the discovery of new groupings beyond classical labels.

\subsection{Implications, challenges and Future Work}

Traditional diagnostics for identifying AGN face two main challenges: different techniques often select different AGN samples, leading to incomplete and inconsistent catalogs; and these methods typically rely on detailed spectral analysis, which is difficult to scale to the large volumes of data produced by modern surveys \citep{AGN}. While ML algorithms can address scalability, supervised methods risk propagating the incompleteness by relying on labels based on arbitrary decisions \citep{ball, Baron2019}. Unsupervised representation learning overcomes this issue by extracting information from the entire spectrum, not just specific lines, and does so without requiring a priori definitions. However, these models do not directly provide galaxy classifications or quantitative metrics.

Our methodology addresses this limitation by combining an autoencoder with a k-d tree nearest-neighbor search in the latent space to assign labels. Although the final predictions still depend on predefined labels—and thus may inherit their biases—our findings in Section~\ref{sec:out} show that the method can recover sources missed by the original diagnostics but supported by alternative ones not used during labeling, resulting in more complete catalogs. Additionally, the approach offers flexibility and interpretability: the k-d tree algorithm can be applied to different label definitions without retraining the network, and the latent space visualization provides insight into physical trends and ambiguous regions.

Nevertheless, to truly address the challenge of incompleteness, we plan to integrate unsupervised spectroscopic algorithms with multiwavelength data, such as mid-IR, X-ray, or photometric observations. This integration will be essential for the building of more comprehensive AGN catalogs and to capture the continuous transition between AGN-dominated and host-dominated spectral features \citep{LINER, AGN_obscure, MAC}. A key challenge will be managing missing data, as not all sources have matched observations across all wavelengths. To address this, we will explore multi-branch architectures \citep{multiibranch}, feature-level fusion \citep{fusion, fusion_2, fusion_3}, and strategies to handle missing data such as masking, imputation, or robust models like XGBoost \citep{XGBoost}.

The interpretability of the latent space opens several scientific avenues for future work, such as exploring how it traces dust and stellar population age, and whether slope changes reflect reddening or intrinsic AGN properties. A promising but challenging extension would be to disentangle AGN and host contributions, particularly in low-mass galaxies. Such analyses could clarify whether AGN preferentially inhabit star-forming or evolved systems, with implications for feedback and quenching \citep{Piotrowska2022}.

Moreover, as mentioned in Section \ref{sec:Intro}, broad emission lines can also be produced by transient stellar phenomena, particularly in low-mass galaxies, and follow-up spectroscopy is often required to disentangle these cases from genuine AGN activity \citep{Baldassare_2016, Panda}. In future work, we plan to investigate whether our model can help break this degeneracy and distinguish BL signatures originating from AGN from those produced by stellar processes using single-epoch spectra. Specifically, training in more carefully curated samples may be required to fully address this question.

Finally, based on the promising results of this study, we are now applying our methodology to a subset of DESI DR2 \citep{DR2}, comprising over 14 million galaxies and QSOs from three years of observations. This is nearly 280 times larger than the 50,000 sources analyzed in our current sample, enabling significantly improved statistical precision and more robust constraints on AGN populations and their connection to galaxy evolution. The flexibility of our framework, together with the straightforward preprocessing and computational efficiency of the \software{SPENDER} model, makes it well suited for scaling to large datasets. Training on larger samples is expected to further improve accuracy and robustness, enabling the construction of more complete AGN catalogs. Furthermore, this approach provides a pathway to investigate the physical processes underlying AGN activity and their connection to host-galaxy properties.

\section{Conclusions}
\label{sec:conclusions}

Traditional AGN identification methods often yield incomplete and inconsistent samples, as different diagnostics select different subsets of sources and require detailed spectral analysis that does not scale well with modern surveys. Our results demonstrate that advanced, data-driven approaches can help mitigate these biases. We show that autoencoders can learn latent representations of galaxy spectra that correlate strongly with physical galaxy classes. Unlike previous studies that relied on visual inspection of the latent space \citep{Portillo}, our work introduces a semi-supervised framework for galaxy classification that combines unsupervised feature extraction with a k-d tree classifier, using labels derived from \software{FastSpecFit}.

Our main conclusions can be summarized as follows:
\begin{itemize}
    \item The proposed model achieves high performance in AGN identification in large spectroscopic surveys, with an accuracy of 0.952 and an AUC of 0.976 in a three-class scheme.
    \item The model reliably recovers AGN in low-S/N spectra, overcoming the limitations of traditional diagnostics that depend on strict emission-line thresholds, as confirmed by stacking analyses and multi-diagnostic validation.
    \item BL AGN are effectively identified (accuracy 0.965 and AUC 0.973), including sources missed by classical methods due to weak or undetected broad components in individual spectra, highlighting the sensitivity of the approach to BL features.
    \item Stacking analyses and cross-matching with independent diagnostics show that many sources predicted as AGN but unlabeled by classical criteria are bona fide AGN; in particular, 93.5\% of these sources are confirmed by at least one diagnostic in the AGN/Galaxy VAC. This implies that the reported performance metrics represent conservative lower limits on the model’s true capability.
    \item Conversely, the model reduces contamination in the AGN class by identifying sources whose emission is consistent with retired or weakly ionized galaxies, a regime where line-ratio diagnostics alone can be misleading.
    \item The learned latent space encodes physically meaningful structure, with clear correlations with galaxy classes and properties such as stellar mass and SFR, providing interpretability and insight into ambiguous or transition populations, and revealing, for instance, two distinct populations of passive galaxies.
    \item The framework is robust to biases inherent to specific labeling schemes and is validated using independent diagnostics from the DESI AGN/Galaxy VAC.
\end{itemize}

Overall, this work demonstrates that unsupervised spectral representation learning, combined with flexible labeling strategies, provides a powerful and scalable pathway toward more complete and reliable AGN catalogs. By leveraging the full spectral information, this approach overcomes key limitations of traditional emission-line diagnostics, particularly in low-S/N and ambiguous regimes. The semi-supervised framework is well suited for integration with multiwavelength data and for application to upcoming DESI data releases, enabling improved AGN demographics and more robust constraints on the connection between nuclear activity and galaxy evolution.

\begin{acknowledgements}
M.S. acknowledges support by the State Research Agency of the Spanish Ministry of Science and Innovation under the grants 'Galaxy Evolution with Artificial Intelligence' (PGC2018-100852-A-I00) and 'BASALT' (PID2021-126838NB-I00) and the Polish National Agency for Academic Exchange (Bekker grant BPN/BEK/2021/1/00298/DEC/1). This work was partially supported by the European Union's Horizon 2020 Research and Innovation program under the Maria Sklodowska-Curie grant agreement (No. 754510). M.M. acknowledges support from the Spanish Ministry of Science and Innovation through the project PID2021-124243NBC22. This work was partially supported by the program Unidad de Excelencia Mar\'ia de Maeztu CEX2020-001058-M. M.E. acknowledges support from the Spanish Ministry of Science and Innovation through the project PID2023-152069NA-I00. This work has received funding from the ‘Severo Ochoa Centres of Excellence’ programme (grant CEX2024-001441-S), funded by MICIU/AEI/10.13039/501100011033. R.P. is currently supported by the University of Utah, and was previously supported by the University of Arizona and, in part, by U.S. NSF NOIRLab. The research of S.J. is supported by U.S. NSF NOIRLab, which is operated by the Association of Universities for Research in Astronomy (AURA) under a cooperative agreement with the National Science Foundation. S.P. is supported by the International Gemini Observatory, a program of NSF NOIRLab, which is managed by AURA under a cooperative agreement with the U.S. National Science Foundation on behalf of the Gemini partnership of Argentina, Brazil, Canada, Chile, the Republic of Korea, and the United States of America.

This material is based upon work supported by the U.S. Department of Energy (DOE), Office of Science, Office of High-Energy Physics, under Contract No. DE–AC02–05CH11231, and by the National Energy Research Scientific Computing Center, a DOE Office of Science User Facility under the same contract. Additional support for DESI was provided by the U.S. National Science Foundation (NSF), Division of Astronomical Sciences under Contract No. AST-0950945 to the NSF’s National Optical-Infrared Astronomy Research Laboratory; the Science and Technology Facilities Council of the United Kingdom; the Gordon and Betty Moore Foundation; the Heising-Simons Foundation; the French Alternative Energies and Atomic Energy Commission (CEA); the National Council of Humanities, Science and Technology of Mexico (CONAHCYT); the Ministry of Science, Innovation and Universities of Spain (MICIU/AEI/10.13039/501100011033), and by the DESI Member Institutions: \href{https://www.desi.lbl.gov/collaborating-institutions}{https://www.desi.lbl.gov/collaborating-institutions}. Any opinions, findings, and conclusions or recommendations expressed in this material are those of the author(s) and do not necessarily reflect the views of the U. S. National Science Foundation, the U. S. Department of Energy, or any of the listed funding agencies.

The authors are honored to be permitted to conduct scientific research on I'oligam Du'ag (Kitt Peak), a mountain with particular significance to the Tohono O’odham Nation.
\end{acknowledgements}

\section*{Data availability}

The data supporting the findings of this study are publicly available on Zenodo (DOI:
\href{https://doi.org/10.5281/zenodo.20490011}{10.5281/zenodo.20490011}).

The code used to reproduce the analyses presented in this work is publicly available on Zenodo (DOI:
\href{https://doi.org/10.5281/zenodo.21069095}{10.5281/zenodo.21069095}).

\bibliography{biblio}

@article{spender,
author = {Peter Melchior and Yan Liang and ChangHoon Hahn and Andy Goulding},
title = {Autoencoding Galaxy Spectra. I. Architecture},
journal = {The Astronomical Journal},
year = {2023},
volume  = {166},
number  = {2},
}

@ARTICLE{WHAN,
       author = {{Cid Fernandes}, R. and {Stasi{\'n}ska}, G. and {Mateus}, A. and {Vale Asari}, N.},
        title = "{A comprehensive classification of galaxies in the Sloan Digital Sky Survey: how to tell true from fake AGN?}",
      journal = {\mnras},
         year = 2011,
        month = may,
       volume = {413},
       number = {3},
        pages = {1687-1699},
          doi = {10.1111/j.1365-2966.2011.18244.x},
archivePrefix = {arXiv},
       eprint = {1012.4426},
 primaryClass = {astro-ph.CO},
       adsurl = {https://ui.adsabs.harvard.edu/abs/2011MNRAS.413.1687C}
}

@ARTICLE{MEx,
       author = {{Juneau}, St{\'e}phanie and {Bournaud}, Fr{\'e}d{\'e}ric and {Charlot}, St{\'e}phane and {Daddi}, Emanuele and {Elbaz}, David and {Trump}, Jonathan R. and {Brinchmann}, Jarle and {Dickinson}, Mark and {Duc}, Pierre-Alain and {Gobat}, Raphael and {Jean-Baptiste}, Ingrid and {Le Floc'h}, {\'E}meric and {Lehnert}, M.~D. and {Pacifici}, Camilla and {Pannella}, Maurilio and {Schreiber}, Corentin},
        title = "{Active Galactic Nuclei Emission Line Diagnostics and the Mass-Metallicity Relation up to Redshift z \raisebox{-0.5ex}\textasciitilde 2: The Impact of Selection Effects and Evolution}",
      journal = {\apj},
         year = 2014,
        month = jun,
       volume = {788},
       number = {1},
          eid = {88},
        pages = {88},
          doi = {10.1088/0004-637X/788/1/88},
archivePrefix = {arXiv},
       eprint = {1403.6832},
 primaryClass = {astro-ph.GA},
       adsurl = {https://ui.adsabs.harvard.edu/abs/2014ApJ...788...88J}
}

@article{spender2,
doi = {10.3847/1538-3881/ace100},
url = {https://dx.doi.org/10.3847/1538-3881/ace100},
year = {2023},
month = {jul},
publisher = {The American Astronomical Society},
volume = {166},
number = {2},
pages = {75},
author = {Yan Liang and Peter Melchior and Sicong Lu and Andy Goulding and Charlotte Ward},
title = {Autoencoding Galaxy Spectra. II. Redshift Invariance and Outlier Detection},
journal = {The Astronomical Journal}}

@article{spenderDESI,
doi = {10.3847/2041-8213/acfa03},
url = {https://doi.org/10.3847/2041-8213/acfa03},
year = {2023},
month = {oct},
publisher = {The American Astronomical Society},
volume = {956},
number = {1},
pages = {L6},
author = {Liang, Yan and Melchior, Peter and Hahn, ChangHoon and Shen, Jeff and Goulding, Andy and Ward, Charlotte},
title = {Outlier Detection in the DESI Bright Galaxy Survey},
journal = {The Astrophysical Journal Letters}
}

@ARTICLE{WISE1,
       author = {{Jarrett}, T.~H. and {Cohen}, M. and {Masci}, F. and {Wright}, E. and {Stern}, D. and {Benford}, D. and {Blain}, A. and {Carey}, S. and {Cutri}, R.~M. and {Eisenhardt}, P. and {Lonsdale}, C. and {Mainzer}, A. and {Marsh}, K. and {Padgett}, D. and {Petty}, S. and {Ressler}, M. and {Skrutskie}, M. and {Stanford}, S. and {Surace}, J. and {Tsai}, C.~W. and {Wheelock}, S. and {Yan}, D.~L.},
        title = "{The Spitzer-WISE Survey of the Ecliptic Poles}",
      journal = {\apj},
         year = 2011,
        month = jul,
       volume = {735},
       number = {2},
          eid = {112},
        pages = {112},
          doi = {10.1088/0004-637X/735/2/112},
       adsurl = {https://ui.adsabs.harvard.edu/abs/2011ApJ...735..112J}
}

@ARTICLE{WISE2,
       author = {{Mateos}, S. and {Alonso-Herrero}, A. and {Carrera}, F.~J. and {Blain}, A. and {Watson}, M.~G. and {Barcons}, X. and {Braito}, V. and {Severgnini}, P. and {Donley}, J.~L. and {Stern}, D.},
        title = "{Using the Bright Ultrahard XMM-Newton survey to define an IR selection of luminous AGN based on WISE colours}",
      journal = {\mnras},
         year = 2012,
        month = nov,
       volume = {426},
       number = {4},
        pages = {3271-3281},
          doi = {10.1111/j.1365-2966.2012.21843.x},
archivePrefix = {arXiv},
       eprint = {1208.2530},
 primaryClass = {astro-ph.CO},
       adsurl = {https://ui.adsabs.harvard.edu/abs/2012MNRAS.426.3271M}
}

@ARTICLE{WISE3,
       author = {{Stern}, Daniel and {Assef}, Roberto J. and {Benford}, Dominic J. and {Blain}, Andrew and {Cutri}, Roc and {Dey}, Arjun and {Eisenhardt}, Peter and {Griffith}, Roger L. and {Jarrett}, T.~H. and {Lake}, Sean and {Masci}, Frank and {Petty}, Sara and {Stanford}, S.~A. and {Tsai}, Chao-Wei and {Wright}, E.~L. and {Yan}, Lin and {Harrison}, Fiona and {Madsen}, Kristin},
        title = "{Mid-infrared Selection of Active Galactic Nuclei with the Wide-Field Infrared Survey Explorer. I. Characterizing WISE-selected Active Galactic Nuclei in COSMOS}",
      journal = {\apj},
         year = 2012,
        month = jul,
       volume = {753},
       number = {1},
          eid = {30},
        pages = {30},
          doi = {10.1088/0004-637X/753/1/30},
archivePrefix = {arXiv},
       eprint = {1205.0811},
 primaryClass = {astro-ph.CO},
       adsurl = {https://ui.adsabs.harvard.edu/abs/2012ApJ...753...30S}
}

@ARTICLE{WISE4,
       author = {{Assef}, R.~J. and {Stern}, D. and {Noirot}, G. and {Jun}, H.~D. and {Cutri}, R.~M. and {Eisenhardt}, P.~R.~M.},
        title = "{The WISE AGN Catalog}",
      journal = {\apjs},
         year = 2018,
        month = feb,
       volume = {234},
       number = {2},
          eid = {23},
        pages = {23},
          doi = {10.3847/1538-4365/aaa00a},
archivePrefix = {arXiv},
       eprint = {1706.09901},
 primaryClass = {astro-ph.GA},
       adsurl = {https://ui.adsabs.harvard.edu/abs/2018ApJS..234...23A}
}

@ARTICLE{WISE5,
       author = {{Yao}, H.~F.~M. and {Jarrett}, T.~H. and {Cluver}, M.~E. and {Marchetti}, L. and {Taylor}, Edward N. and {Santos}, M.~G. and {Owers}, Matt S. and {Lopez-Sanchez}, Angel R. and {Gordon}, Y.~A. and {Brown}, M.~J.~I. and {Brough}, S. and {Phillipps}, S. and {Holwerda}, B.~W. and {Hopkins}, A.~M. and {Wang}, L.},
        title = "{Galaxy and Mass Assembly (GAMA): A WISE Study of the Activity of Emission-line Systems in G23}",
      journal = {\apj},
         year = 2020,
        month = nov,
       volume = {903},
       number = {2},
          eid = {91},
        pages = {91},
          doi = {10.3847/1538-4357/abba1a},
archivePrefix = {arXiv},
       eprint = {2009.05981},
 primaryClass = {astro-ph.GA},
       adsurl = {https://ui.adsabs.harvard.edu/abs/2020ApJ...903...91Y}
}

@article{Carvalho,
    author = {Carvalho, S P and Dors, O L and Cardaci, M V and Hägele, G F and Krabbe, A C and Pérez-Montero, E and Monteiro, A F and Armah, M and Freitas-Lemes, P},
    title = {Chemical abundances of Seyfert 2 AGNs – II. N2 metallicity calibration based on SDSS},
    journal = {Monthly Notices of the Royal Astronomical Society},
    volume = {492},
    number = {4},
    pages = {5675-5683},
    year = {2020},
    month = {02},
    issn = {0035-8711},
    doi = {10.1093/mnras/staa193},
    url = {https://doi.org/10.1093/mnras/staa193},
    eprint = {https://academic.oup.com/mnras/article-pdf/492/4/5675/32429141/staa193.pdf},
}

@ARTICLE{blue1,
       author = {{Lamareille}, F. and {Mouhcine}, M. and {Contini}, T. and {Lewis}, I. and {Maddox}, S.},
        title = "{The luminosity-metallicity relation in the local Universe from the 2dF Galaxy Redshift Survey}",
      journal = {\mnras},
         year = 2004,
        month = may,
       volume = {350},
       number = {2},
        pages = {396-406},
          doi = {10.1111/j.1365-2966.2004.07697.x},
archivePrefix = {arXiv},
       eprint = {astro-ph/0401615},
 primaryClass = {astro-ph},
       adsurl = {https://ui.adsabs.harvard.edu/abs/2004MNRAS.350..396L}
}

@ARTICLE{blue2,
       author = {{Lamareille}, F.},
        title = "{Spectral classification of emission-line galaxies from the Sloan Digital Sky Survey. I. An improved classification for high-redshift galaxies}",
      journal = {\aap},
         year = 2010,
        month = jan,
       volume = {509},
          eid = {A53},
        pages = {A53},
          doi = {10.1051/0004-6361/200913168},
archivePrefix = {arXiv},
       eprint = {0910.4814},
 primaryClass = {astro-ph.CO},
       adsurl = {https://ui.adsabs.harvard.edu/abs/2010A&A...509A..53L}
}

@ARTICLE{NeV,
       author = {{Schmidt}, M. and {Hasinger}, G. and {Gunn}, J. and {Schneider}, D. and {Burg}, R. and {Giacconi}, R. and {Lehmann}, I. and {MacKenty}, J. and {Trumper}, J. and {Zamorani}, G.},
        title = "{The ROSAT deep survey. II. Optical identification, photometry and spectra of X-ray sources in the Lockman field}",
      journal = {\aap},
         year = 1998,
        month = jan,
       volume = {329},
        pages = {495-503},
          doi = {10.48550/arXiv.astro-ph/9709144},
archivePrefix = {arXiv},
       eprint = {astro-ph/9709144},
 primaryClass = {astro-ph},
       adsurl = {https://ui.adsabs.harvard.edu/abs/1998A&A...329..495S}
}

@ARTICLE{KE,
       author = {{Zhang}, Kai and {Hao}, Lei},
        title = "{A New Diagnostic Diagram of Ionization Sources for High-redshift Emission Line Galaxies}",
      journal = {\apj},
         year = 2018,
        month = apr,
       volume = {856},
       number = {2},
          eid = {171},
        pages = {171},
          doi = {10.3847/1538-4357/aab207},
       adsurl = {https://ui.adsabs.harvard.edu/abs/2018ApJ...856..171Z}
}

@ARTICLE{heII,
       author = {{Shirazi}, Maryam and {Brinchmann}, Jarle},
        title = "{Strongly star forming galaxies in the local Universe with nebular He II{\ensuremath{\lambda}}4686 emission}",
      journal = {\mnras},
         year = 2012,
        month = apr,
       volume = {421},
       number = {2},
        pages = {1043-1063},
          doi = {10.1111/j.1365-2966.2012.20439.x},
archivePrefix = {arXiv},
       eprint = {1201.1290},
 primaryClass = {astro-ph.CO},
       adsurl = {https://ui.adsabs.harvard.edu/abs/2012MNRAS.421.1043S}
}

@ARTICLE{sIIbpt1,
       author = {{Law}, David R. and {Ji}, Xihan and {Belfiore}, Francesco and {Bershady}, Matthew A. and {Cappellari}, Michele and {Westfall}, Kyle B. and {Yan}, Renbin and {Bizyaev}, Dmitry and {Brownstein}, Joel R. and {Drory}, Niv and {Andrews}, Brett H.},
        title = "{SDSS-IV MaNGA: Refining Strong Line Diagnostic Classifications Using Spatially Resolved Gas Dynamics}",
      journal = {\apj},
         year = 2021,
        month = jul,
       volume = {915},
       number = {1},
          eid = {35},
        pages = {35},
          doi = {10.3847/1538-4357/abfe0a},
archivePrefix = {arXiv},
       eprint = {2011.06012},
 primaryClass = {astro-ph.GA},
       adsurl = {https://ui.adsabs.harvard.edu/abs/2021ApJ...915...35L}
}

@article{StorchiBergmann1998,
  author = {T. Storchi-Bergmann and R. A. Schmitt and H. R. Schmitt and O. L. Dors},
  title = {Emission-line diagnostics of low-metallicity active galactic nuclei},
  journal = {Monthly Notices of the Royal Astronomical Society},
  volume = {371},
  issue = {4},
  pages = {1559-1569},
  year = {2006},
  doi = {10.1111/j.1365-2966.2006.10812.x}
}

@misc{quasarnet,
      title={QuasarNET: Human-level spectral classification and redshifting with Deep Neural Networks}, 
      author={Nicolas Busca and Christophe Balland},
      year={2018},
      eprint={1808.09955},
      archivePrefix={arXiv},
      primaryClass={astro-ph.IM},
      url={https://arxiv.org/abs/1808.09955}, 
}

@article{Schawinski2007,
  author       = {Schawinski, K. and Thomas, D. and Sarzi, M. and Maraston, C. and Kaviraj, S. and Joo, S.-J. and Yi, S. K. and Silk, J.},
  title        = {Observational evidence for AGN feedback in early-type galaxies},
  journal      = {Monthly Notices of the Royal Astronomical Society},
  volume       = {382},
  number       = {4},
  pages        = {1415--1431},
  year         = {2007},
  doi          = {10.1111/j.1365-2966.2007.12487.x},
  url          = {https://academic.oup.com/mnras/article/382/4/1415/1141702}
}

@ARTICLE{WISE6,
       author = {{Hviding}, Raphael E. and {Hainline}, Kevin N. and {Rieke}, Marcia and {Juneau}, St{\'e}phanie and {Lyu}, Jianwei and {Pucha}, Ragadeepika},
        title = "{A New Infrared Criterion for Selecting Active Galactic Nuclei to Lower Luminosities}",
      journal = {\aj},
         year = 2022,
        month = may,
       volume = {163},
       number = {5},
          eid = {224},
        pages = {224},
          doi = {10.3847/1538-3881/ac5e33},
archivePrefix = {arXiv},
       eprint = {2203.11217},
 primaryClass = {astro-ph.GA},
       adsurl = {https://ui.adsabs.harvard.edu/abs/2022AJ....163..224H}
}

@article{AGN,
author = {P. Padovani and  D.M. Alexander and R.J. Assef and et al. },
title = {Active galactic nuclei: what’s in a name?},
journal = {The Astronomy and Astrophysics Review},
year = {2017},
volume  = {25},
number  = {2},
}

@ARTICLE{ball,
       author = {{Ball}, Nicholas M. and {Brunner}, Robert J.},
        title = "{Data Mining and Machine Learning in Astronomy}",
      journal = {International Journal of Modern Physics D},
         year = 2010,
        month = jan,
       volume = {19},
       number = {7},
        pages = {1049-1106},
          doi = {10.1142/S0218271810017160},
archivePrefix = {arXiv},
       eprint = {0906.2173},
 primaryClass = {astro-ph.IM},
       adsurl = {https://ui.adsabs.harvard.edu/abs/2010IJMPD..19.1049B}
}

@article{Baron2019,
  author = {Baron, D.},
  title = {Machine learning in astronomy: a practical overview},
  journal = {Monthly Notices of the Royal Astronomical Society},
  volume = {487},
  number = {4},
  pages = {5450-5469},
  year = {2019},
  doi = {https://doi.org/10.48550/arXiv.1904.07248}
}

@INPROCEEDINGS{MAC,
       author = {{Pfeifle}, Ryan and {Weaver}, Kimberly and {Secrest}, Nathan and {Rothberg}, Barry},
        title = {The Multi-AGN Catalog: The First Literature-Complete Accounting of Multi-AGN Systems},
    booktitle = {American Astronomical Society Meeting Abstracts},
         year = {2024},
       series = {American Astronomical Society Meeting Abstracts},
       volume = {243},
        month = feb,
          eid = {218.03},
        pages = {218.03},
       adsurl = {https://ui.adsabs.harvard.edu/abs/2024AAS...24321803P}
}

@inbook{beckmann,
author = {Volker Beckmann and Chris R. Shrader},
publisher = {John Wiley \& Sons, Ltd},
isbn = {9783527666829},
title = {The Central Engine},
booktitle = {Active Galactic Nuclei},
chapter = {3},
pages = {35-88},
doi = {https://doi.org/10.1002/9783527666829.ch3},
url = {https://onlinelibrary.wiley.com/doi/abs/10.1002/9783527666829.ch3},
eprint = {https://onlinelibrary.wiley.com/doi/pdf/10.1002/9783527666829.ch3},
year = {2012}
}

@ARTICLE{VIMOS_b,
       author = {Siudek, M. and {Ma{\l}ek}, K. and {Pollo}, A. and {Granett}, B.~R. and {Scodeggio}, M. and {Moutard}, T. and {Iovino}, A. and {Guzzo}, L. and {Garilli}, B. and {Bolzonella}, M. and {de la Torre}, S. and {Abbas}, U. and {Adami}, C. and {Bottini}, D. and {Cappi}, A. and {Cucciati}, O. and {Davidzon}, I. and {Franzetti}, P. and {Fritz}, A. and {Krywult}, J. and {Le Brun}, V. and {Le F{\`e}vre}, O. and {Maccagni}, D. and {Marulli}, F. and {Polletta}, M. and {Tasca}, L.~A.~M. and {Tojeiro}, R. and {Vergani}, D. and {Zanichelli}, A. and {Arnouts}, S. and {Bel}, J. and {Branchini}, E. and {Coupon}, J. and {De Lucia}, G. and {Ilbert}, O. and {Moscardini}, L. and {Zamorani}, G. and {Takeuchi}, T.~T.},
        title = "{The VIMOS Public Extragalactic Redshift Survey (VIPERS). Unsupervised classification with photometric redshifts: a method to accurately classify large galaxy samples without spectroscopic information}",
      journal = {arXiv e-prints},
         year = 2018,
          eid = {arXiv:1805.09905},
        pages = {arXiv:1805.09905},
          doi = {10.48550/arXiv.1805.09905},
archivePrefix = {arXiv},
       eprint = {1805.09905},
 primaryClass = {astro-ph.GA},
       adsurl = {https://ui.adsabs.harvard.edu/abs/2018arXiv180509905S}
}

@article{VIMOS_a,
	author = {Siudek, M. and {Małek, K.} and {Pollo, A.} and {Krakowski, T.} and {Iovino, A.} and {Scodeggio, M.} and {Moutard, T.} and {Zamorani, G.} and {Guzzo, L.} and {Garilli, B.} and {Granett, B. R.} and {Bolzonella, M.} and {de la Torre, S.} and {Abbas, U.} and {Adami, C.} and {Bottini, D.} and {Cappi, A.} and {Cucciati, O.} and {Davidzon, I.} and {Franzetti, P.} and {Fritz, A.} and {Krywult, J.} and {Le Brun, V.} and {Le Fèvre, O.} and {Maccagni, D.} and {Marulli, F.} and {Polletta, M.} and {Tasca, L. A.M.} and {Tojeiro, R.} and {Vergani, D.} and {Zanichelli, A.} and {Arnouts, S.} and {Bel, J.} and {Branchini, E.} and {Coupon, J.} and {De Lucia, G.} and {Ilbert, O.} and {Haines, C. P.} and {Moscardini, L.} and {Takeuchi, T. T.}},
	title = {The VIMOS Public Extragalactic Redshift Survey (VIPERS) - The complexity of galaxy populations at 0.4 < z < 1.3 revealed with unsupervised machine-learning algorithms},
	DOI= "10.1051/0004-6361/201832784",
	url= "https://doi.org/10.1051/0004-6361/201832784",
	journal = {A\&A},
	year = 2018,
	volume = 617,
	pages = "A70",
}

@misc{Spender_wise,
      title={The optical and infrared are connected}, 
      author={Christian K. Jespersen and Peter Melchior and David N. Spergel and Andy D. Goulding and ChangHoon Hahn and Kartheik G. Iyer},
      year={2025},
      eprint={2503.03816},
      archivePrefix={arXiv},
      primaryClass={astro-ph.GA},
      url={https://arxiv.org/abs/2503.03816}, 
}

@misc{spenderQ,
      title={Reconstructing Quasar Spectra and Measuring the Ly$\alpha$ Forest with ${\rm S{\scriptsize pender}Q}$}, 
      author={ChangHoon Hahn and Satya Gontcho A Gontcho and Peter Melchior and Hiram K. Herrera-Alcantar and Jessica Nicole Aguilar and Steven Ahlen and Davide Bianchi and David Brooks and Todd Claybaugh and Axel de la Macorra and Arjun Dey and Peter Doel and Jaime E. Forero-Romero and Gaston Gutierrez and Mustapha Ishak and Stephanie Juneau and David Kirkby and Theodore Kisner and Anthony Kremin and Andrew Lambert and Martin Landriau and Laurent Le Guillou and Marc Manera and Ramon Miquel and John Moustakas and Adam D. Myers and Gustavo Niz and Nathalie Palanque-Delabrouille and Claire Poppett and Francisco Prada and Ignasi Pérez-Ràfols and Graziano Rossi and Eusebio Sanchez and David Schlegel and Michael Schubnell and Hee-Jong Seo and David Sprayberry and Gregory Tarlé and Benjamin A. Weaver and Hu Zou},
      year={2025},
      eprint={2506.18986},
      archivePrefix={arXiv},
      primaryClass={astro-ph.CO},
      url={https://arxiv.org/abs/2506.18986}, 
}

@ARTICLE{DR1,
       author = {{DESI Collaboration} and {Karim}, M. Abdul and {Adame}, A.~G. and {Aguado}, D. and {Aguilar}, J. and {Ahlen}, S. and {Alam}, S. and {Aldering}, G. and {Alexander}, D.~M. and {Alfarsy}, R. and {Allen}, L. and {Allende Prieto}, C. and {Alves}, O. and {Anand}, A. and {Andrade}, U. and {Armengaud}, E. and {Avila}, S. and {Aviles}, A. and {Awan}, H. and {Bailey}, S. and {Baleato Lizancos}, A. and {Ballester}, O. and {Bault}, A. and {Bautista}, J. and {Bean}, R. and {Behera}, J. and {BenZvi}, S. and {Beraldo e Silva}, L. and {Bermejo-Climent}, J.~R. and {Beutler}, F. and {Bianchi}, D. and {Blake}, C. and {Blum}, R. and {Bolton}, A.~S. and {Bonici}, M. and {Brieden}, S. and {Brodzeller}, A. and {Brooks}, D. and {Buckley-Geer}, E. and {Burtin}, E. and {Bystr{\"o}m}, A. and {Canning}, R. and {Carnero Rosell}, A. and {Carr}, A. and {Carrilho}, P. and {Casas}, L. and {Castander}, F.~J. and {Cereskaite}, R. and {Cervantes-Cota}, J.~L. and {Chaussidon}, E. and {Chaves-Montero}, J. and {Chen}, S. and {Chen}, X. and {Circosta}, C. and {Claybaugh}, T. and {Cole}, S. and {Cooper}, A.~P. and {Cousinou}, M.-C. and {Cuceu}, A. and {Davis}, T.~M. and {Dawson}, K.~S. and {de Belsunce}, R. and {de la Cruz}, R. and {de la Macorra}, A. and {de Mattia}, A. and {Deiosso}, N. and {Della Costa}, J. and {Demina}, R. and {Demirbozan}, U. and {DeRose}, J. and {Dey}, A. and {Dey}, B. and {Ding}, J. and {Ding}, Z. and {Doel}, P. and {Douglass}, K. and {Dowicz}, M. and {Ebina}, H. and {Edelstein}, J. and {Eisenstein}, D.~J. and {Elbers}, W. and {Emas}, N. and {Escoffier}, S. and {Fagrelius}, P. and {Fan}, X. and {Fanning}, K. and {Favole}, G. and {Fawcett}, V.~A. and {Fern{\'a}ndez-Garc{\'\i}a}, E. and {Ferraro}, S. and {Findlay}, N. and {Font-Ribera}, A. and {Forero-Romero}, J.~E. and {Forero-S{\'a}nchez}, D. and {Frenk}, C.~S. and {G{\"a}nsicke}, B.~T. and {Galbany}, L. and {Garc{\'\i}a-Bellido}, J. and {Garcia-Quintero}, C. and {Garrison}, L.~H. and {Gazta{\~n}aga}, E. and {Gil-Mar{\'\i}n}, H. and {Gloudemans}, A. and {Gnedin}, O.~Y. and {Gontcho}, S. Gontcho A and {Gonzalez}, D. and {Gonzalez-Morales}, A.~X. and {Gonzalez-Perez}, V. and {Gordon}, C. and {Graur}, O. and {Green}, D. and {Gruen}, D. and {Gsponer}, R. and {Guandalin}, C. and {Gutierrez}, G. and {Guy}, J. and {Hahn}, C. and {Han}, J.~J. and {Han}, J. and {He}, S. and {Herrera-Alcantar}, H.~K. and {Heydenreich}, S. and {Honscheid}, K. and {Hou}, J. and {Howlett}, C. and {Huterer}, D. and {Ir{\v{s}}i{\v{c}}}, V. and {Ishak}, M. and {Jacques}, A. and {Jiang}, L. and {Jimenez}, J. and {Jing}, Y.~P. and {Joachimi}, B. and {Joudaki}, S. and {Joyce}, R. and {Jullo}, E. and {Juneau}, S. and {Kara{\c{c}}ayl{\i}}, N.~G. and {Karim}, T. and {Kehoe}, R. and {Kent}, S. and {Khederlarian}, A. and {Kirkby}, D. and {Kisner}, T. and {Kitaura}, F.-S. and {Kizhuprakkat}, N. and {Kong}, H. and {Koposov}, S.~E. and {Kremin}, A. and {Krolewski}, A. and {Lahav}, O. and {Lai}, Y. and {Lamman}, C. and {Lan}, T.-W. and {Landriau}, M. and {Lang}, D. and {Lange}, J.~U. and {Lasker}, J. and {Le Goff}, J.~M. and {Le Guillou}, L. and {Leauthaud}, A. and {Levi}, M.~E. and {Li}, S. and {Li}, T.~S. and {Liu}, W. and {Lodha}, K. and {Lokken}, M. and {Luo}, Y. and {Luo}, Y. and {Magneville}, C. and {Manera}, M. and {Manser}, C.~J. and {Margala}, D. and {Martini}, P. and {Maus}, M. and {McCullough}, J. and {McDonald}, P. and {Medina}, G.~E. and {Medina-Varela}, L. and {Meisner}, A. and {Mena-Fern{\'a}ndez}, J. and {Menegas}, A. and {Meneses-Rizo}, J. and {Mezcua}, M. and {Miquel}, R. and {Montero-Camacho}, P. and {Moon}, J. and {Moustakas}, J. and {Mu{\~n}oz-Guti{\'e}rrez}, A. and {Mu{\~n}oz-Santos}, D. and {Myers}, A.~D. and {Myles}, J. and {Nadathur}, S. and {Najita}, J. and {Napolitano}, L. and {Newman}, J.~A. and {Nikakhtar}, F. and {Nikutta}, R. and {Niz}, G. and {Noriega}, H.~E.},
        title = "{Data Release 1 of the Dark Energy Spectroscopic Instrument}",
      journal = {arXiv e-prints},
         year = 2025,
        month = mar,
          eid = {arXiv:2503.14745},
        pages = {arXiv:2503.14745},
          doi = {10.48550/arXiv.2503.14745},
archivePrefix = {arXiv},
       eprint = {2503.14745},
 primaryClass = {astro-ph.CO},
       adsurl = {https://ui.adsabs.harvard.edu/abs/2025arXiv250314745D}
}

@article{Hviding2024,
  author = {Hviding, Raphael E. and Hainline, Kevin N. and Goulding, Andy D. and Greene, Jenny E.},
  title = {Spectroscopic Confirmation of Obscured AGN Populations from Unsupervised Machine Learning},
  journal = {The Astronomical Journal},
  volume = {167},
  number = {4},
  pages = {169},
  year = {2024},
  doi = {10.3847/1538-3881/ad28b4},
  url = {https://doi.org/10.3847/1538-3881/ad28b4}
}

@article{perez-diaz,
    author = {Pérez-Díaz, Víctor Samuel and Martínez-Galarza, Juan Rafael and Caicedo, Alexander and D’Abrusco, Raffaele},
    title = {Unsupervised machine learning for the classification of astrophysical X-ray sources},
    journal = {Monthly Notices of the Royal Astronomical Society},
    volume = {528},
    number = {3},
    pages = {4852-4871},
    year = {2024},
    month = {01},
    issn = {0035-8711},
    doi = {10.1093/mnras/stae260},
    url = {https://doi.org/10.1093/mnras/stae260},
    eprint = {https://academic.oup.com/mnras/article-pdf/528/3/4852/56670604/stae260.pdf},
}

@ARTICLE{lisiecki,
       author = {{Lisiecki}, Krzysztof and {Ma{\l}ek}, Katarzyna and {Siudek}, Ma{\l}gorzata and {Pollo}, Agnieszka and {Krywult}, Janusz and {Karska}, Agata and {Junais}},
        title = "{The first catalogue of spectroscopically confirmed red nuggets at z {\ensuremath{\sim}} 0.7 from the VIPERS survey. Linking high-z red nuggets and local relics}",
      journal = {\aap},
         year = 2023,
        month = jan,
       volume = {669},
          eid = {A95},
        pages = {A95},
          doi = {10.1051/0004-6361/202243616},
archivePrefix = {arXiv},
       eprint = {2208.04601},
 primaryClass = {astro-ph.GA},
       adsurl = {https://ui.adsabs.harvard.edu/abs/2023A&A...669A..95L}
}

@article{siudek_22,
    author = {Siudek, M and Mezcua, M and Krywult, J},
    title = {The environment of AGN dwarf galaxies at z ∼ 0.7 from the VIPERS survey},
    journal = {Monthly Notices of the Royal Astronomical Society},
    volume = {518},
    number = {1},
    pages = {724-741},
    year = {2022},
    month = {10},
    issn = {0035-8711},
    doi = {10.1093/mnras/stac3092},
    url = {https://doi.org/10.1093/mnras/stac3092},
    eprint = {https://academic.oup.com/mnras/article-pdf/518/1/724/47103798/stac3092.pdf},
}

@article{Mezcua_2023,
   title={Overmassive Black Holes in Dwarf Galaxies Out to z ∼ 0.9 in the VIPERS Survey},
   volume={943},
   ISSN={2041-8213},
   url={http://dx.doi.org/10.3847/2041-8213/acae25},
   DOI={10.3847/2041-8213/acae25},
   number={1},
   journal={The Astrophysical Journal Letters},
   publisher={American Astronomical Society},
   author={Mezcua, Mar and Siudek, Malgorzata and Suh, Hyewon and Valiante, Rosa and Spinoso, Daniele and Bonoli, Silvia},
   year={2023},
   month=jan, pages={L5} }

@ARTICLE{Siudek_springer,
       author = {{Siudek}, M. and {Lisiecki}, K. and {Mezcua}, M. and {Ma{\l}ek}, K. and {Pollo}, A. and {Krywult}, J. and {Karska}, A. and {Junais}},
        title = "{Unsupervised classification reveals new evolutionary pathways}",
      journal = {arXiv e-prints},
         year = 2022,
        month = nov,
          eid = {arXiv:2211.11792},
        pages = {arXiv:2211.11792},
          doi = {10.48550/arXiv.2211.11792},
archivePrefix = {arXiv},
       eprint = {2211.11792},
 primaryClass = {astro-ph.GA},
       adsurl = {https://ui.adsabs.harvard.edu/abs/2022arXiv221111792S}
}

@article{shen2019,
doi = {10.3847/2041-8213/ab4b54},
url = {https://dx.doi.org/10.3847/2041-8213/ab4b54},
year = {2019},
month = {oct},
publisher = {The American Astronomical Society},
volume = {885},
number = {1},
pages = {L4},
author = {Shen, Yue and Hwang, Hsiang-Chih and Zakamska, Nadia and Liu, Xin},
title = {Varstrometry for Off-nucleus and Dual Sub-Kpc AGN (VODKA): How Well Centered Are Low-z AGN?},
journal = {The Astrophysical Journal Letters}
}

@ARTICLE{netzer2015,
       author = {{Netzer}, Hagai},
        title = {Revisiting the Unified Model of Active Galactic Nuclei},
      journal = {\araa},
         year = 2015,
        month = aug,
       volume = {53},
        pages = {365-408},
          doi = {10.1146/annurev-astro-082214-122302},
archivePrefix = {arXiv},
       eprint = {1505.00811},
 primaryClass = {astro-ph.GA}
}

@misc{Pat,
      title={Reconstructing and Classifying SDSS DR16 Galaxy Spectra with Machine-Learning and Dimensionality Reduction Algorithms}, 
      author={Felix Pat and Stéphanie Juneau and Vanessa Böhm and Ragadeepika Pucha and A. G. Kim and A. S. Bolton and Cleo Lepart and Dylan Green and Adam D. Myers},
      year={2022},
      eprint={2211.11783},
      archivePrefix={arXiv},
      primaryClass={astro-ph.GA},
      url={https://arxiv.org/abs/2211.11783}, 
}

@article{Portillo,
   title={Dimensionality Reduction of SDSS Spectra with Variational Autoencoders},
   volume={160},
   ISSN={1538-3881},
   url={http://dx.doi.org/10.3847/1538-3881/ab9644},
   DOI={10.3847/1538-3881/ab9644},
   number={1},
   journal={The Astronomical Journal},
   publisher={American Astronomical Society},
   author={Portillo, Stephen K. N. and Parejko, John K. and Vergara, Jorge R. and Connolly, Andrew J.},
   year={2020},
   month=jun, pages={45} }

@article{Teimoorinia,
   title={Mapping the Diversity of Galaxy Spectra with Deep Unsupervised Machine Learning},
   volume={163},
   ISSN={1538-3881},
   url={http://dx.doi.org/10.3847/1538-3881/ac4039},
   DOI={10.3847/1538-3881/ac4039},
   number={2},
   journal={The Astronomical Journal},
   publisher={American Astronomical Society},
   author={Teimoorinia, Hossen and Archinuk, Finn and Woo, Joanna and Shishehchi, Sara and Bluck, Asa F. L.},
   year={2022},
   month=jan, pages={71} }

@misc{umap,
      title={UMAP: Uniform Manifold Approximation and Projection for Dimension Reduction}, 
      author={Leland McInnes and John Healy and James Melville},
      year={2020},
      eprint={1802.03426},
      archivePrefix={arXiv}
}

@article{umap-software,
  title={UMAP: Uniform Manifold Approximation and Projection},
  author={McInnes, Leland and Healy, John and Saul, Nathaniel and Grossberger, Lukas},
  journal={The Journal of Open Source Software},
  volume={3},
  number={29},
  pages={861},
  year={2018}
}

@article{scikit-learn,
 title={Scikit-learn: Machine Learning in {P}ython},
 author={Pedregosa, F. and Varoquaux, G. and Gramfort, A. and Michel, V.
         and Thirion, B. and Grisel, O. and Blondel, M. and Prettenhofer, P.
         and Weiss, R. and Dubourg, V. and Vanderplas, J. and Passos, A. and
         Cournapeau, D. and Brucher, M. and Perrot, M. and Duchesnay, E.},
 journal={Journal of Machine Learning Research},
 volume={12},
 pages={2825--2830},
 year={2011}
}

@MISC{fastspecfit23,
  author = {{Moustakas}, John},
  title = "{FastSpecFit: Fast spectral synthesis and emission-line fitting of DESI spectra}",
  howpublished = {Astrophysics Source Code Library, record ascl:2308.005},
  year = {2023},
  month = aug,
  eid = {ascl:2308.005},
  pages = {ascl:2308.005},
  archivePrefix = {ascl},
  eprint = {2308.005},
  adsurl = {https://ui.adsabs.harvard.edu/abs/2023ascl.soft08005M}
}

@article{Guy_2023,
title={The Spectroscopic Data Processing Pipeline for the Dark Energy Spectroscopic Instrument},
volume={165},
ISSN={1538-3881},
url={http://dx.doi.org/10.3847/1538-3881/acb212},
DOI={10.3847/1538-3881/acb212},
number={4},
journal={The Astronomical Journal},
publisher={American Astronomical Society},
author={Guy, J. and Bailey, S. and Kremin, A. and Alam, Shadab and Alexander, D. M. and Allende Prieto, C. and BenZvi, S. and Bolton, A. S. and Brooks, D. and Chaussidon, E. and Cooper, A. P. and Dawson, K. and de la Macorra, A. and Dey, A. and Dey, Biprateep and Dhungana, G. and Eisenstein, D. J. and Font-Ribera, A. and Forero-Romero, J. E. and Gaztañaga, E. and Gontcho A Gontcho, S. and Green, D. and Honscheid, K. and Ishak, M. and Kehoe, R. and Kirkby, D. and Kisner, T. and Koposov, Sergey E. and Lan, Ting-Wen and Landriau, M. and Le Guillou, L. and Levi, Michael E. and Magneville, C. and Manser, Christopher J. and Martini, P. and Meisner, Aaron M. and Miquel, R. and Moustakas, J. and Myers, Adam D. and Newman, Jeffrey A. and Nie, Jundan and Palanque-Delabrouille, N. and Percival, W. J. and Poppett, C. and Prada, F. and Raichoor, A. and Ravoux, C. and Ross, A. J. and Schlafly, E. F. and Schlegel, D. and Schubnell, M. and Sharples, Ray M. and Tarlé, Gregory and Weaver, B. A. and Yéche, Christophe and Zhou, Rongpu and Zhou, Zhimin and Zou, H.},
year={2023},
month=mar, 
pages={144} }

@book{DL_goodfellow,
    title={Deep Learning},
    author={Ian Goodfellow and Yoshua Bengio and Aaron Courville},
    publisher={MIT Press},
note={\url{http://www.deeplearningbook.org}},
    year={2016}
}

@ARTICLE{WEAVE,
       author = {{Jin}, Shoko and {Trager}, Scott C. and {Dalton}, Gavin B. and {Aguerri}, J. Alfonso L. and {Drew}, J.~E. and {Falc{\'o}n-Barroso}, Jes{\'u}s and {G{\"a}nsicke}, Boris T. and {Hill}, Vanessa and {Iovino}, Angela and {Pieri}, Matthew M. and {Poggianti}, Bianca M. and {Smith}, D.~J.~B. and {Vallenari}, Antonella and {Abrams}, Don Carlos and {Aguado}, David S. and {Antoja}, Teresa and {Arag{\'o}n-Salamanca}, Alfonso and {Ascasibar}, Yago and {Babusiaux}, Carine and {Balcells}, Marc and {Barrena}, R. and {Battaglia}, Giuseppina and {Belokurov}, Vasily and {Bensby}, Thomas and {Bonifacio}, Piercarlo and {Bragaglia}, Angela and {Carrasco}, Esperanza and {Carrera}, Ricardo and {Cornwell}, Daniel J. and {Dom{\'\i}nguez-Palmero}, Lilian and {Duncan}, Kenneth J. and {Famaey}, Benoit and {Fari{\~n}a}, Cecilia and {Gonzalez}, Oscar A. and {Guest}, Steve and {Hatch}, Nina A. and {Hess}, Kelley M. and {Hoskin}, Matthew J. and {Irwin}, Mike and {Knapen}, Johan H. and {Koposov}, Sergey E. and {Kuchner}, Ulrike and {Laigle}, Clotilde and {Lewis}, Jim and {Longhetti}, Marcella and {Lucatello}, Sara and {M{\'e}ndez-Abreu}, Jairo and {Mercurio}, Amata and {Molaeinezhad}, Alireza and {Mongui{\'o}}, Maria and {Morrison}, Sean and {Murphy}, David N.~A. and {Peralta de Arriba}, Luis and {P{\'e}rez}, Isabel and {P{\'e}rez-R{\`a}fols}, Ignasi and {Pic{\'o}}, Sergio and {Raddi}, Roberto and {Romero-G{\'o}mez}, Merc{\`e} and {Royer}, Fr{\'e}d{\'e}ric and {Siebert}, Arnaud and {Seabroke}, George M. and {Som}, Debopam and {Terrett}, David and {Thomas}, Guillaume and {Wesson}, Roger and {Worley}, C. Clare and {Alfaro}, Emilio J. and {Allende Prieto}, Carlos and {Alonso-Santiago}, Javier and {Amos}, Nicholas J. and {Ashley}, Richard P. and {Balaguer-N{\'u}{\~n}ez}, Lola and {Balbinot}, Eduardo and {Bellazzini}, Michele and {Benn}, Chris R. and {Berlanas}, Sara R. and {Bernard}, Edouard J. and {Best}, Philip and {Bettoni}, Daniela and {Bianco}, Andrea and {Bishop}, Georgia and {Blomqvist}, Michael and {Boeche}, Corrado and {Bolzonella}, Micol and {Bonoli}, Silvia and {Bosma}, Albert and {Britavskiy}, Nikolay and {Busarello}, Gianni and {Caffau}, Elisabetta and {Cantat-Gaudin}, Tristan and {Castro-Ginard}, Alfred and {Couto}, Guilherme and {Carbajo-Hijarrubia}, Juan and {Carter}, David and {Casamiquela}, Laia and {Conrado}, Ana M. and {Corcho-Caballero}, Pablo and {Costantin}, Luca and {Deason}, Alis and {de Burgos}, Abel and {De Grandi}, Sabrina and {Di Matteo}, Paola and {Dom{\'\i}nguez-G{\'o}mez}, Jes{\'u}s and {Dorda}, Ricardo and {Drake}, Alyssa and {Dutta}, Rajeshwari and {Erkal}, Denis and {Feltzing}, Sofia and {Ferr{\'e}-Mateu}, Anna and {Feuillet}, Diane and {Figueras}, Francesca and {Fossati}, Matteo and {Franciosini}, Elena and {Frasca}, Antonio and {Fumagalli}, Michele and {Gallazzi}, Anna and {Garc{\'\i}a-Benito}, Rub{\'e}n and {Gentile Fusillo}, Nicola and {Gebran}, Marwan and {Gilbert}, James and {Gledhill}, T.~M. and {Gonz{\'a}lez Delgado}, Rosa M. and {Greimel}, Robert and {Guarcello}, Mario Giuseppe and {Guerra}, Jose and {Gullieuszik}, Marco and {Haines}, Christopher P. and {Hardcastle}, Martin J. and {Harris}, Amy and {Haywood}, Misha and {Helmi}, Amina and {Hernandez}, Nauzet and {Herrero}, Artemio and {Hughes}, Sarah and {Ir{\v{s}}i{\v{c}}}, Vid and {Jablonka}, Pascale and {Jarvis}, Matt J. and {Jordi}, Carme and {Kondapally}, Rohit and {Kordopatis}, Georges and {Krogager}, Jens-Kristian and {La Barbera}, Francesco and {Lam}, Man I. and {Larsen}, S{\o}ren S. and {Lemasle}, Bertrand and {Lewis}, Ian J. and {Lhom{\'e}}, Emilie and {Lind}, Karin and {Lodi}, Marcello and {Longobardi}, Alessia and {Lonoce}, Ilaria and {Magrini}, Laura and {Ma{\'\i}z Apell{\'a}niz}, Jes{\'u}s and {Marchal}, Olivier and {Marco}, Amparo and {Martin}, Nicolas F. and {Matsuno}, Tadafumi and {Maurogordato}, Sophie and {Merluzzi}, Paola and {Miralda-Escud{\'e}}, Jordi and {Molinari}, Emilio and {Monari}, Giacomo and {Morelli}, Lorenzo and {Mottram}, Christopher J. and {Naylor}, Tim and {Negueruela}, Ignacio and {O{\~n}orbe}, Jose and {Pancino}, Elena and {Peirani}, S{\'e}bastien and {Peletier}, Reynier F. and {Pozzetti}, Lucia and {Rainer}, Monica and {Ramos}, Pau and {Read}, Shaun C. and {Rossi}, Elena Maria and {R{\"o}ttgering}, Huub J.~A. and {Rubi{\~n}o-Mart{\'\i}n}, Jose Alberto and {Sabater}, Jose and {San Juan}, Jos{\'e} and {Sanna}, Nicoletta and {Schallig}, Ellen and {Schiavon}, Ricardo P. and {Schultheis}, Mathias and {Serra}, Paolo and {Shimwell}, Timothy W. and {Sim{\'o}n-D{\'\i}az}, Sergio and {Smith}, Russell J. and {Sordo}, Rosanna and {Sorini}, Daniele and {Soubiran}, Caroline and {Starkenburg}, Else and {Steele}, Iain A. and {Stott}, John and {Stuik}, Remko and {Tolstoy}, Eline and {Tortora}, Crescenzo and {Tsantaki}, Maria and {Van der Swaelmen}, Mathieu and {van Weeren}, Reinout J. and {Vergani}, Daniela},
        title = "{The wide-field, multiplexed, spectroscopic facility WEAVE: Survey design, overview, and simulated implementation}",
      journal = {\mnras},
         year = 2024,
        month = may,
       volume = {530},
       number = {3},
        pages = {2688-2730},
          doi = {10.1093/mnras/stad557},
archivePrefix = {arXiv},
       eprint = {2212.03981},
 primaryClass = {astro-ph.IM},
       adsurl = {https://ui.adsabs.harvard.edu/abs/2024MNRAS.530.2688J}
}

@article{Chaussidon2023,
doi = {10.3847/1538-4357/acb3c2},
url = {https://doi.org/10.3847/1538-4357/acb3c2},
year = {2023},
month = {feb},
publisher = {The American Astronomical Society},
volume = {944},
number = {1},
pages = {107},
author = {Chaussidon, Edmond and Yèche, Christophe and Palanque-Delabrouille, Nathalie and Alexander, David M. and Yang, Jinyi and Ahlen, Steven and Bailey, Stephen and Brooks, David and Cai, Zheng and Chabanier, Solène and Davis, Tamara M. and Dawson, Kyle and de laMacorra, Axel and Dey, Arjun and Dey, Biprateep and Eftekharzadeh, Sarah and Eisenstein, Daniel J. and Fanning, Kevin and Font-Ribera, Andreu and Gaztañaga, Enrique and A Gontcho, Satya Gontcho and Gonzalez-Morales, Alma X. and Guy, Julien and Herrera-Alcantar, Hiram K. and Honscheid, Klaus and Ishak, Mustapha and Jiang, Linhua and Juneau, Stephanie and Kehoe, Robert and Kisner, Theodore and Kovács, Andras and Kremin, Anthony and Lan, Ting-Wen and Landriau, Martin and Le Guillou, Laurent and Levi, Michael E. and Magneville, Christophe and Martini, Paul and Meisner, Aaron M. and Moustakas, John and Muñoz-Gutiérrez, Andrea and Myers, Adam D. and Newman, Jeffrey A. and Nie, Jundan and Percival, Will J. and Poppett, Claire and Prada, Francisco and Raichoor, Anand and Ravoux, Corentin and Ross, Ashley J. and Schlafly, Edward and Schlegel, David and Tan, Ting and Tarlé, Gregory and Zhou, Rongpu and Zhou, Zhimin and Zou, Hu},
title = {Target Selection and Validation of DESI Quasars},
journal = {The Astrophysical Journal}
}

@ARTICLE{EDR,
       author = {{DESI Collaboration} and {Adame}, A.~G. and {Aguilar}, J. and {Ahlen}, S. and {Alam}, S. and {Aldering}, G. and {Alexander}, D.~M. and {Alfarsy}, R. and {Allende Prieto}, C. and {Alvarez}, M. and {Alves}, O. and {Anand}, A. and {Andrade-Oliveira}, F. and {Armengaud}, E. and {Asorey}, J. and {Avila}, S. and {Aviles}, A. and {Bailey}, S. and {Balaguera-Antol{\'\i}nez}, A. and {Ballester}, O. and {Baltay}, C. and {Bault}, A. and {Bautista}, J. and {Behera}, J. and {Beltran}, S.~F. and {BenZvi}, S. and {Beraldo e Silva}, L. and {Bermejo-Climent}, J.~R. and {Berti}, A. and {Besuner}, R. and {Beutler}, F. and {Bianchi}, D. and {Blake}, C. and {Blum}, R. and {Bolton}, A.~S. and {Brieden}, S. and {Brodzeller}, A. and {Brooks}, D. and {Brown}, Z. and {Buckley-Geer}, E. and {Burtin}, E. and {Cabayol-Garcia}, L. and {Cai}, Z. and {Canning}, R. and {Cardiel-Sas}, L. and {Carnero Rosell}, A. and {Castander}, F.~J. and {Cervantes-Cota}, J.~L. and {Chabanier}, S. and {Chaussidon}, E. and {Chaves-Montero}, J. and {Chen}, S. and {Chen}, X. and {Chuang}, C. and {Claybaugh}, T. and {Cole}, S. and {Cooper}, A.~P. and {Cuceu}, A. and {Davis}, T.~M. and {Dawson}, K. and {de Belsunce}, R. and {de la Cruz}, R. and {de la Macorra}, A. and {Della Costa}, J. and {de Mattia}, A. and {Demina}, R. and {Demirbozan}, U. and {DeRose}, J. and {Dey}, A. and {Dey}, B. and {Dhungana}, G. and {Ding}, J. and {Ding}, Z. and {Doel}, P. and {Doshi}, R. and {Douglass}, K. and {Edge}, A. and {Eftekharzadeh}, S. and {Eisenstein}, D.~J. and {Elliott}, A. and {Ereza}, J. and {Escoffier}, S. and {Fagrelius}, P. and {Fan}, X. and {Fanning}, K. and {Fawcett}, V.~A. and {Ferraro}, S. and {Flaugher}, B. and {Font-Ribera}, A. and {Forero-Romero}, J.~E. and {Forero-S{\'a}nchez}, D. and {Frenk}, C.~S. and {G{\"a}nsicke}, B.~T. and {Garc{\'\i}a}, L. {\'A}. and {Garc{\'\i}a-Bellido}, J. and {Garcia-Quintero}, C. and {Garrison}, L.~H. and {Gil-Mar{\'\i}n}, H. and {Golden-Marx}, J. and {Gontcho A Gontcho}, S. and {Gonzalez-Morales}, A.~X. and {Gonzalez-Perez}, V. and {Gordon}, C. and {Graur}, O. and {Green}, D. and {Gruen}, D. and {Guy}, J. and {Hadzhiyska}, B. and {Hahn}, C. and {Han}, J.~J. and {Hanif}, M.~M.~S. and {Herrera-Alcantar}, H.~K. and {Honscheid}, K. and {Hou}, J. and {Howlett}, C. and {Huterer}, D. and {Ir{\v{s}}i{\v{c}}}, V. and {Ishak}, M. and {Jacques}, A. and {Jana}, A. and {Jiang}, L. and {Jimenez}, J. and {Jing}, Y.~P. and {Joudaki}, S. and {Joyce}, R. and {Jullo}, E. and {Juneau}, S. and {Kara{\c{c}}ayl{\i}}, N.~G. and {Karim}, T. and {Kehoe}, R. and {Kent}, S. and {Khederlarian}, A. and {Kim}, S. and {Kirkby}, D. and {Kisner}, T. and {Kitaura}, F. and {Kizhuprakkat}, N. and {Kneib}, J. and {Koposov}, S.~E. and {Kov{\'a}cs}, A. and {Kremin}, A. and {Krolewski}, A. and {L'Huillier}, B. and {Lahav}, O. and {Lambert}, A. and {Lamman}, C. and {Lan}, T.-W. and {Landriau}, M. and {Lang}, D. and {Lange}, J.~U. and {Lasker}, J. and {Leauthaud}, A. and {Le Guillou}, L. and {Levi}, M.~E. and {Li}, T.~S. and {Linder}, E. and {Lyons}, A. and {Magneville}, C. and {Manera}, M. and {Manser}, C.~J. and {Margala}, D. and {Martini}, P. and {McDonald}, P. and {Medina}, G.~E. and {Medina-Varela}, L. and {Meisner}, A. and {Mena-Fern{\'a}ndez}, J. and {Meneses-Rizo}, J. and {Mezcua}, M. and {Miquel}, R. and {Montero-Camacho}, P. and {Moon}, J. and {Moore}, S. and {Moustakas}, J. and {Mueller}, E. and {Mundet}, J. and {Mu{\~n}oz-Guti{\'e}rrez}, A. and {Myers}, A.~D. and {Nadathur}, S. and {Napolitano}, L. and {Neveux}, R. and {Newman}, J.~A. and {Nie}, J. and {Nikutta}, R. and {Niz}, G. and {Norberg}, P. and {Noriega}, H.~E. and {Paillas}, E. and {Palanque-Delabrouille}, N. and {Palmese}, A. and {Pan}, Z. and {Parkinson}, D. and {Penmetsa}, S. and {Percival}, W.~J. and {P{\'e}rez-Fern{\'a}ndez}, A. and {P{\'e}rez-R{\`a}fols}, I. and {Pieri}, M. and {Poppett}, C. and {Porredon}, A. and {Pothier}, S.},
        title = "{The Early Data Release of the Dark Energy Spectroscopic Instrument}",
      journal = {\aj},
         year = 2024,
        month = aug,
       volume = {168},
       number = {2},
          eid = {58},
        pages = {58},
          doi = {10.3847/1538-3881/ad3217},
archivePrefix = {arXiv},
       eprint = {2306.06308},
 primaryClass = {astro-ph.CO},
       adsurl = {https://ui.adsabs.harvard.edu/abs/2024AJ....168...58D}
}

@article{kdtree,
  title={Multidimensional binary search trees used for associative searching},
  author={Bentley, Jon Louis},
  journal={Communications of the ACM},
  volume={18},
  number={9},
  pages={509--517},
  year={1975},
  publisher={ACM}
}

@misc{SpecPT,
      title={SpecPT (Spectroscopy Pre-trained Transformer) Model for Extragalactic Spectroscopy: I. Architecture and Automated Redshift Measurement}, 
      author={Rohan Pattnaik and Jeyhan S. Kartaltepe and Clive Binu},
      year={2025},
      eprint={2501.01070},
      archivePrefix={arXiv},
      primaryClass={astro-ph.IM},
      url={https://arxiv.org/abs/2501.01070}, 
}

@article{GasNetII,
   title={Galaxy Spectra neural Network (GaSNet). II. Using deep learning for spectral classification and redshift predictions},
   volume={532},
   ISSN={1365-2966},
   url={http://dx.doi.org/10.1093/mnras/stae1461},
   DOI={10.1093/mnras/stae1461},
   number={1},
   journal={Monthly Notices of the Royal Astronomical Society},
   publisher={Oxford University Press (OUP)},
   author={Zhong, Fucheng and Napolitano, Nicola R and Heneka, Caroline and Li, Rui and Bauer, Franz Erik and Bouche, Nicolas and Comparat, Johan and Kim, Young-Lo and Krogager, Jens-Kristian and Longhetti, Marcella and Loveday, Jonathan and Roukema, Boudewijn F and Rouse, Benedict L and Salvato, Mara and Tortora, Crescenzo and Assef, Roberto J and Cassarà, Letizia P and Costantin, Luca and Croom, Scott M and Davies, Luke J M and Fritz, Alexander and Guiglion, Guillaume and Humphrey, Andrew and Pompei, Emanuela and Ricci, Claudio and Sifón, Cristóbal and Tempel, Elmo and Zafar, Tayyaba},
   year={2024},
   month=jun, pages={643–665} }

@article{raga2025,
author = {Pucha, Ragadeepika and Juneau, S. and Dey, Arjun and Siudek, M. and Mezcua, M. and Moustakas, J. and BenZvi, S. and Hainline, K. and Hviding, R. and Mao, Yao-Yuan and Alexander, D. and Alfarsy, R. and Circosta, Chiara and Guo, Wei-Jian and Manwadkar, Viraj and Martini, P. and Weaver, B. and Aguilar, Janette and Ahlen, S. and Zou, H.},
year = {2025},
month = {03},
pages = {10},
title = {Tripling the Census of Dwarf AGN Candidates Using DESI Early Data},
volume = {982},
journal = {The Astrophysical Journal},
doi = {10.3847/1538-4357/adb1dd}
}

@article{Siudek_2025,
   title={Beyond traditional diagnostics: Identifying active galactic nuclei using spectral energy distribution fitting in DESI data},
   volume={700},
   ISSN={1432-0746},
   url={http://dx.doi.org/10.1051/0004-6361/202555463},
   DOI={10.1051/0004-6361/202555463},
   journal={Astronomy \& Astrophysics},
   publisher={EDP Sciences},
   author={Siudek, M. and Mezcua, M. and Circosta, C. and Maraston, C. and Moustakas, J. and Zou, H. and Aguilar, J. and Ahlen, S. and Bianchi, D. and Brooks, D. and Claybaugh, T. and Dawson, K. S. and de la Macorra, A. and Dey, A. and Doel, P. and Forero-Romero, J. E. and Gaztañaga, E. and Gontcho A Gontcho, S. and Gutierrez, G. and Ishak, M. and Juneau, S. and Kirkby, D. and Kisner, T. and Kremin, A. and Lambert, A. and Landriau, M. and Le Guillou, L. and Meisner, A. and Miquel, R. and Prada, F. and Pérez-Ràfols, I. and Rossi, G. and Sanchez, E. and Schlegel, D. and Schubnell, M. and Seo, H. and Sprayberry, D. and Tarlé, G. and Weaver, B. A.},
   year={2025},
   month=aug, pages={A209} }

@ARTICLE{Pan2025,
       author = {{Pan}, Zhiwei and {Jiang}, Linhua and {Guo}, Wei-Jian and {Sun}, Shengxiu and {Siudek}, Ma{\l}gorzata and {Aguilar}, Jessica Nicole and {Ahlen}, Steven and {Brooks}, David and {Claybaugh}, Todd and {de la Macorra}, Axel and {Doel}, Peter and {Gazta{\~n}aga}, Enrique and {Gontcho A Gontcho}, Satya and {Juneau}, Stephanie and {Kisner}, Theodore and {Lambert}, Andrew and {Landriau}, Martin and {Le Guillou}, Laurent and {Manera}, Marc and {Martini}, Paul and {Meisner}, Aaron and {Miquel}, Ramon and {Moustakas}, John and {Myers}, Adam and {Poppett}, Claire and {Prada}, Francisco and {Rossi}, Graziano and {Sanchez}, Eusebio and {Schubnell}, Michael and {Seo}, Hee-Jong and {Sprayberry}, David and {Tarl{\'e}}, Gregory and {Weaver}, Benjamin Alan and {Zou}, Hu},
        title = "{Iron-corrected Single-epoch Black Hole Masses of DESI Quasars at Low Redshift}",
      journal = {\apj},
         year = 2025,
        month = jul,
       volume = {987},
       number = {1},
          eid = {48},
        pages = {48},
          doi = {10.3847/1538-4357/add7dd},
archivePrefix = {arXiv},
       eprint = {2502.03684},
 primaryClass = {astro-ph.GA},
       adsurl = {https://ui.adsabs.harvard.edu/abs/2025ApJ...987...48P}
}

@ARTICLE{Nicolaou,
       author = {{Nicolaou}, C. and {Nathan}, R.~P. and {Lahav}, O. and {Palmese}, A. and {Saintonge}, A. and {Aguilar}, J. and {Ahlen}, S. and {Allende Prieto}, C. and {Bailey}, S. and {BenZvi}, S. and {Bianchi}, D. and {Brodzeller}, A. and {Brooks}, D. and {Claybaugh}, T. and {de la Macorra}, A. and {Della Costa}, J. and {Dey}, Arjun and {Doel}, P. and {Forero-Romero}, J.~E. and {Gazta{\~n}aga}, E. and {Gontcho}, S. Gontcho A and {Gutierrez}, G. and {Honscheid}, K. and {Howlett}, C. and {Ishak}, M. and {Kehoe}, R. and {Kirkby}, D. and {Kisner}, T. and {Kremin}, A. and {Lambert}, A. and {Landriau}, M. and {Le Guillou}, L. and {Meisner}, A. and {Miquel}, R. and {Moustakas}, J. and {Nadathur}, S. and {Prada}, F. and {P{\'e}rez-R{\`a}fols}, I. and {Rossi}, G. and {Sanchez}, E. and {Schubnell}, M. and {Siudek}, M. and {Sprayberry}, D. and {Tarl{\'e}}, G. and {Weaver}, B.~A. and {Zou}, H.},
        title = "{Identifying Anomalous DESI Galaxy Spectra with a Variational Autoencoder}",
      journal = {arXiv e-prints},
         year = 2025,
        month = jun,
          eid = {arXiv:2506.17376},
        pages = {arXiv:2506.17376},
          doi = {10.48550/arXiv.2506.17376},
archivePrefix = {arXiv},
       eprint = {2506.17376},
 primaryClass = {astro-ph.IM},
       adsurl = {https://ui.adsabs.harvard.edu/abs/2025arXiv250617376N}
}

@ARTICLE{suhBH,
       author = {{Suh}, Hyewon and {Civano}, Francesca and {Trakhtenbrot}, Benny and {Shankar}, Francesco and {Hasinger}, G{\"u}nther and {Sanders}, David B. and {Allevato}, Viola},
        title = "{No Significant Evolution of Relations between Black Hole Mass and Galaxy Total Stellar Mass Up to z {\ensuremath{\sim}} 2.5}",
      journal = {\apj},
         year = 2020,
        month = jan,
       volume = {889},
       number = {1},
          eid = {32},
        pages = {32},
          doi = {10.3847/1538-4357/ab5f5f},
archivePrefix = {arXiv},
       eprint = {1912.02824},
 primaryClass = {astro-ph.GA},
       adsurl = {https://ui.adsabs.harvard.edu/abs/2020ApJ...889...32S}
}

@ARTICLE{MoranBH,
       author = {{Moran}, Edward C. and {Shahinyan}, Karlen and {Sugarman}, Hannah R. and {V{\'e}lez}, Darik O. and {Eracleous}, Michael},
        title = "{Black Holes At the Centers of Nearby Dwarf Galaxies}",
      journal = {\aj},
         year = 2014,
        month = dec,
       volume = {148},
       number = {6},
          eid = {136},
        pages = {136},
          doi = {10.1088/0004-6256/148/6/136},
archivePrefix = {arXiv},
       eprint = {1408.4451},
 primaryClass = {astro-ph.GA},
       adsurl = {https://ui.adsabs.harvard.edu/abs/2014AJ....148..136M}
}

@article{Miller_2024,
doi = {10.3847/1538-3881/ad45fe},
url = {https://dx.doi.org/10.3847/1538-3881/ad45fe},
year = {2024},
month = {jul},
publisher = {The American Astronomical Society},
volume = {168},
number = {2},
pages = {95},
author = {Miller, Timothy N. and Doel, Peter and Gutierrez, Gaston and Besuner, Robert and Brooks, David and Gallo, Giuseppe and Heetderks, Henry and Jelinsky, Patrick and Kent, Stephen M. and Lampton, Michael and Levi, Michael E. and Liang, Ming and Meisner, Aaron and Sholl, Michael J. and Silber, Joseph Harry and Sprayberry, David and Aguilar, Jessica Nicole and de la Macorra, Axel and Eisenstein, Daniel and Fanning, Kevin and Font-Ribera, Andreu and Gaztañaga, Enrique and Gontcho A Gontcho, Satya and Honscheid, Klaus and Jimenez, Jorge and Joyce, Dick and Kehoe, Robert and Kisner, Theodore and Kremin, Anthony and Landriau, Martin and Le Guillou, Laurent and Magneville, Christophe and Martini, Paul and Miquel, Ramon and Moustakas, John and Nie, Jundan and Percival, Will and Poppett, Claire and Prada, Francisco and Rossi, Graziano and Schlegel, David and Schubnell, Michael and Seo, Hee-Jong and Sharples, Ray and Tarlé, Gregory and Vargas-Magaña, Mariana and Zhou, Zhimin and the DESI Collaboration},
title = {The Optical Corrector for the Dark Energy Spectroscopic Instrument},
journal = {The Astronomical Journal}
}

@ARTICLE{DESI2016b.Instr,
author = {{DESI Collaboration} and {Aghamousa}, Amir and {Aguilar}, Jessica and {Ahlen}, Steve and {Alam}, Shadab and {Allen}, Lori E. and {Allende Prieto}, Carlos and {Annis}, James and {Bailey}, Stephen and {Balland}, Christophe and {Ballester}, Otger and {Baltay}, Charles and {Beaufore}, Lucas and {Bebek}, Chris and {Beers}, Timothy C. and {Bell}, Eric F. and {Bernal}, Jos{\'e} Luis and {Besuner}, Robert and {Beutler}, Florian and {Blake}, Chris and {Bleuler}, Hannes and {Blomqvist}, Michael and {Blum}, Robert and {Bolton}, Adam S. and {Briceno}, Cesar and {Brooks}, David and {Brownstein}, Joel R. and {Buckley-Geer}, Elizabeth and {Burden}, Angela and {Burtin}, Etienne and {Busca}, Nicolas G. and {Cahn}, Robert N. and {Cai}, Yan-Chuan and {Cardiel-Sas}, Laia and {Carlberg}, Raymond G. and {Carton}, Pierre-Henri and {Casas}, Ricard and {Castander}, Francisco J. and {Cervantes-Cota}, Jorge L. and {Claybaugh}, Todd M. and {Close}, Madeline and {Coker}, Carl T. and {Cole}, Shaun and {Comparat}, Johan and {Cooper}, Andrew P. and {Cousinou}, M. -C. and {Crocce}, Martin and {Cuby}, Jean-Gabriel and {Cunningham}, Daniel P. and {Davis}, Tamara M. and {Dawson}, Kyle S. and {de la Macorra}, Axel and {De Vicente}, Juan and {Delubac}, Timoth{\'e}e and {Derwent}, Mark and {Dey}, Arjun and {Dhungana}, Govinda and {Ding}, Zhejie and {Doel}, Peter and {Duan}, Yutong T. and {Ealet}, Anne and {Edelstein}, Jerry and {Eftekharzadeh}, Sarah and {Eisenstein}, Daniel J. and {Elliott}, Ann and {Escoffier}, St{\'e}phanie and {Evatt}, Matthew and {Fagrelius}, Parker and {Fan}, Xiaohui and {Fanning}, Kevin and {Farahi}, Arya and {Farihi}, Jay and {Favole}, Ginevra and {Feng}, Yu and {Fernandez}, Enrique and {Findlay}, Joseph R. and {Finkbeiner}, Douglas P. and {Fitzpatrick}, Michael J. and {Flaugher}, Brenna and {Flender}, Samuel and {Font-Ribera}, Andreu and {Forero-Romero}, Jaime E. and {Fosalba}, Pablo and {Frenk}, Carlos S. and {Fumagalli}, Michele and {Gaensicke}, Boris T. and {Gallo}, Giuseppe and {Garcia-Bellido}, Juan and {Gaztanaga}, Enrique and {Pietro Gentile Fusillo}, Nicola and {Gerard}, Terry and {Gershkovich}, Irena and {Giannantonio}, Tommaso and {Gillet}, Denis and {Gonzalez-de-Rivera}, Guillermo and {Gonzalez-Perez}, Violeta and {Gott}, Shelby and {Graur}, Or and {Gutierrez}, Gaston and {Guy}, Julien and {Habib}, Salman and {Heetderks}, Henry and {Heetderks}, Ian and {Heitmann}, Katrin and {Hellwing}, Wojciech A. and {Herrera}, David A. and {Ho}, Shirley and {Holland}, Stephen and {Honscheid}, Klaus and {Huff}, Eric and {Hutchinson}, Timothy A. and {Huterer}, Dragan and {Hwang}, Ho Seong and {Illa Laguna}, Joseph Maria and {Ishikawa}, Yuzo and {Jacobs}, Dianna and {Jeffrey}, Niall and {Jelinsky}, Patrick and {Jennings}, Elise and {Jiang}, Linhua and {Jimenez}, Jorge and {Johnson}, Jennifer and {Joyce}, Richard and {Jullo}, Eric and {Juneau}, St{\'e}phanie and {Kama}, Sami and {Karcher}, Armin and {Karkar}, Sonia and {Kehoe}, Robert and {Kennamer}, Noble and {Kent}, Stephen and {Kilbinger}, Martin and {Kim}, Alex G. and {Kirkby}, David and {Kisner}, Theodore and {Kitanidis}, Ellie and {Kneib}, Jean-Paul and {Koposov}, Sergey and {Kovacs}, Eve and {Koyama}, Kazuya and {Kremin}, Anthony and {Kron}, Richard and {Kronig}, Luzius and {Kueter-Young}, Andrea and {Lacey}, Cedric G. and {Lafever}, Robin and {Lahav}, Ofer and {Lambert}, Andrew and {Lampton}, Michael and {Landriau}, Martin and {Lang}, Dustin and {Lauer}, Tod R. and {Le Goff}, Jean-Marc and {Le Guillou}, Laurent and {Le Van Suu}, Auguste and {Lee}, Jae Hyeon and {Lee}, Su-Jeong and {Leitner}, Daniela and {Lesser}, Michael and {Levi}, Michael E. and {L'Huillier}, Benjamin and {Li}, Baojiu and {Liang}, Ming and {Lin}, Huan and {Linder}, Eric and {Loebman}, Sarah R. and {Luki{\'c}}, Zarija and {Ma}, Jun and {MacCrann}, Niall and {Magneville}, Christophe and {Makarem}, Laleh and {Manera}, Marc and {Manser}, Christopher J. and {Marshall}, Robert and {Martini}, Paul and {Massey}, Richard and {Matheson}, Thomas and {McCauley}, Jeremy and {McDonald}, Patrick and {McGreer}, Ian D. and {Meisner}, Aaron and {Metcalfe}, Nigel and {Miller}, Timothy N. and {Miquel}, Ramon and {Moustakas}, John and {Myers}, Adam and {Naik}, Milind and {Newman}, Jeffrey A. and {Nichol}, Robert C. and {Nicola}, Andrina and {Nicolati da Costa}, Luiz and {Nie}, Jundan and {Niz}, Gustavo and {Norberg}, Peder and {Nord}, Brian and {Norman}, Dara and {Nugent}, Peter and {O'Brien}, Thomas and {Oh}, Minji and {Olsen}, Knut A.~G. and {Padilla}, Cristobal and {Padmanabhan}, Hamsa and {Padmanabhan}, Nikhil and {Palanque-Delabrouille}, Nathalie and {Palmese}, Antonella and {Pappalardo}, Daniel and {P{\^a}ris}, Isabelle and {Park}, Changbom and {Patej}, Anna and {Peacock}, John A. and {Peiris}, Hiranya V. and {Peng}, Xiyan and {Percival}, Will J. and {Perruchot}, Sandrine and {Pieri}, Matthew M. and {Pogge}, Richard and {Pollack}, Jennifer E. and {Poppett}, Claire and {Prada}, Francisco and {Prakash}, Abhishek and {Probst}, Ronald G. and {Rabinowitz}, David and {Raichoor}, Anand and {Ree}, Chang Hee and {Refregier}, Alexandre and {Regal}, Xavier and {Reid}, Beth and {Reil}, Kevin and {Rezaie}, Mehdi and {Rockosi}, Constance M. and {Roe}, Natalie and {Ronayette}, Samuel and {Roodman}, Aaron and {Ross}, Ashley J. and {Ross}, Nicholas P. and {Rossi}, Graziano and {Rozo}, Eduardo and {Ruhlmann-Kleider}, Vanina and {Rykoff}, Eli S. and {Sabiu}, Cristiano and {Samushia}, Lado and {Sanchez}, Eusebio and {Sanchez}, Javier and {Schlegel}, David J. and {Schneider}, Michael and {Schubnell}, Michael and {Secroun}, Aur{\'e}lia and {Seljak}, Uros and {Seo}, Hee-Jong and {Serrano}, Santiago and {Shafieloo}, Arman and {Shan}, Huanyuan and {Sharples}, Ray and {Sholl}, Michael J. and {Shourt}, William V. and {Silber}, Joseph H. and {Silva}, David R. and {Sirk}, Martin M. and {Slosar}, Anze and {Smith}, Alex and {Smoot}, George F. and {Som}, Debopam and {Song}, Yong-Seon and {Sprayberry}, David and {Staten}, Ryan and {Stefanik}, Andy and {Tarle}, Gregory and {Sien Tie}, Suk and {Tinker}, Jeremy L. and {Tojeiro}, Rita and {Valdes}, Francisco and {Valenzuela}, Octavio and {Valluri}, Monica and {Vargas-Magana}, Mariana and {Verde}, Licia and {Walker}, Alistair R. and {Wang}, Jiali and {Wang}, Yuting and {Weaver}, Benjamin A. and {Weaverdyck}, Curtis and {Wechsler}, Risa H. and {Weinberg}, David H. and {White}, Martin and {Yang}, Qian and {Yeche}, Christophe and {Zhang}, Tianmeng and {Zhao}, Gong-Bo and {Zheng}, Yi and {Zhou}, Xu and {Zhou}, Zhimin and {Zhu}, Yaling and {Zou}, Hu and {Zu}, Ying},
title = "{The DESI Experiment Part II: Instrument Design}",
journal = {arXiv e-prints},
year = 2016,
month = oct,
eid = {arXiv:1611.00037},
pages = {arXiv:1611.00037},
archivePrefix = {arXiv},
eprint = {1611.00037},
primaryClass = {astro-ph.IM},
adsurl = {https://ui.adsabs.harvard.edu/abs/2016arXiv161100037D}
	}

@article{Anand_2024_redrock,
doi = {10.3847/1538-3881/ad60c2},
url = {https://dx.doi.org/10.3847/1538-3881/ad60c2},
year = {2024},
month = {aug},
publisher = {The American Astronomical Society},
volume = {168},
number = {3},
pages = {124},
author = {Anand, Abhijeet and Guy, Julien and Bailey, Stephen and Moustakas, John and Aguilar, J. and Ahlen, S. and Bolton, A. S. and Brodzeller, A. and Brooks, D. and Claybaugh, T. and Cole, S. and de la Macorra, A. and Dey, Biprateep and Fanning, K. and Forero-Romero, J. E. and Gaztañaga, E. and Gontcho A Gontcho, S. and Gutierrez, G. and Honscheid, K. and Howlett, C. and Juneau, S. and Kirkby, D. and Kisner, T. and Kremin, A. and Lambert, A. and Landriau, M. and Le Guillou, L. and Manera, M. and Meisner, A. and Miquel, R. and Mueller, E. and Niz, G. and Palanque-Delabrouille, N. and Percival, W. J. and Poppett, C. and Prada, F. and Raichoor, A. and Rezaie, M. and Rossi, G. and Sanchez, E. and Schlafly, E. F. and Schlegel, D. and Schubnell, M. and Sprayberry, D. and Tarlé, G. and Warner, C. and Weaver, B. A. and Zhou, R. and Zou, H.},
title = {Archetype-based Redshift Estimation for the Dark Energy Spectroscopic Instrument Survey},
journal = {The Astronomical Journal}
}

@inproceedings{XGBoost,
author = {Chen, Tianqi and Guestrin, Carlos},
title = {XGBoost: A Scalable Tree Boosting System},
year = {2016},
isbn = {9781450342322},
publisher = {Association for Computing Machinery},
address = {New York, NY, USA},
url = {https://doi.org/10.1145/2939672.2939785},
doi = {10.1145/2939672.2939785},
booktitle = {Proceedings of the 22nd ACM SIGKDD International Conference on Knowledge Discovery and Data Mining},
pages = {785–794},
numpages = {10},
location = {San Francisco, California, USA},
series = {KDD '16}
}

@article {AGN_obscure,
	title = {Obscured Active Galactic Nuclei},
	doi = {10.1146/annurev-astro-081817-051803},
	issn = {0066-4146},
	journal = {Annual Review of Astronomy and Astrophysics},
	pages = {625-671},
	publicationstatus = {Published},
	publisher = {Annual Reviews},
	url = {https://durham-repository.worktribe.com/output/1222869},
	volume = {56},
	year = {2018},
	author = {Hickox, R.C and Alexander, D.M.}
}

@article{cid2,
    author = {Cid Fernandes, R. and Stasińska, G. and Mateus, A. and Vale Asari, N.},
    title = {A comprehensive classification of galaxies in the Sloan Digital Sky Survey: how to tell true from fake AGN?},
    journal = {Monthly Notices of the Royal Astronomical Society},
    volume = {413},
    number = {3},
    pages = {1687-1699},
    year = {2011},
    month = {05},
    issn = {0035-8711},
    doi = {10.1111/j.1365-2966.2011.18244.x},
    url = {https://doi.org/10.1111/j.1365-2966.2011.18244.x},
    eprint = {https://academic.oup.com/mnras/article-pdf/413/3/1687/2873756/mnras0413-1687.pdf},
}

@ARTICLE{Unified1,
       author = {{Antonucci}, Robert},
        title = "{Unified models for active galactic nuclei and quasars.}",
      journal = {\araa},
         year = 1993,
        month = jan,
       volume = {31},
        pages = {473-521},
          doi = {10.1146/annurev.aa.31.090193.002353},
       adsurl = {https://ui.adsabs.harvard.edu/abs/1993ARA&A..31..473A}
}

@article{Panda,
doi = {10.3847/1538-4365/ad344f},
url = {https://doi.org/10.3847/1538-4365/ad344f},
year = {2024},
month = {apr},
publisher = {The American Astronomical Society},
volume = {272},
number = {1},
pages = {13},
author = {Panda, Swayamtrupta and Śniegowska, Marzena},
title = {Changing-look Active Galactic Nuclei. I. Tracking the Transition on the Main Sequence of Quasars},
journal = {The Astrophysical Journal Supplement Series}
}

@ARTICLE{Unified2,
       author = {{Urry}, C. Megan and {Padovani}, Paolo},
        title = "{Unified Schemes for Radio-Loud Active Galactic Nuclei}",
      journal = {\pasp},
         year = 1995,
        month = sep,
       volume = {107},
        pages = {803},
          doi = {10.1086/133630},
archivePrefix = {arXiv},
       eprint = {astro-ph/9506063},
 primaryClass = {astro-ph},
       adsurl = {https://ui.adsabs.harvard.edu/abs/1995PASP..107..803U}
}

@ARTICLE{PAE,
       author = {{B{\"o}hm}, Vanessa and {Kim}, Alex G. and {Juneau}, St{\'e}phanie},
        title = "{Fast and efficient identification of anomalous galaxy spectra with neural density estimation}",
      journal = {\mnras},
         year = 2023,
        month = dec,
       volume = {526},
       number = {2},
        pages = {3072-3087},
          doi = {10.1093/mnras/stad2773},
archivePrefix = {arXiv},
       eprint = {2308.00752},
 primaryClass = {astro-ph.IM},
       adsurl = {https://ui.adsabs.harvard.edu/abs/2023MNRAS.526.3072B}
}

@ARTICLE{trump,
       author = {{Trump}, Jonathan R. and {Sun}, Mouyuan and {Zeimann}, Gregory R. and {Luck}, Cuyler and {Bridge}, Joanna S. and {Grier}, Catherine J. and {Hagen}, Alex and {Juneau}, Stephanie and {Montero-Dorta}, Antonio and {Rosario}, David J. and {Brandt}, W. Niel and {Ciardullo}, Robin and {Schneider}, Donald P.},
        title = "{The Biases of Optical Line-Ratio Selection for Active Galactic Nuclei and the Intrinsic Relationship between Black Hole Accretion and Galaxy Star Formation}",
      journal = {\apj},
         year = 2015,
        month = sep,
       volume = {811},
       number = {1},
          eid = {26},
        pages = {26},
          doi = {10.1088/0004-637X/811/1/26},
archivePrefix = {arXiv},
       eprint = {1501.02801},
 primaryClass = {astro-ph.GA},
       adsurl = {https://ui.adsabs.harvard.edu/abs/2015ApJ...811...26T}
}

@ARTICLE{Juneau2013,
       author = {{Juneau}, St{\'e}phanie and {Dickinson}, Mark and {Bournaud}, Fr{\'e}d{\'e}ric and {Alexander}, David M. and {Daddi}, Emanuele and {Mullaney}, James R. and {Magnelli}, Benjamin and {Kartaltepe}, Jeyhan S. and {Hwang}, Ho Seong and {Willner}, S.~P. and {Coil}, Alison L. and {Rosario}, David J. and {Trump}, Jonathan R. and {Weiner}, Benjamin J. and {Willmer}, Christopher N.~A. and {Cooper}, Michael C. and {Elbaz}, David and {Faber}, S.~M. and {Frayer}, David T. and {Kocevski}, Dale D. and {Laird}, Elise S. and {Monkiewicz}, Jacqueline A. and {Nandra}, Kirpal and {Newman}, Jeffrey A. and {Salim}, Samir and {Symeonidis}, Myrto},
        title = "{Widespread and Hidden Active Galactic Nuclei in Star-forming Galaxies at Redshift >0.3}",
      journal = {\apj},
         year = 2013,
        month = feb,
       volume = {764},
       number = {2},
          eid = {176},
        pages = {176},
          doi = {10.1088/0004-637X/764/2/176},
archivePrefix = {arXiv},
       eprint = {1211.6436},
 primaryClass = {astro-ph.CO},
       adsurl = {https://ui.adsabs.harvard.edu/abs/2013ApJ...764..176J}
}

@ARTICLE{cid,
       author = {{Cid Fernandes}, R. and {Stasi{\'n}ska}, G. and {Schlickmann}, M.~S. and {Mateus}, A. and {Vale Asari}, N. and {Schoenell}, W. and {Sodr{\'e}}, L.},
        title = "{Alternative diagnostic diagrams and the `forgotten' population of weak line galaxies in the SDSS}",
      journal = {\mnras},
         year = 2010,
        month = apr,
       volume = {403},
       number = {2},
        pages = {1036-1053},
          doi = {10.1111/j.1365-2966.2009.16185.x},
archivePrefix = {arXiv},
       eprint = {0912.1643},
 primaryClass = {astro-ph.CO},
       adsurl = {https://ui.adsabs.harvard.edu/abs/2010MNRAS.403.1036C}
}

@ARTICLE{sanchez,
       author = {{S{\'a}nchez}, S.~F. and {Mu{\~n}oz-Tu{\~n}{\'o}n}, C. and {S{\'a}nchez Almeida}, J. and {Gonz{\'a}lez-Mart{\'\i}n}, O. and {P{\'e}rez}, E.},
        title = "{Beyond diagnostic-diagrams: A critical exploration of the classification of ionization processes}",
      journal = {\aap},
         year = 2025,
        month = dec,
       volume = {704},
          eid = {A145},
        pages = {A145},
          doi = {10.1051/0004-6361/202556809},
archivePrefix = {arXiv},
       eprint = {2510.07256},
 primaryClass = {astro-ph.GA},
       adsurl = {https://ui.adsabs.harvard.edu/abs/2025A&A...704A.145S}
}

@article{Kristensen,
    author = {Kristensen, Mikkel T and Pimbblet, Kevin and Penny, Samantha},
    title = {Environments of dwarf galaxies with optical AGN characteristics},
    journal = {Monthly Notices of the Royal Astronomical Society},
    volume = {496},
    number = {3},
    pages = {2577-2590},
    year = {2020},
    month = {06},
    issn = {0035-8711},
    doi = {10.1093/mnras/staa1719},
    url = {https://doi.org/10.1093/mnras/staa1719},
    eprint = {https://academic.oup.com/mnras/article-pdf/496/3/2577/41143027/staa1719.pdf},
}

@article{Cann_2019,
doi = {10.3847/2041-8213/aaf88d},
url = {https://doi.org/10.3847/2041-8213/aaf88d},
year = {2019},
month = {jan},
publisher = {The American Astronomical Society},
volume = {870},
number = {1},
pages = {L2},
author = {Cann, Jenna M. and Satyapal, Shobita and Abel, Nicholas P. and Blecha, Laura and Mushotzky, Richard F. and Reynolds, Christopher S. and Secrest, Nathan J.},
title = {The Limitations of Optical Spectroscopic Diagnostics in Identifying Active Galactic Nuclei in the Low-mass Regime},
journal = {The Astrophysical Journal Letters}
}

@article{andrianomena2023,
  author = {Andrianomena, S. and Tang, H.},
  title = {Radio Galaxy Zoo: Leveraging latent space representations from variational autoencoder},
  journal = {arXiv e-prints},
  year = {2023},
  eprint = {2311.08331},
  archivePrefix = {arXiv},
  primaryClass = {astro-ph.GA}
}

@article{Agostino_2019,
doi = {10.3847/1538-4357/ab1094},
url = {https://doi.org/10.3847/1538-4357/ab1094},
year = {2019},
month = {apr},
publisher = {The American Astronomical Society},
volume = {876},
number = {1},
pages = {12},
author = {Agostino, Christopher J. and Salim, Samir},
title = {Crossing the Line: Active Galactic Nuclei in the Star-forming Region of the BPT Diagram},
journal = {The Astrophysical Journal}
}

@misc{pyqsofit,
       author = {{Guo}, Hengxiao and {Shen}, Yue and {Wang}, Shu},
        title = "{PyQSOFit: Python code to fit the spectrum of quasars}",
 howpublished= {Astrophysics Source Code Library, record ascl:1809.008},
         year = 2018,
        month = sep,
          eid = {ascl:1809.008},
archivePrefix = {ascl},
       eprint = {1809.008},
       adsurl = {https://ui.adsabs.harvard.edu/abs/2018ascl.soft09008G}
}

@article{Baldassare_2016,
   title={Multi-epoch Spectroscopy of Dwarf Galaxies with AGN Signatures: Identifying Sources with Persistent Broad Hα emission},
   volume={829},
   ISSN={1538-4357},
   url={http://dx.doi.org/10.3847/0004-637X/829/1/57},
   DOI={10.3847/0004-637x/829/1/57},
   number={1},
   journal={The Astrophysical Journal},
   publisher={American Astronomical Society},
   author={Baldassare, Vivienne F. and Reines, Amy E. and Gallo, Elena and Greene, Jenny E. and Graur, Or and Geha, Marla and Hainline, Kevin and Carroll, Christopher M. and Hickox, Ryan C.},
   year={2016},
   month=sep, pages={57} }

@ARTICLE{Silber_2023,
 author = {{Silber}, Joseph Harry and {Fagrelius}, Parker and {Fanning}, Kevin and {Schubnell}, Michael and {Aguilar}, Jessica Nicole and {Ahlen}, Steven and {Ameel}, Jon and {Ballester}, Otger and {Baltay}, Charles and {Bebek}, Chris and {Benton Beard}, Dominic and {Besuner}, Robert and {Cardiel-Sas}, Laia and {Casas}, Ricard and {Castander}, Francisco Javier and {Claybaugh}, Todd and {Dobson}, Carl and {Duan}, Yutong and {Dunlop}, Patrick and {Edelstein}, Jerry and {Emmet}, William T. and {Elliott}, Ann and {Evatt}, Matthew and {Gershkovich}, Irena and {Guy}, Julien and {Harris}, Stu and {Heetderks}, Henry and {Heetderks}, Ian and {Honscheid}, Klaus and {Illa}, Jose Maria and {Jelinsky}, Patrick and {Jelinsky}, Sharon R. and {Jimenez}, Jorge and {Karcher}, Armin and {Kent}, Stephen and {Kirkby}, David and {Kneib}, Jean-Paul and {Lambert}, Andrew and {Lampton}, Mike and {Leitner}, Daniela and {Levi}, Michael and {McCauley}, Jeremy and {Meisner}, Aaron and {Miller}, Timothy N. and {Miquel}, Ramon and {Mundet}, Juli{\'a} and {Poppett}, Claire and {Rabinowitz}, David and {Reil}, Kevin and {Roman}, David and {Schlegel}, David and {Serrano}, Santiago and {Van Shourt}, William and {Sprayberry}, David and {Tarl{\'e}}, Gregory and {Tie}, Suk Sien and {Weaverdyck}, Curtis and {Zhang}, Kai and {Azzaro}, Marco and {Bailey}, Stephen and {Becerril}, Santiago and {Blackwell}, Tami and {Bouri}, Mohamed and {Brooks}, David and {Buckley-Geer}, Elizabeth and {Castro}, Jose Pe{\~n}ate and {Derwent}, Mark and {Dey}, Arjun and {Dhungana}, Govinda and {Doel}, Peter and {Eisenstein}, Daniel J. and {Fahim}, Nasib and {Garcia-Bellido}, Juan and {Gazta{\~n}aga}, Enrique and {A Gontcho}, Satya Gontcho and {Gutierrez}, Gaston and {H{\"o}rler}, Philipp and {Kehoe}, Robert and {Kisner}, Theodore and {Kremin}, Anthony and {Kronig}, Luzius and {Landriau}, Martin and {Le Guillou}, Laurent and {Martini}, Paul and {Moustakas}, John and {Palanque-Delabrouille}, Nathalie and {Peng}, Xiyan and {Percival}, Will and {Prada}, Francisco and {Allende Prieto}, Carlos and {de Rivera}, Guillermo Gonzalez and {Sanchez}, Eusebio and {Sanchez}, Justo and {Sharples}, Ray and {Soares-Santos}, Marcelle and {Schlafly}, Edward and {Weaver}, Benjamin Alan and {Zhou}, Zhimin and {Zhu}, Yaling and {Zou}, Hu and {DESI Collaboration}},
title = "{The Robotic Multiobject Focal Plane System of the Dark Energy Spectroscopic Instrument (DESI)}",
journal = {\aj},
year = 2023,
month = jan,
volume = {165},
 number = {1},
eid = {9},
pages = {9},
doi = {10.3847/1538-3881/ac9ab1},
archivePrefix = {arXiv},
eprint = {2205.09014},
primaryClass = {astro-ph.IM},
adsurl = {https://ui.adsabs.harvard.edu/abs/2023AJ....165....9S}
}

@article{DESI_Collaboration_2022,
   title={Overview of the Instrumentation for the Dark Energy Spectroscopic Instrument},
   volume={164},
   ISSN={1538-3881},
   url={http://dx.doi.org/10.3847/1538-3881/ac882b},
   DOI={10.3847/1538-3881/ac882b},
   number={5},
   journal={The Astronomical Journal},
   publisher={American Astronomical Society},
   author={{DESI Collaboration} and Abareshi, B. and Aguilar, J. and Ahlen, S. and Alam, Shadab and Alexander, David M. and Alfarsy, R. and Allen, L. and Allende Prieto, C. and Alves, O. and Ameel, J. and Armengaud, E. and Asorey, J. and Aviles, Alejandro and Bailey, S. and Balaguera-Antolínez, A. and Ballester, O. and Baltay, C. and Bault, A. and Beltran, S. F. and Benavides, B. and BenZvi, S. and Berti, A. and Besuner, R. and Beutler, Florian and Bianchi, D. and Blake, C. and Blanc, P. and Blum, R. and Bolton, A. and Bose, S. and Bramall, D. and Brieden, S. and Brodzeller, A. and Brooks, D. and Brownewell, C. and Buckley-Geer, E. and Cahn, R. N. and Cai, Z. and Canning, R. and Capasso, R. and Carnero Rosell, A. and Carton, P. and Casas, R. and Castander, F. J. and Cervantes-Cota, J. L. and Chabanier, S. and Chaussidon, E. and Chuang, C. and Circosta, C. and Cole, S. and Cooper, A. P. and da Costa, L. and Cousinou, M.-C. and Cuceu, A. and Davis, T. M. and Dawson, K. and de la Cruz-Noriega, R. and de la Macorra, A. and de Mattia, A. and Della Costa, J. and Demmer, P. and Derwent, M. and Dey, A. and Dey, B. and Dhungana, G. and Ding, Z. and Dobson, C. and Doel, P. and Donald-McCann, J. and Donaldson, J. and Douglass, K. and Duan, Y. and Dunlop, P. and Edelstein, J. and Eftekharzadeh, S. and Eisenstein, D. J. and Enriquez-Vargas, M. and Escoffier, S. and Evatt, M. and Fagrelius, P. and Fan, X. and Fanning, K. and Fawcett, V. A. and Ferraro, S. and Ereza, J. and Flaugher, B. and Font-Ribera, A. and Forero-Romero, J. E. and Frenk, C. S. and Fromenteau, S. and Gänsicke, B. T. and Garcia-Quintero, C. and Garrison, L. and Gaztañaga, E. and Gerardi, F. and Gil-Marín, H. and Gontcho A Gontcho, S. and Gonzalez-Morales, Alma X. and Gonzalez-de-Rivera, G. and Gonzalez-Perez, V. and Gordon, C. and Graur, O. and Green, D. and Grove, C. and Gruen, D. and Gutierrez, G. and Guy, J. and Hahn, C. and Harris, S. and Herrera, D. and Herrera-Alcantar, Hiram K. and Honscheid, K. and Howlett, C. and Huterer, D. and Iršič, V. and Ishak, M. and Jelinsky, P. and Jiang, L. and Jimenez, J. and Jing, Y. P. and Joyce, R. and Jullo, E. and Juneau, S. and Karaçaylı, N. G. and Karamanis, M. and Karcher, A. and Karim, T. and Kehoe, R. and Kent, S. and Kirkby, D. and Kisner, T. and Kitaura, F. and Koposov, S. E. and Kovács, A. and Kremin, A. and Krolewski, Alex and L’Huillier, B. and Lahav, O. and Lambert, A. and Lamman, C. and Lan, Ting-Wen and Landriau, M. and Lane, S. and Lang, D. and Lange, J. U. and Lasker, J. and Le Guillou, L. and Leauthaud, A. and Le Van Suu, A. and Levi, Michael E. and Li, T. S. and Magneville, C. and Manera, M. and Manser, Christopher J. and Marshall, B. and Martini, Paul and McCollam, W. and McDonald, P. and Meisner, Aaron M. and Mena-Fernández, J. and Meneses-Rizo, J. and Mezcua, M. and Miller, T. and Miquel, R. and Montero-Camacho, P. and Moon, J. and Moustakas, J. and Mueller, E. and Muñoz-Gutiérrez, Andrea and Myers, Adam D. and Nadathur, S. and Najita, J. and Napolitano, L. and Neilsen, E. and Newman, Jeffrey A. and Nie, J. D. and Ning, Y. and Niz, G. and Norberg, P. and Noriega, Hernán E. and O’Brien, T. and Obuljen, A. and Palanque-Delabrouille, N. and Palmese, A. and Zhiwei, P. and Pappalardo, D. and PENG, X. and Percival, W. J. and Perruchot, S. and Pogge, R. and Poppett, C. and Porredon, A. and Prada, F. and Prochaska, J. and Pucha, R. and Pérez-Fernández, A. and Pérez-Ràfols, I. and Rabinowitz, D. and Raichoor, A. and Ramirez-Solano, S. and Ramírez-Pérez, César and Ravoux, C. and Reil, K. and Rezaie, M. and Rocher, A. and Rockosi, C. and Roe, N. A. and Roodman, A. and Ross, A. J. and Rossi, G. and Ruggeri, R. and Ruhlmann-Kleider, V. and Sabiu, C. G. and Gaines, S. and Said, K. and Saintonge, A. and Salas Catonga, Javier and Samushia, L. and Sanchez, E. and Saulder, C. and Schaan, E. and Schlafly, E. and Schlegel, D. and Schmoll, J. and Scholte, D. and Schubnell, M. and Secroun, A. and Seo, H. and Serrano, S. and Sharples, Ray M. and Sholl, Michael J. and Silber, Joseph Harry and Silva, D. R. and Sirk, M. and Siudek, M. and Smith, A. and Sprayberry, D. and Staten, R. and Stupak, B. and Tan, T. and Tarlé, Gregory and Tie, Suk Sien and Tojeiro, R. and Ureña-López, L. A. and Valdes, F. and Valenzuela, O. and Valluri, M. and Vargas-Magaña, M. and Verde, L. and Walther, M. and Wang, B. and Wang, M. S. and Weaver, B. A. and Weaverdyck, C. and Wechsler, R. and Wilson, Michael J. and Yang, J. and Yu, Y. and Yuan, S. and Yèche, Christophe and Zhang, H. and Zhang, K. and Zhao, Cheng and Zhou, Rongpu and Zhou, Zhimin and Zou, H. and Zou, J. and Zou, S. and Zu, Y.},
   year={2022},
   month=oct, pages={207} }

@article{Dey_2019,
doi = {10.3847/1538-3881/ab089d},
url = {https://doi.org/10.3847/1538-3881/ab089d},
year = {2019},
month = {apr},
publisher = {The American Astronomical Society},
volume = {157},
number = {5},
pages = {168},
author = {Dey, Arjun and Schlegel, David J. and Lang, Dustin and Blum, Robert and Burleigh, Kaylan and Fan, Xiaohui and Findlay, Joseph R. and Finkbeiner, Doug and Herrera, David and Juneau, Stéphanie and Landriau, Martin and Levi, Michael and McGreer, Ian and Meisner, Aaron and Myers, Adam D. and Moustakas, John and Nugent, Peter and Patej, Anna and Schlafly, Edward F. and Walker, Alistair R. and Valdes, Francisco and Weaver, Benjamin A. and Yèche, Christophe and Zou, Hu and Zhou, Xu and Abareshi, Behzad and Abbott, T. M. C. and Abolfathi, Bela and Aguilera, C. and Alam, Shadab and Allen, Lori and Alvarez, A. and Annis, James and Ansarinejad, Behzad and Aubert, Marie and Beechert, Jacqueline and Bell, Eric F. and BenZvi, Segev Y. and Beutler, Florian and Bielby, Richard M. and Bolton, Adam S. and Briceño, César and Buckley-Geer, Elizabeth J. and Butler, Karen and Calamida, Annalisa and Carlberg, Raymond G. and Carter, Paul and Casas, Ricard and Castander, Francisco J. and Choi, Yumi and Comparat, Johan and Cukanovaite, Elena and Delubac, Timothée and DeVries, Kaitlin and Dey, Sharmila and Dhungana, Govinda and Dickinson, Mark and Ding, Zhejie and Donaldson, John B. and Duan, Yutong and Duckworth, Christopher J. and Eftekharzadeh, Sarah and Eisenstein, Daniel J. and Etourneau, Thomas and Fagrelius, Parker A. and Farihi, Jay and Fitzpatrick, Mike and Font-Ribera, Andreu and Fulmer, Leah and Gänsicke, Boris T. and Gaztanaga, Enrique and George, Koshy and Gerdes, David W. and A Gontcho, Satya Gontcho and Gorgoni, Claudio and Green, Gregory and Guy, Julien and Harmer, Diane and Hernandez, M. and Honscheid, Klaus and Huang, Lijuan (Wendy) and James, David J. and Jannuzi, Buell T. and Jiang, Linhua and Joyce, Richard and Karcher, Armin and Karkar, Sonia and Kehoe, Robert and Kneib, Jean-Paul and Kueter-Young, Andrea and Lan, Ting-Wen and Lauer, Tod R. and Guillou, Laurent Le and Van Suu, Auguste Le and Lee, Jae Hyeon and Lesser, Michael and Levasseur, Laurence Perreault and Li, Ting S. and Mann, Justin L. and Marshall, Robert and Martínez-Vázquez, C. E. and Martini, Paul and du Mas des Bourboux, Hélion and McManus, Sean and Meier, Tobias Gabriel and Ménard, Brice and Metcalfe, Nigel and Muñoz-Gutiérrez, Andrea and Najita, Joan and Napier, Kevin and Narayan, Gautham and Newman, Jeffrey A. and Nie, Jundan and Nord, Brian and Norman, Dara J. and Olsen, Knut A. G. and Paat, Anthony and Palanque-Delabrouille, Nathalie and Peng, Xiyan and Poppett, Claire L. and Poremba, Megan R. and Prakash, Abhishek and Rabinowitz, David and Raichoor, Anand and Rezaie, Mehdi and Robertson, A. N. and Roe, Natalie A. and Ross, Ashley J. and Ross, Nicholas P. and Rudnick, Gregory and Gaines, Sasha and Saha, Abhijit and Sánchez, F. Javier and Savary, Elodie and Schweiker, Heidi and Scott, Adam and Seo, Hee-Jong and Shan, Huanyuan and Silva, David R. and Slepian, Zachary and Soto, Christian and Sprayberry, David and Staten, Ryan and Stillman, Coley M. and Stupak, Robert J. and Summers, David L. and Tie, Suk Sien and Tirado, H. and Vargas-Magaña, Mariana and Vivas, A. Katherina and Wechsler, Risa H. and Williams, Doug and Yang, Jinyi and Yang, Qian and Yapici, Tolga and Zaritsky, Dennis and Zenteno, A. and Zhang, Kai and Zhang, Tianmeng and Zhou, Rongpu and Zhou, Zhimin},
title = {Overview of the DESI Legacy Imaging Surveys},
journal = {The Astronomical Journal}
}

@ARTICLE{poppett,
       author = {{Poppett}, Claire and {Tyas}, Luke and {Aguilar}, J. and {Bebek}, Christopher and {Bramall}, D. and {Claybaugh}, T. and {Edelstein}, J. and {Fagrelius}, P. and {Heetderks}, H. and {Jelinsky}, P. and {Jelinsky}, S. and {Lafever}, Robin and {Lambert}, A. and {Lampton}, M. and {Levi}, Michael E. and {Martini}, P. and {Rockosi}, C. and {Schmoll}, J. and {Sharples}, Ray M. and {Sirk}, Martin and {Wishnow}, Edward and {Yu}, Jiaxi and {Ahlen}, S. and {Bault}, A. and {BenZvi}, S. and {Brooks}, D. and {Cole}, S. and {de la Macorra}, A. and {Dey}, Arjun and {Doel}, P. and {Fanning}, K. and {Font-Ribera}, A. and {Forero-Romero}, J.~E. and {Gazta{\~n}aga}, E. and {Gontcho A Gontcho}, S. and {Gonzalez-Morales}, A.~X. and {Hahn}, C. and {Honscheid}, K. and {Jimenez}, J. and {Juneau}, S. and {Kirkby}, D. and {Kremin}, A. and {Landriau}, M. and {Le Guillou}, L. and {Manera}, M. and {Meisner}, A. and {Miquel}, R. and {Moustakas}, J. and {Mueller}, E. and {Mu{\~n}oz-Guti{\'e}rrez}, A. and {Myers}, A.~D. and {Nie}, J. and {Niz}, G. and {Palanque-Delabrouille}, N. and {Percival}, W.~J. and {Prada}, F. and {Rabinowitz}, D. and {Rezaie}, M. and {Rossi}, G. and {Sanchez}, E. and {Schlafly}, Edward F. and {Schlegel}, D. and {Schubnell}, M. and {Seo}, H. and {Sprayberry}, D. and {Tarl{\'e}}, G. and {Vargas-Maga{\~n}a}, M. and {Weaver}, B.~A. and {Zhou}, R.},
        title = "{Overview of the Fiber System for the Dark Energy Spectroscopic Instrument}",
      journal = {\aj},
         year = 2024,
        month = dec,
       volume = {168},
       number = {6},
          eid = {245},
        pages = {245},
          doi = {10.3847/1538-3881/ad76a4},
       adsurl = {https://ui.adsabs.harvard.edu/abs/2024AJ....168..245P}
}

@misc{schlafly,
      title={Survey Operations for the Dark Energy Spectroscopic Instrument}, 
      author={E. F. Schlafly and D. Kirkby and D. J. Schlegel and A. D. Myers and A. Raichoor and K. Dawson and J. Aguilar and C. Allende Prieto and S. Bailey and S. BenZvi and J. Bermejo-Climent and D. Brooks and A. de la Macorra and Arjun Dey and P. Doel and K. Fanning and A. Font-Ribera and J. E. Forero-Romero and J. García-Bellido and S. Gontcho A Gontcho and J. Guy and C. Hahn and K. Honscheid and M. Ishak and S. Juneau and R. Kehoe and T. Kisner and A. Kremin and M. Landriau and D. A. Lang and J. Lasker and M. E. Levi and C. Magneville and C. J. Manser and P. Martini and A. M. Meisner and R. Miquel and J. Moustakas and J. A. Newman and Jundan Nie and N. Palanque-Delabrouille and W. J. Percival and C. Poppett and C. Rockosi and A. J. Ross and G. Rossi and G. Tarlé and B. A. Weaver and C. Yèche and R. Zhou},
      year={2024},
      eprint={2306.06309},
      archivePrefix={arXiv},
      primaryClass={astro-ph.CO},
      url={https://arxiv.org/abs/2306.06309}
}

@ARTICLE{cigale_boquien,
       author = {{Boquien}, M. and {Burgarella}, D. and {Roehlly}, Y. and {Buat}, V. and {Ciesla}, L. and {Corre}, D. and {Inoue}, A.~K. and {Salas}, H.},
        title = "{CIGALE: a python Code Investigating GALaxy Emission}",
      journal = {\aap},
         year = 2019,
        month = feb,
       volume = {622},
          eid = {A103},
        pages = {A103},
          doi = {10.1051/0004-6361/201834156},
archivePrefix = {arXiv},
       eprint = {1811.03094},
 primaryClass = {astro-ph.GA},
       adsurl = {https://ui.adsabs.harvard.edu/abs/2019A&A...622A.103B}
}

@ARTICLE{greeneBH,
       author = {{Greene}, Jenny E. and {Ho}, Luis C.},
        title = "{Estimating Black Hole Masses in Active Galaxies Using the H{\ensuremath{\alpha}} Emission Line}",
      journal = {\apj},
         year = 2005,
        month = sep,
       volume = {630},
       number = {1},
        pages = {122-129},
          doi = {10.1086/431897},
archivePrefix = {arXiv},
       eprint = {astro-ph/0508335},
 primaryClass = {astro-ph},
       adsurl = {https://ui.adsabs.harvard.edu/abs/2005ApJ...630..122G}
}

@article{Piotrowska2022,
  author = {Piotrowska, J. M. and Bluck, A. F. L. and Maiolino, R. and Peng, Y.},
  title = {On the quenching of star formation in observed and simulated central galaxies: evidence for the role of integrated AGN feedback},
  journal = {Monthly Notices of the Royal Astronomical Society},
  volume = {512},
  issue = {1},
  pages = {1052-1090},
  year = {2022},
  doi = {10.1093/mnras/stab3673},
  url = {https://academic.oup.com/mnras/article/512/1/1052/6482843}
}

@InProceedings{multiibranch,
  title = 	 {Deep Neural Networks with Multi-Branch Architectures Are Intrinsically Less Non-Convex},
  author =       {Zhang, Hongyang and Shao, Junru and Salakhutdinov, Ruslan},
  booktitle = 	 {Proceedings of the Twenty-Second International Conference on Artificial Intelligence and Statistics},
  pages = 	 {1099--1109},
  year = 	 {2019},
  editor = 	 {Chaudhuri, Kamalika and Sugiyama, Masashi},
  volume = 	 {89},
  series = 	 {Proceedings of Machine Learning Research},
  month = 	 {16--18 Apr},
  publisher =    {PMLR},
  url = 	 {https://proceedings.mlr.press/v89/zhang19d.html}
}

@article{DR2,
  title = {DESI DR2 results. I. Baryon acoustic oscillations from the Lyman alpha forest},
  author = {Abdul Karim, M. and Aguilar, J. and Ahlen, S. and Allende Prieto, C. and Alves, O. and Anand, A. and Andrade, U. and Armengaud, E. and Aviles, A. and Bailey, S. and Bault, A. and Behera, J. and BenZvi, S. and Bianchi, D. and Blake, C. and Brodzeller, A. and Brooks, D. and Buckley-Geer, E. and Burtin, E. and Calderon, R. and Canning, R. and Carnero Rosell, A. and Carrilho, P. and Casas, L. and Castander, F. J. and Cereskaite, R. and Charles, M. and Chaussidon, E. and Chaves-Montero, J. and Chebat, D. and Claybaugh, T. and Cole, S. and Cooper, A. P. and Cuceu, A. and Dawson, K. S. and de Belsunce, R. and de la Macorra, A. and de Mattia, A. and Deiosso, N. and Della Costa, J. and Dey, A. and Dey, B. and Ding, Z. and Doel, P. and Edelstein, J. and Eisenstein, D. J. and Elbers, W. and Fagrelius, P. and Fanning, K. and Ferraro, S. and Font-Ribera, A. and Forero-Romero, J. E. and Garcia-Quintero, C. and Garrison, L. H. and Gazta\~naga, E. and Gil-Mar\'{\i}n, H. and Gontcho, S. Gontcho A. and Gonzalez-Morales, A. X. and Gordon, C. and Green, D. and Gutierrez, G. and Guy, J. and Hahn, C. and Herbold, M. and Herrera-Alcantar, H. K. and Ho, M. and Ho, M.-F. and Honscheid, K. and Howlett, C. and Huterer, D. and Ishak, M. and Juneau, S. and Kara\ifmmode \mbox{\c{c}}\else \c{c}\fi{}ayl��, N. G. and Kehoe, R. and Kent, S. and Kirkby, D. and Kisner, T. and Kitaura, F.-S. and Koposov, S. E. and Kremin, A. and Lahav, O. and Lamman, C. and Landriau, M. and Lang, D. and Lasker, J. and Le Goff, J. M. and Le Guillou, L. and Leauthaud, A. and Levi, M. E. and Li, Q. and Li, T. S. and Lodha, K. and Lokken, M. and Magneville, C. and Manera, M. and Martini, P. and Matthewson, W. L. and McDonald, P. and Meisner, A. and Mena-Fern\'andez, J. and Miquel, R. and Moustakas, J. and Mu\~noz-Guti\'errez, A. and Mu\~noz-Santos, D. and Myers, A. D. and Newman, J. A. and Niz, G. and Noriega, H. E. and Paillas, E. and Palanque-Delabrouille, N. and Pan, J. and Percival, W. J. and P\'erez-R\`afols, I. and Pieri, M. M. and Poppett, C. and Prada, F. and Rabinowitz, D. and Raichoor, A. and Ram\'{\i}rez-P\'erez, C. and Rashkovetskyi, M. and Ravoux, C. and Rich, J. and Rockosi, C. and Ross, A. J. and Rossi, G. and Ruhlmann-Kleider, V. and Sanchez, E. and Sanders, N. and Satyavolu, S. and Schlegel, D. and Schubnell, M. and Seo, H. and Shafieloo, A. and Sharples, R. and Silber, J. and Sinigaglia, F. and Sprayberry, D. and Tan, T. and Tarl\'e, G. and Taylor, P. and Turner, W. and Valdes, F. and Vargas-Maga\~na, M. and Walther, M. and Weaver, B. A. and Wolfson, M. and Y\`eche, C. and Zarrouk, P. and Zhou, R. and Zou, H.},
  collaboration = {DESI Collaboration},
  journal = {Phys. Rev. D},
  volume = {112},
  issue = {8},
  pages = {083514},
  numpages = {28},
  year = {2025},
  month = {Oct},
  publisher = {American Physical Society},
  doi = {10.1103/2wwn-xjm5},
  url = {https://link.aps.org/doi/10.1103/2wwn-xjm5}
}

@misc{OneCycleLR,
  author = {{PyTorch}},
  title = {{torch.optim.lrscheduler.OneCycleLR} Documentation},
  year = {2025},
  howpublished = {\url{https://docs.pytorch.org/docs/stable/generated/torch.optim.lr_scheduler.OneCycleLR.html}}
}

@article{fusion_3,
title = {LFDT-Fusion: A latent feature-guided diffusion Transformer model for general image fusion},
journal = {Information Fusion},
volume = {113},
pages = {102639},
year = {2025},
issn = {1566-2535},
doi = {https://doi.org/10.1016/j.inffus.2024.102639},
url = {https://www.sciencedirect.com/science/article/pii/S1566253524004172},
author = {Bo Yang and Zhaohui Jiang and Dong Pan and Haoyang Yu and Gui Gui and Weihua Gui}}

@ARTICLE{fusion_2,
  author={Sharma, Anshul and Varman, Utkarsh and Bharti, Vandana and Kumar, Abhinav and Singh, Amit Kumar and Singh, Sanjay Kumar},
  journal={IEEE Journal of Biomedical and Health Informatics}, 
  title={ILAM: Cross-Fusion of Latent and Attention Features for Explainable Medical Image Classification}, 
  year={2025},
  volume={},
  number={},
  pages={1-11},
  doi={10.1109/JBHI.2025.3561024}}

@INPROCEEDINGS{fusion,
       author = {{Ross}, Arun A. and {Govindarajan}, Rohin},
        title = "{Feature level fusion of hand and face biometrics}",
    booktitle = {Biometric Technology for Human Identification II},
         year = 2005,
       editor = {{Jain}, Anil K. and {Ratha}, Nalini K.},
       series = {Society of Photo-Optical Instrumentation Engineers (SPIE) Conference Series},
       volume = {5779},
        month = mar,
        pages = {196-204},
          doi = {10.1117/12.606093},
       adsurl = {https://ui.adsabs.harvard.edu/abs/2005SPIE.5779..196R}
}

@ARTICLE{LINER,
       author = {{Kewley}, Lisa J. and {Groves}, Brent and {Kauffmann}, Guinevere and {Heckman}, Tim},
        title = "{The host galaxies and classification of active galactic nuclei}",
      journal = {\mnras},
         year = 2006,
        month = nov,
       volume = {372},
       number = {3},
        pages = {961-976},
          doi = {10.1111/j.1365-2966.2006.10859.x},
archivePrefix = {arXiv},
       eprint = {astro-ph/0605681},
 primaryClass = {astro-ph},
       adsurl = {https://ui.adsabs.harvard.edu/abs/2006MNRAS.372..961K}
}

@article{Tous2025,
  title={Fully comprehensive diagnostic of galaxy activity using principal components of visible spectra: implementation on nearby S0s},
  author={Tous, J. L. and Solanes, J. M. and Perea, J. D.},
  journal={Monthly Notices of the Royal Astronomical Society},
  volume={537},
  number={2},
  pages={1459--1469},
  year={2025},
  doi={10.1093/mnras/staf084},
  url={https://doi.org/10.1093/mnras/staf084}
}

@article{NUV,
    author = {Chilingarian, Igor V. and Zolotukhin, Ivan Yu.},
    title = {A universal ultraviolet–optical colour–colour–magnitude relation of galaxies},
    journal = {Monthly Notices of the Royal Astronomical Society},
    volume = {419},
    number = {2},
    pages = {1727-1739},
    year = {2011},
    month = {12},
issn = {0035-8711},
    doi = {10.1111/j.1365-2966.2011.19837.x},
    url = {https://doi.org/10.1111/j.1365-2966.2011.19837.x},
    eprint = {https://academic.oup.com/mnras/article-pdf/419/2/1727/3129713/mnras0419-1727.pdf},
}

@ARTICLE{kewely,
       author = {{Kewley}, L.~J. and {Dopita}, M.~A. and {Sutherland}, R.~S. and {Heisler}, C.~A. and {Trevena}, J.},
        title = "{Theoretical Modeling of Starburst Galaxies}",
      journal = {The Astrophysical Journal},
         year = {2001},
       volume = {556},
       number = {1},
        pages = {121-140},
          doi = {10.1086/321545},
archivePrefix = {arXiv},
       eprint = {astro-ph/0106324},
 primaryClass = {astro-ph},
       adsurl = {https://ui.adsabs.harvard.edu/abs/2001ApJ...556..121K}
}

@ARTICLE{kaufman,
       author = {{Kauffmann}, Guinevere and {Heckman}, Timothy M. and {Tremonti}, Christy and {Brinchmann}, Jarle and {Charlot}, St{\'e}phane and {White}, Simon D.~M. and {Ridgway}, Susan E. and {Brinkmann}, Jon and {Fukugita}, Masataka and {Hall}, Patrick B. and {Ivezi{\'c}}, {\v{Z}}eljko and {Richards}, Gordon T. and {Schneider}, Donald P.},
        title = "{The host galaxies of active galactic nuclei}",
      journal = {\mnras},
         year = {2003},
       volume = {346},
       number = {4},
        pages = {1055-1077},
          doi = {10.1111/j.1365-2966.2003.07154.x},
archivePrefix = {arXiv},
       eprint = {astro-ph/0304239},
 primaryClass = {astro-ph},
       adsurl = {https://ui.adsabs.harvard.edu/abs/2003MNRAS.346.1055K},
}

@article{stasinska,
    author = {Stasińska, G. and Costa-Duarte, M. V. and Vale Asari, N. and Cid Fernandes, R. and Sodré, L., Jr},
    title = {Retired galaxies: not to be forgotten in the quest of the star formation – AGN connection},
    journal = {Monthly Notices of the Royal Astronomical Society},
    volume = {449},
    number = {1},
    pages = {559-573},
    year = {2015},
    month = {03},
    issn = {0035-8711},
    doi = {10.1093/mnras/stv078},
    url = {https://doi.org/10.1093/mnras/stv078},
    eprint = {https://academic.oup.com/mnras/article-pdf/449/1/559/4122969/stv078.pdf},
}

@article{NUV2,
    author = {Johansson, Jonas and Woods, Tyrone E. and Gilfanov, Marat and Sarzi, Marc and Chen, Yan-Mei and Oh, Kyuseok},
    title = {Diffuse gas in retired galaxies: nebular emission templates and constraints on the sources of ionization},
    journal = {Monthly Notices of the Royal Astronomical Society},
    volume = {461},
    number = {4},
    pages = {4505-4516},
    year = {2016},
    month = {07},
   issn = {0035-8711},
    doi = {10.1093/mnras/stw1668},
    url = {https://doi.org/10.1093/mnras/stw1668},
    eprint = {https://academic.oup.com/mnras/article-pdf/461/4/4505/13774112/stw1668.pdf}
}

@ARTICLE{salpeter,
       author = {{Salpeter}, E.~E.},
        title = "{Accretion of Interstellar Matter by Massive Objects.}",
      journal = {\apj},
         year = {1964},
       volume = {140},
        pages = {796-800},
          doi = {10.1086/147973},
       adsurl = {https://ui.adsabs.harvard.edu/abs/1964ApJ...140..796S}
}

@ARTICLE{zel,
       author = {{Zel'dovich}, Ya. B.},
        title = "{The Fate of a Star and the Evolution of Gravitational Energy Upon Accretion}",
      journal = {Soviet Physics Doklady},
         year = {1964},
       volume = {9},
        pages = {195},
       adsurl = {https://ui.adsabs.harvard.edu/abs/1964SPhD....9..195Z}
}

@ARTICLE{agnDM,
       author = {{Risaliti}, G. and {Lusso}, E.},
        title = "{Cosmological Constraints from the Hubble Diagram of Quasars at High Redshifts}",
      journal = {Nature Astronomy},
         year = {2019},
       volume = {3},
        pages = {272-277},
          doi = {10.1038/s41550-018-0657-z}

}

@article{feedback,
   title={A flat trend of star formation rate with X-ray luminosity of galaxies hosting AGN in the SCUBA-2 Cosmology Legacy Survey},
   volume={486},
   ISSN={1365-2966},
   url={http://dx.doi.org/10.1093/mnras/stz1093},
   DOI={10.1093/mnras/stz1093},
   number={3},
   journal={Monthly Notices of the Royal Astronomical Society},
   publisher={Oxford University Press (OUP)},
   author={Ramasawmy, Joanna and Stevens, Jason and Martin, Garreth and Geach, James E},
   year={2019},
   month=apr, pages={4320–4333} }

@article{feedsim,
	author = {Rosito, M. S. and Pedrosa, S. E. and Tissera, P. B. and Chisari, N. E. and Domínguez-Tenreiro, R. and Dubois, Y. and Peirani, S. and Devriendt, J. and Pichon, C. and Slyz, A.},
	title = {The role of AGN feedback in the structure, kinematics, and evolution of ETGs in Horizon simulations},
	DOI= "10.1051/0004-6361/202039976",
	url= "https://doi.org/10.1051/0004-6361/202039976",
	journal = {A\& A},
	year = {2021},
	volume = {652},
	pages = "A44",
}

@article{coev,
author = {Heckman, Timothy M. and Best, Philip N.},
title = {The Coevolution of Galaxies and Supermassive Black Holes: Insights from Surveys of the Contemporary Universe},
journal = {Annual Review of Astronomy and Astrophysics},
volume = {52},
number = {1},
pages = {589-660},
year = {2014},
doi = {10.1146/annurev-astro-081913-035722}}

@article{Mezcua_2017,
   title={Observational evidence for intermediate-mass black holes},
   volume={26},
   ISSN={1793-6594},
   url={http://dx.doi.org/10.1142/S021827181730021X},
   DOI={10.1142/s021827181730021x},
   number={11},
   journal={International Journal of Modern Physics D},
   publisher={World Scientific Pub Co Pte Lt},
   author={Mezcua, Mar},
   year={2017},
   month=sep, pages={1730021} }

@misc{desigal,
  author = {DESI Collaboration},
  title = {desigal},
  url = {https://github.com/desihub/desigal?tab=readme-ov-file},
  version = {0.1.0 },
  year = {2017}
}

@article{harrison,
AUTHOR = {Harrison, Chris M. and Ramos Almeida, Cristina},
TITLE = {Observational Tests of Active Galactic Nuclei Feedback: An Overview of Approaches and Interpretation},
JOURNAL = {Galaxies},
VOLUME = {12},
YEAR = {2024},
NUMBER = {2},
ARTICLE-NUMBER = {17},
URL = {https://www.mdpi.com/2075-4434/12/2/17},
ISSN = {2075-4434},
DOI = {10.3390/galaxies12020017}
}

@article{Vogelsberger_2014,
   title={Properties of galaxies reproduced by a hydrodynamic simulation},
   volume={509},
   ISSN={1476-4687},
   url={http://dx.doi.org/10.1038/nature13316},
   DOI={10.1038/nature13316},
   number={7499},
   journal={Nature},
   publisher={Springer Science and Business Media LLC},
   author={Vogelsberger, M. and Genel, S. and Springel, V. and Torrey, P. and Sijacki, D. and Xu, D. and Snyder, G. and Bird, S. and Nelson, D. and Hernquist, L.},
   year={2014},
   month=may, pages={177–182} }

@inproceedings{
MMU,
title={The Multimodal Universe: Enabling Large-Scale Machine Learning with 100{TB} of Astronomical Scientific Data},
author={Eirini Angeloudi and Jeroen Audenaert and Micah Bowles and Benjamin M. Boyd and David Chemaly and Brian Cherinka and Ioana Ciuca and Miles Cranmer and Aaron Do and Matthew Grayling and Erin Elizabeth Hayes and Tom Hehir and Shirley Ho and Marc Huertas-Company and Kartheik G. Iyer and Maja Jablonska and Francois Lanusse and Henry W. Leung and Kaisey Mandel and Juan Rafael Mart{\'\i}nez-Galarza and Peter Melchior and Lucas Thibaut Meyer and Liam Holden Parker and Helen Qu and Jeff Shen and Michael J. Smith and Connor Stone and Mike Walmsley and John F Wu},
booktitle={The Thirty-eight Conference on Neural Information Processing Systems Datasets and Benchmarks Track},
year={2024},
url={https://openreview.net/forum?id=EWm9zR5Qy1}
}

%\newpage
%\hspace{1cm}
%\newpage
\begin{appendix}
\section{Performance metrics}
\label{sup:metrics}
In this work, the performance metrics are defined as follows:
\begin{itemize}
    \item {Accuracy:} The proportion of correct predictions, both true positives and true negatives, among all predictions:
\begin{equation}
    \text{Accuracy} = \frac{ TP + TN}{ TP + FP +  TN + FN}
\end{equation}

    where \( TP \), \( TN \), \( FP \), and \( FN \) denotes the number of true positives, true negatives, false positives, and false negatives, respectively.

    \item {Precision:} The proportion of true positives among all instances predicted as positive:
\begin{equation}
    \text{Precision} = \frac{TP}{ TP + FP}
\end{equation}

    \item {Recall (also known as true positive rate, TPR):} The proportion of true positives among all actual positive instances:
\begin{equation}
    \text{Recall} = \frac{TP}{ TP + FN}
\end{equation}

    \item {False positive rate (FPR):} The proportion of false positives among all actual negatives:
\begin{equation}
    \text{FPR} = \frac{ FP}{FP + TN}
\end{equation}

    \item {F1 score:} The harmonic mean of precision and recall, providing a balanced measure of both:
\begin{equation}
    F1 = 2 \times \frac{\text{precision} \times \text{recall}}{\text{precision} + \text{recall}}.
\end{equation}

    \item{Receiver Operating Characteristic (ROC) Curve:} A plot of the true positive rate against the false positive rate as the classification threshold is varied. The implementation details are provided in Appendix ~\ref{sec:sup_roc}.
    
    \item{Area Under the Curve (AUC):} A scalar metric representing the area under the ROC curve, which summarizes the model's performance across all classification thresholds. An AUC of 1 indicates perfect classification, while an AUC of 0.5 corresponds to random guessing.

\end{itemize}

\section{ROC curves}
\label{sec:sup_roc}
To construct the ROC curves, since the autoencoder does not provide direct probabilities of belonging to each class, we use the proportion of neighbors in the k-d tree voting for a given class as a proxy for class probability. For instance, if 6 of the neighbors are labeled AGN, we interpret this as a 60\% probability of being AGN. By varying the threshold on this estimated probability, we obtain different TPR–FPR pairs to construct the ROC curve. 

\begin{figure}
\centering
\begin{subfigure}{0.45\textwidth}
    \centering
    \includegraphics[width=\textwidth]{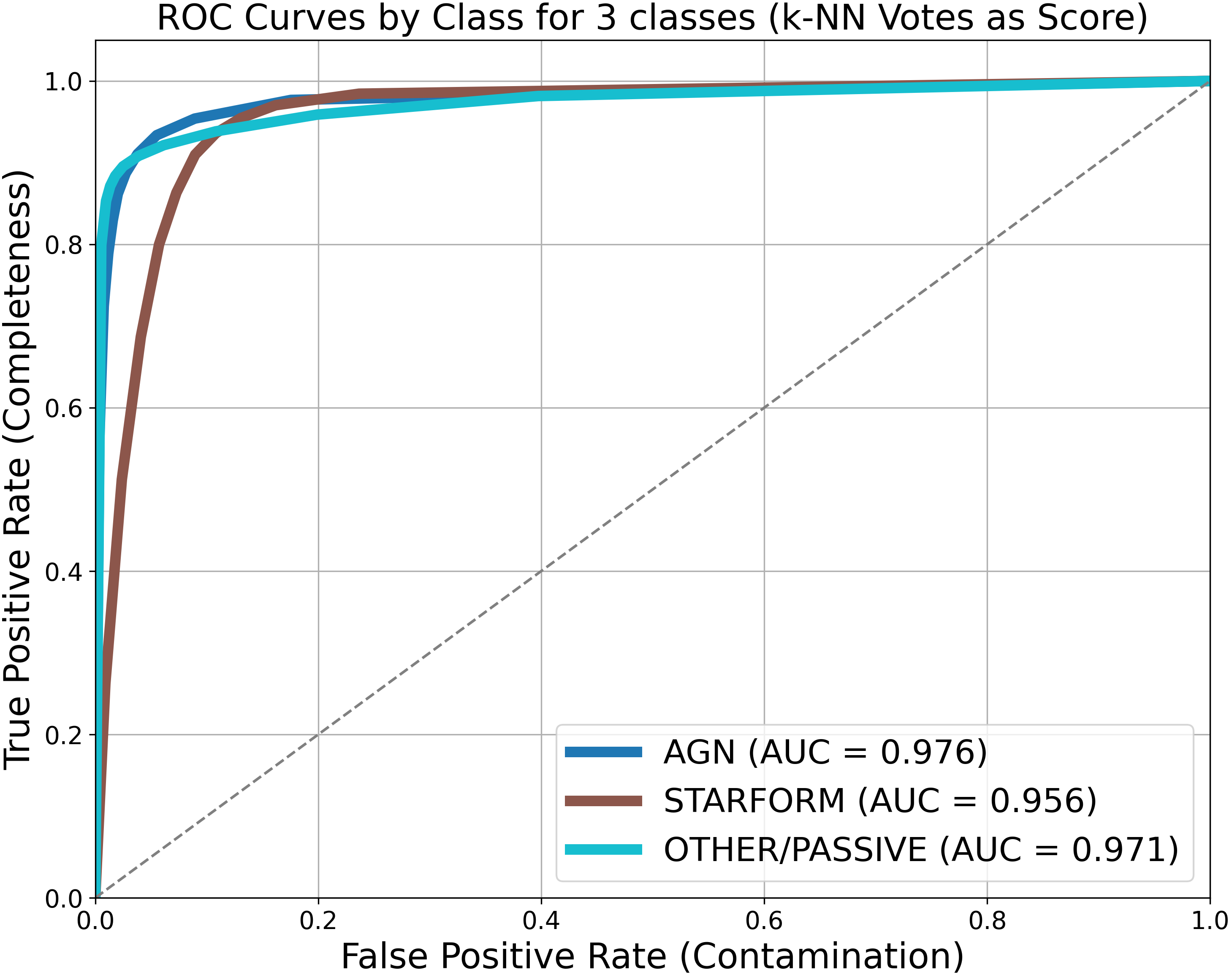}
    \caption{}
\end{subfigure}
\hfill
\begin{subfigure}{0.5\textwidth}
    \centering
    \includegraphics[width=\textwidth]{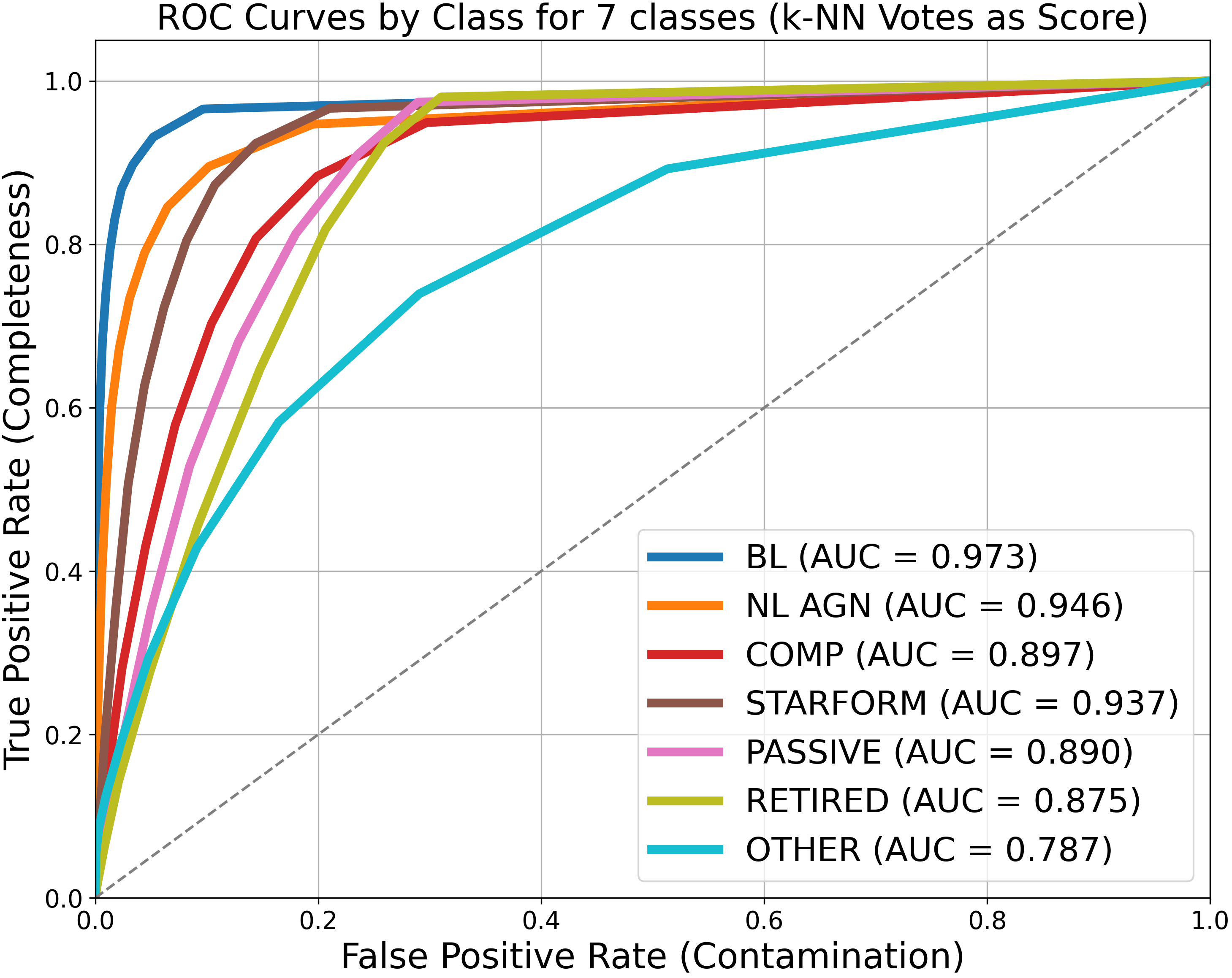}
    \caption{}
\end{subfigure}

\caption{ROC curves for the classifier predictions, with the AUC value for each class given in parentheses.
(a) ROC curves for the three galaxy classes: AGN, star-forming, and Other/passive.
(b) ROC curves for the extended galaxy classes: NL AGN, BL, star-forming, composite, retired, passive, and Other.}
\label{fig:roc_panels}
\end{figure}

In panel (a) of Figure \ref{fig:roc_panels}, which shows the ROC curves for three galaxy definitions, the performance in balancing completeness and contamination remains good across different thresholds for all classes, with the AGN curve performing slightly better (higher AUC). In panel (b), where the extended galaxy classification is used, there is more variation between classes. BL performs the best, followed by NL AGN and star-forming, while the Other class shows the curve closest to the diagonal. Passive and retired galaxies, which are often misclassified between each other, reach high completeness only at the expense of considerable contamination, with their ROC curves peaking further to the right compared to the other classes.

\section{S/N Limitations of BPT Diagnostics}
\label{sec:S/N}

Among the 286 sources  labeled as Other/passive but predicted as AGN examined in Section \ref{subsec:bpt}, 144 (50.3\%) fail the S/N $>3$ requirement in at least one of the four BPT lines. Notably, 121 of these 144 sources (84.0\%) fail the S/N threshold in either [O\,III] and/or H$\beta$, highlighting the dominant role of the vertical BPT axis. Individually, H$\beta$ is the most frequently limiting line, falling below the S/N threshold in 104 sources (36.4\%), followed by [O\,III] in 81 sources (28.3\%), [N\,II] in 72 sources (25.2\%), and H$\alpha$ in 61 sources (21.3\%). %This confirms that diagnostics relying on the [O\,III]/H$\beta$ ratio are particularly prone to incompleteness, even when the overall spectral quality is otherwise sufficient.

Several alternative optical diagnostic diagrams have been proposed to mitigate the limitations of the classical BPT, particularly those associated with the reliance on the [OIII]/H$\beta$ ratio. Among them, the WHAN diagram \citep{cid, cid2}, which combines the equivalent width of H$\alpha$ with the [NII]/H$\alpha$ line ratio, was developed explicitly to address the large population of emission-line galaxies left unclassified by BPT selections. By requiring only two emission lines and incorporating equivalent width information, WHAN is more sensitive to weak AGN and low-luminosity systems, for which H$\beta$ is often undetectable or intrinsically weak \citep{Cann_2019, sanchez}.

However, studies comparing BPT and WHAN classifications have shown that while most BPT-selected AGN are also identified as AGN in WHAN, the converse is not true: a substantial fraction of WHAN-selected AGN are classified as star-forming or composite in the BPT diagram \citep[e.g.,][]{Kristensen}. This does not imply that the WHAN classification is unreliable; rather, it reflects the fact that WHAN is designed to probe weaker or diluted AGN, which may be missed by the more stringent four-line requirements of the BPT. Such systems can be affected by star-formation dilution, intrinsically weak emission lines, or environmental effects, all of which tend to suppress classical AGN signatures \citep{cid, cid2, sanchez}.

Previous studies have shown that a significant fraction of AGN cannot be robustly classified using BPT diagnostics even when the spectroscopy is sufficiently deep, as a result of intrinsically weak or obscured narrow-line regions, radiatively inefficient accretion flows, or short AGN duty cycles; indeed, \citet{Agostino_2019} find that $\sim$40\% of X-ray–selected AGN exhibit emission lines too weak to be placed on the BPT diagram and are often associated with low specific SFRs.

The above considerations highlight that no single optical diagnostic diagram can provide a complete and unbiased AGN census \citep{Juneau2013, AGN, sanchez}. This motivates the use of approaches that exploit the full spectral information, rather than relying on a small number of emission-line ratios.

\section{Stack of sources with discrepant prediction–label combinations}

\begin{figure*}
\centering
 \includegraphics[width=\hsize]{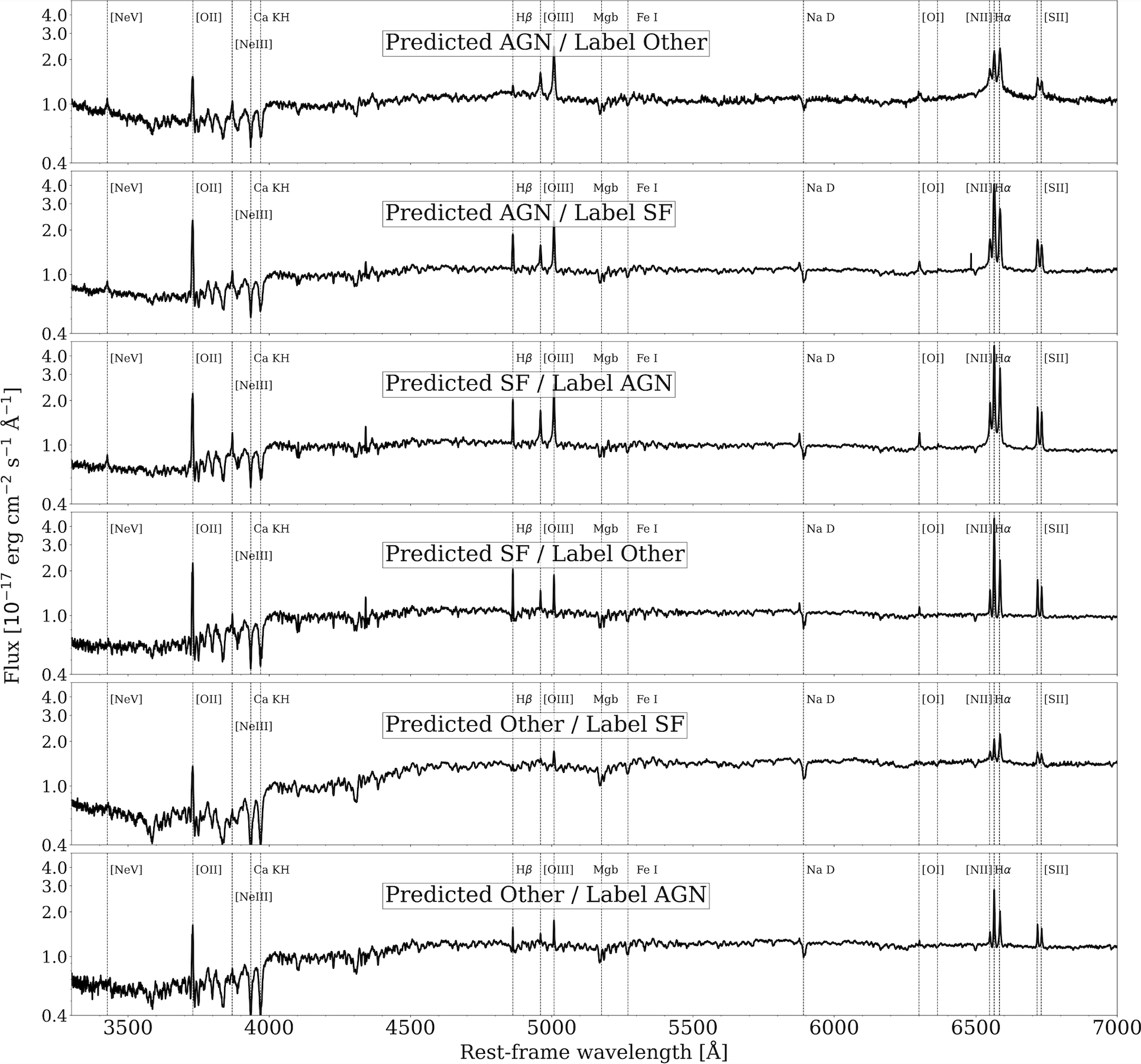}
\caption{Stacked rest-frame optical spectra corresponding to the different sources with discrepant prediction–label combinations analyzed in this work.
Key emission and absorption features are indicated with vertical dashed lines.}
 \label{fig:stack}
\end{figure*}

\begin{figure}[ht]
\centering
 \includegraphics[width=\hsize]{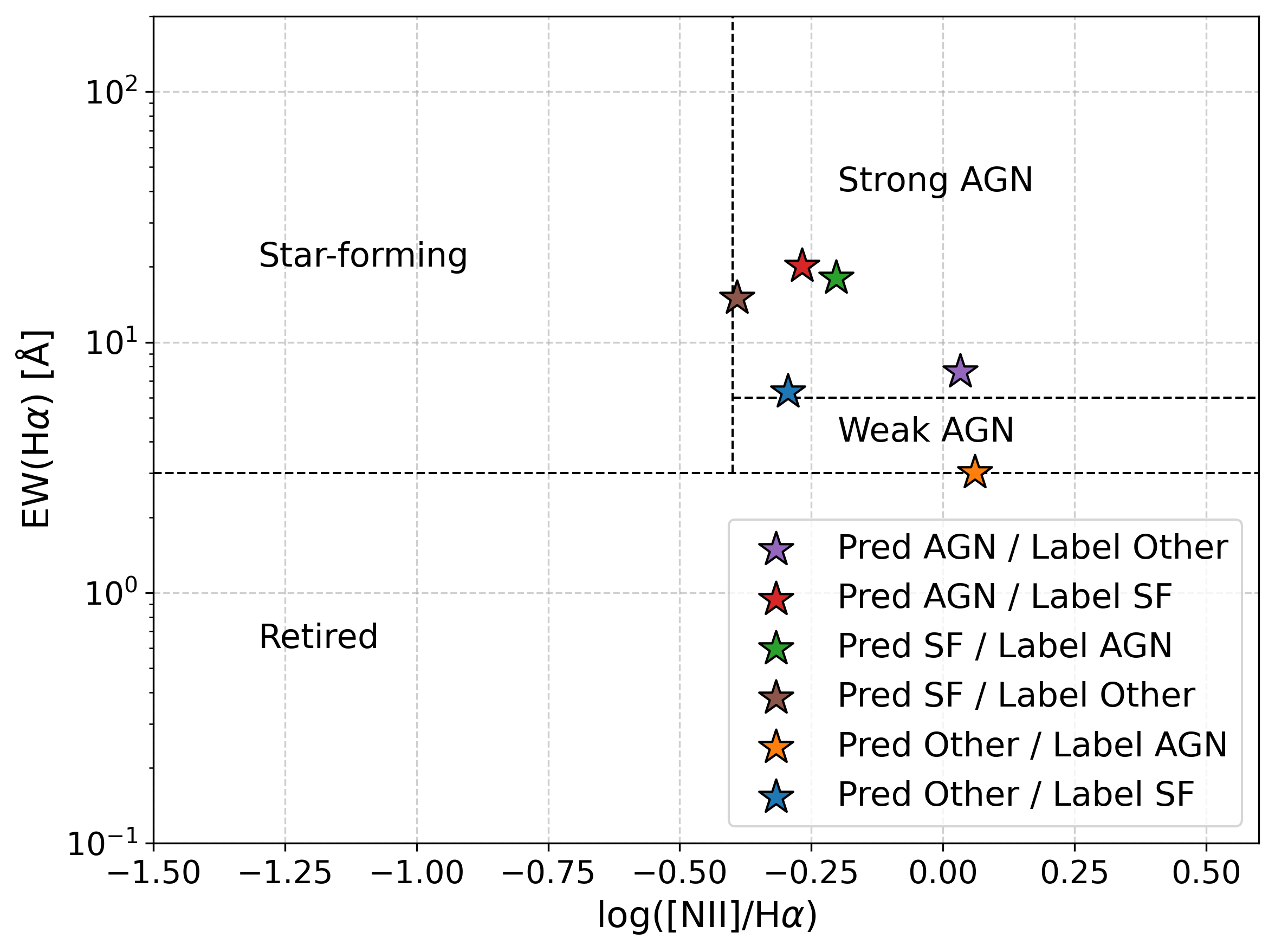}
\caption{Location on the WHAN diagram of the stacked spectra corresponding to sources with discrepant classifications.
Each symbol represents the emission-line measurements derived from the \software{FastSpecFit} model of the corresponding stacked spectrum. The classical demarcation lines separating star-forming, strong and weak AGN, and retired galaxies are shown for reference.
}

 \label{fig:whan}
\end{figure}

Building on the analysis presented in Section~\ref{subsec:bpt}, this section extends the study to all discrepant prediction--label combinations. While the main text focuses on sources predicted as AGN but labeled as Other/passive, here we examine the remaining cases. Figure~\ref{fig:stack} shows the stacked spectra for each off-diagonal population in the confusion matrix. To complement the BPT-based classification discussed in the main text (Figure~\ref{pic:bpt}), we also place the line measurements of the stacked spectra on the WHAN diagram (Fig.~\ref{fig:whan}), which incorporates H$\alpha$ equivalent width and enables a more robust distinction between weak AGN and retired galaxies in low-ionization regimes. Table~\ref{tab:stack_summary} summarizes the classification of each stack according to both diagnostics. The following subsections examine each prediction--label combination in detail, providing a consistent interpretation of their physical nature.
\label{sec:stack}

\begin{table*}
\centering
\begin{tabular}{l l l l p{6cm}}
\toprule
Original label & Predicted & BPT (stack) & WHAN (stack) & Notes \\
\midrule\midrule
AGN & Other/passive & LINER & Weak AGN--Retired  &
Line ratios consistent with LINER-like ionization; low H$\alpha$ equivalent width places the stack close to the retired region in the WHAN diagram. \\

AGN & Star-forming & \makecell[l]{Seyfert \\ (almost LINER/composite)} & Strong AGN &
Clear AGN signatures in both diagnostics despite the star-forming prediction; high equivalent width indicates ongoing nuclear activity. \\

Star-forming & Other/passive & Composite & \makecell[l]{Strong AGN \\(almost weak)} &
Composite location in the BPT diagram but classified as strong AGN in WHAN. \\

Star-forming & AGN & Composite &  Strong AGN  &
Classified as composite by the BPT but robust AGN classification in WHAN, likely driven by high H$\alpha$ equivalent width. \\

Other/passive & AGN & Seyfert & Strong AGN &
Clear Seyfert classification in both BPT and WHAN; the presence of a broad H$\alpha$ component confirms AGN activity. \\

Other/passive & Star-forming & \makecell[l]{Star-forming \\(almost composite)} & \makecell[l]{Strong AGN\\ (almost SF)} &
Located close to the star-forming/composite boundary; the WHAN classification suggests a transition regime between star formation and AGN activity. \\
\bottomrule
\end{tabular}
\caption{Summary of the stacked spectra corresponding to sources with discrepant prediction--label combinations. For each stack, we report the original label, the model prediction, and the classification inferred from the BPT and WHAN diagnostic diagrams applied to the stacked spectrum.}
\label{tab:stack_summary}
\end{table*}

\subsection{Predicted as Other/passive - labeled as AGN}
\label{sup:passive-AGN}
For sources originally labeled as AGN but predicted as Other/passive by our model, the stacked spectrum lies in the LINER region of the [NII]-BPT diagram, while its location in the WHAN diagram corresponds to the weak-AGN regime, very close to the retired boundary. In particular, the measured H$\alpha$ equivalent width of the stacked spectrum is ${\rm EW(H\alpha)} \simeq 3.0$\AA, which marks the transition between weak AGN and retired galaxies.

This indicates that a fraction of the sources in this population may not host ongoing nuclear activity, but are instead ionized by old stellar populations. As emphasized by \citet{stasinska}, emission-line ratios alone can misclassify retired galaxies as LINERs if line equivalent widths are not taken into account, as the ionization in these systems is driven by hot low-mass evolved stars (HOLMES). By explicitly incorporating equivalent widths, the WHAN diagram provides a more robust classification in this low-ionization regime. In this context, the model prediction may effectively reduce contamination of the AGN class by retired galaxies.

\subsection{Predicted as star-forming - labeled as Other/passive}
Similarly, stacking the 2011 sources predicted as star-forming by our model (including both star-forming and composite classes) but labeled as Other/passive, and placing the resulting spectrum on the [NII]-BPT diagram (Figure~\ref{fig:stack}), shows that the stack lies firmly within the star-forming region. Stacking effectively recovers the characteristic star-formation-driven line ratios that are buried in the individual low-S/N spectra. 
The stacked spectrum is located very close to the \citet{kewely} demarcation line separating star-forming and composite regions, indicating that, on average, these sources occupy the transition between pure star formation and composite excitation. 

In the WHAN diagram, the stack is placed in the strong AGN but falls close to the star-forming boundary. As discussed by \citet{sanchez}, late-type galaxies that are genuinely dominated by star formation can be classified as AGN in equivalent-width-based diagnostics under certain conditions, despite their overall ionization being driven primarily by star formation. Taken together, the BPT and WHAN diagnostics suggest that this population is predominantly star-forming, consistent with the model prediction.

\subsection{Predicted as star-forming - labeled as AGN}

The stacked spectrum of galaxies predicted as star-forming but labeled as AGN shows clear AGN signatures in both diagnostic diagrams. In the [NII]-BPT diagram, the stack lies very close to the boundaries separating the Seyfert, LINER, and composite regions, while the WHAN diagram classifies it as a strong AGN. It is important to note that our star-forming class also includes composite galaxies; therefore, part of this apparent disagreement may reflect sources with mixed ionization. 

\subsection{Predicted as AGN - labeled as star-forming}

For sources originally labeled as star-forming but predicted as AGN by our model, the stacked spectrum occupies the composite region of the [NII]-BPT diagram, while the WHAN diagram classifies it as a strong AGN. This behavior is consistent with well-known limitations of the BPT diagram, which can be biased against AGN identification when narrow-line AGN signatures are overpowered by host galaxy star formation. This effect is particularly relevant in low-mass, blue galaxies with high specific star-formation rates, where AGN emission lines can be diluted by stellar processes \citep{Agostino_2019, trump}. In addition, BPT line ratios are sensitive to metallicity, which can further bias the classification toward the star-forming or composite regions in such systems \citep{StorchiBergmann1998, Carvalho}.

Therefore, despite their composite BPT classification, these sources are likely genuine AGN. This interpretation is supported not only by the WHAN classification of the stacked spectrum, but also by the additional multi-diagnostic analysis presented in Section~\ref{subsec:vac}, where several independent indicators consistently favor an AGN interpretation.

\subsection{Predicted as Other/passive - labeled as star-forming}
For sources originally labeled as star-forming and predicted as Other/passive by our model, the stacked spectrum lies in the composite region of the [NII]-BPT diagram, while the WHAN diagram places it close to the strong-AGN regime. However, this stacked spectrum shows the weakest emission-line strengths among all off-diagonal populations, indicating that the classification is intrinsically uncertain. This population likely represents a transition regime, where low-level ionization and star formation coexist, and small differences in line strength can lead to discrepant diagnostic classifications.

\section{BL signatures in the stacked spectra}
\label{sec:broad_stack}

\begin{figure*}
\centering
 \includegraphics[width=\hsize]{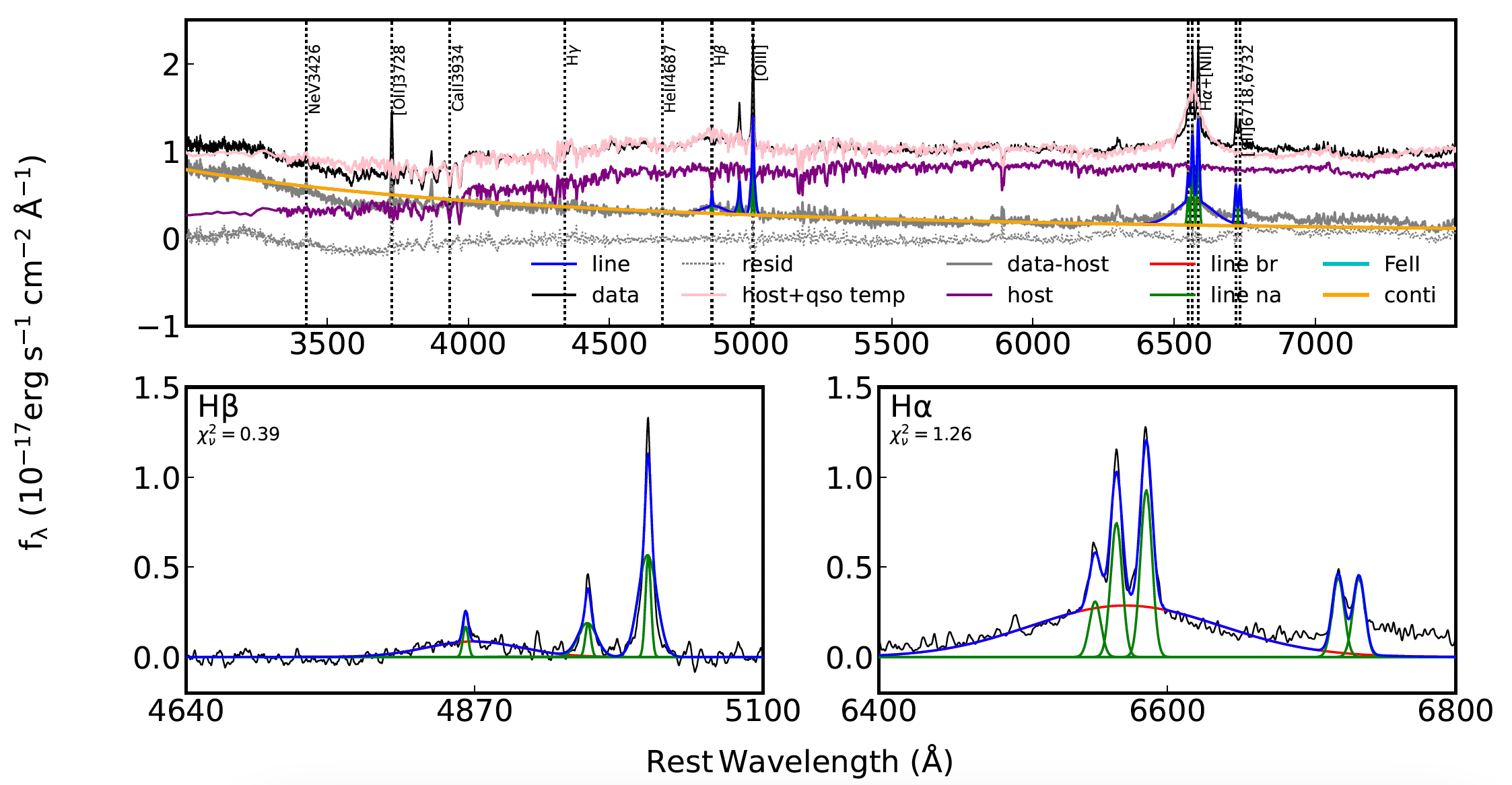}
\caption{Top panel: stacked rest-frame optical spectrum of the 286 sources predicted as AGN by our model but labeled as Other/passive by traditional diagnostics. Bottom panels: zoom-in views of the H$\beta$ (left) and H$\alpha$ (right) regions, with the \software{PyQSOFit} model overplotted. The fits highlight the separation between narrow and broad emission-line components, providing independent evidence for the presence of BL emission in the stacked spectrum.
}

 \label{fig:broad_fit}
\end{figure*}

To illustrate the presence of broad emission-line components in the stacked spectra of galaxies predicted as AGN but labeled as Other/passive, we performed a multi-Gaussian decomposition using the \software{PyQSOFit} package \citep{pyqsofit}. While the spectral measurements used throughout this work are obtained with \software{FastSpecFit}, which provides automated emission-line fits for the full sample, \software{PyQSOFit} is employed here solely to perform a more detailed decomposition of the stacked spectrum and to visually separate the narrow and broad components of the Balmer lines in Fig.~\ref{fig:broad_fit}. \software{PyQSOFit} is specifically designed for AGN spectral modeling and allows simultaneous fitting of narrow and broad components, making it well suited for illustrating the line-profile structure in stacked spectra.

The inclusion of a broad component is required to reproduce the extended wings of the Balmer lines and leads to a significantly improved fit, as indicated by the reduced $\chi^2$ values shown in the figure. The resulting broad component exceeds the typical BL AGN threshold of FWHM of $\sim$1200 km\,s$^{-1}$, consistent with the presence of BL emission in the stacked spectrum.

However, we do not attempt to derive precise kinematic measurements such as an exact FWHM or black hole masses from the stacked spectrum. Stacking spectra with different systemic velocities and intrinsic line widths can artificially broaden strong emission lines (such as H$\alpha$) while simultaneously suppressing the wings of weaker lines (such as H$\beta$), potentially biasing quantitative measurements. Therefore, the Gaussian decomposition is presented here only to illustrate the presence of broad emission, rather than to provide detailed BL region properties.

\section{Validation of BL Classifications with the DESI AGN/Galaxy VAC}
\label{app:bl_vac}

To further investigate discrepancies between predicted and reference BL classifications, we cross-matched these sources with the AGN/Galaxy VAC of DESI DR1, which identifies BL sources using broader criteria than those adopted in this work, considering multiple emission lines (H$\alpha$, H$\beta$, and/or MgII) with FWHM $\geq$ 1200 km s$^{-1}$.

Table~\ref{tab:bl_misclass} compares reference labels, model predictions, and the BL flag in the AGN/Galaxy VAC. Among the 906 sources predicted as BL by our model but not labeled as such, 530 (58\%) are identified as BL in the AGN/Galaxy VAC. Similarly, of the 585 sources labeled as NL AGN but predicted as BL, 364 (62\%) are classified as BL in the AGN/Galaxy VAC.

These discrepancies likely arise from the more restrictive BL identification criteria adopted in our labeling scheme, which rely on the AON and S/N of the fitted broad and narrow components only of H$\alpha$ (see Section~\ref{sub:class}), whereas the AGN/Galaxy VAC incorporates broader line-width criteria and multiple emission-line diagnostics. The high level of agreement with the AGN/Galaxy VAC therefore supports the ability of our model to recover BL sources beyond the restrictive definitions adopted in the reference labeling.

\begin{table*}[ht]
  \centering
  \begin{tabular}{l c c}
    \toprule
    Category & Number of Sources & Of which BL in VAC \\
    \midrule
    Labeled as BL but not predicted as BL & 867 & 648 (75\%) \\
    Not labeled as BL but predicted as BL & 906 & 530 (59\%) \\
    \midrule\midrule
    Labeled as BL, predicted as NL AGN & 643 & 505 (79\%) \\
    Labeled as NL AGN, predicted as BL & 585 & 364 (62\%) \\
    \bottomrule
  \end{tabular}
\caption{Comparison between reference labels, model predictions, and the BL flag in the DESI DR1 AGN/Galaxy VAC. The lower block highlights mismatches between BL and NL AGN classifications.}
  \label{tab:bl_misclass}
\end{table*}

\end{appendix}
\end{document}